\documentclass[a4paper,11pt]{article}
\pdfoutput=1 

\usepackage[table, dvipsnames, usenames]{xcolor}
\usepackage{jheppub} 
\usepackage{tikz}
\usepackage{tikz-feynhand}
\usepackage{tikz-feynman}
\usetikzlibrary{arrows.meta,positioning,shadows,calc,shapes.symbols,decorations.markings}
\usepackage{cancel}
\usetikzlibrary{decorations.markings,decorations.pathmorphing}

\usepackage[skins]{tcolorbox}

\usepackage[T1]{fontenc}
 
\usepackage{microtype}
\usepackage{amssymb} 
\usepackage{amsmath}
\usepackage{mathtools}
\usepackage{amsfonts}    
\usepackage{dsfont}
\usepackage{pdfpages}
\usepackage{verbatim}
\usepackage{braket}
\usepackage{bm}
\usepackage{tensor}
\usepackage{upgreek}
\usepackage{subcaption}
\usepackage{simpler-wick}
\usepackage{enumitem}
\usepackage{amsthm}
\usepackage{tcolorbox}
\usepackage{nicematrix}
\usepackage{fontawesome}
\usepackage{framed}
\usepackage{booktabs}
\usepackage{enumitem}
\usepackage{orcidlink}

\usepackage{extarrows}   
\newlength{\arrowlabelwd}
\AtBeginDocument{\settowidth{\arrowlabelwd}{$\scriptstyle\mathrm{propagator}$}}
\newcommand{\xto}[1]{\xlongrightarrow{\mathmakebox[\arrowlabelwd]{#1}}}

\usepackage{listings}
\usetikzlibrary{decorations.markings}

\usepackage{pgfplots}
\usepgfplotslibrary{fillbetween}
\pgfplotsset{compat=1.18}

\usepackage{tocloft}
\usepackage[multiple,hang,flushmargin]{footmisc}
\usepackage{fnpct}

\definecolor{rosewood}{rgb}{0.4, 0.0, 0.04}
\definecolor{pyblue}{RGB}{31, 119, 180}
\definecolor{pyorange}{RGB}{255, 127, 14}
\definecolor{pygreen}{RGB}{44, 160, 44}
\definecolor{pyred}{RGB}{214, 39, 40}
\definecolor{lightgray}{gray}{0.9}
\colorlet{lightpyblue}{pyblue!30!white}
\colorlet{lightpyred}{pyred!30!white}
\colorlet{lightpygreen}{pygreen!30!white}

\definecolor{lmugreen}{RGB}{0, 136, 58}
\definecolor{lmugreenbg}{RGB}{204, 231, 216}

\definecolor{lmuorange}{RGB}{241, 135, 0}
\definecolor{lmuorangebg}{RGB}{248, 207, 153}

\definecolor{lmublue}{RGB}{15, 25, 136}

\definecolor{lmugray_mid}{RGB}{192, 193, 195}

\usepackage{hyperref}

\def \d {\mathrm{d}}

\def \p {\bm{p}}
\def \k {\bm{k}}
\def \x {\bm{x}}
\def \s {\bm{s}}

\def \C {\mathcal{C}}

\def \E {\mathcal{E}}

\def \I {\mathcal{I}}

\def \L {\mathcal{L}}
\def \M {\mathcal{M}}

\def \R {\mathbb{R}}

\def \V {\mathcal{V}}

\def \Z {\mathcal{Z}}

\def \aa {{\sf{a}}}
\def \bb {{\sf{b}}}

\def \dd {{\sf{d}}}

\def \Im {\mathrm{Im}}
\def \Re {\mathrm{Re}}
\def \Disc {\mathrm{Disc}}

\colorlet{rulegrey}{black!30}
\tikzset{
  rule/.style={x=1.2cm, y=1.2cm, line cap=round, baseline=(current bounding box.center)},
  rule vtx/.style={circle, fill=#1, inner sep=0pt, minimum size=2.6pt},
  rule vtx/.default=black,
  rule meet/.style={circle, draw=#1, fill=white, semithick, inner sep=0pt,
    minimum size=4.6pt,
    path picture={\draw[#1]
      (path picture bounding box.north west) -- (path picture bounding box.south east)
      (path picture bounding box.south west) -- (path picture bounding box.north east);}},
  rule meet/.default=black,
  dressing/.style={dash pattern=on 2pt off 1.3pt},
  dots/.style={line width=0.6pt, dash pattern=on 0pt off 1.15pt},
  rule lab/.style={font=\scriptsize, inner sep=1.5pt},
  rule flow/.style={postaction=decorate, decoration={markings,
      mark=at position 0.56 with {\arrow[#1]{Stealth[length=4pt,width=4pt]}}}},
}
\newcommand{\ruleDd}{0.9pt}   
\newcommand{\ruleDress}[3]{%
  \draw[line width=0.45pt, dressing, #3]
    ($(#1)!\ruleDd!90:(#2)$) -- ($(#2)!\ruleDd!-90:(#1)$);
  \draw[line width=0.45pt, dots, #3]
    ($(#1)!\ruleDd!-90:(#2)$) -- ($(#2)!\ruleDd!90:(#1)$);}
 
\newcommand{\ruleDressCurve}[5]{%
  \draw[line width=0.45pt, dressing, #5]
    let \p1=($(#2)-(#1)$), \n1={0.3915*veclen(\x1,\y1)} in
    ($(#1)+(#3+90:\ruleDd)$)
    .. controls ($(#1)+(#3:\n1)+(#3+90:\ruleDd)$)
            and ($(#2)+(#4:\n1)+(#4-90:\ruleDd)$) ..
    ($(#2)+(#4-90:\ruleDd)$);
  \draw[line width=0.45pt, dots, #5]
    let \p1=($(#2)-(#1)$), \n1={0.3915*veclen(\x1,\y1)} in
    ($(#1)+(#3-90:\ruleDd)$)
    .. controls ($(#1)+(#3:\n1)+(#3-90:\ruleDd)$)
            and ($(#2)+(#4:\n1)+(#4+90:\ruleDd)$) ..
    ($(#2)+(#4+90:\ruleDd)$);}
 
\newcommand{\ruleSetup}{%
  \coordinate (v1) at (-0.6,0);    \coordinate (v2) at (0.6,0);
  \coordinate (e1) at (-1.1, 0.45); \coordinate (e2) at (-1.1,-0.45);
  \coordinate (e3) at ( 1.1, 0.45); \coordinate (e4) at ( 1.1,-0.45);
  \coordinate (m)  at (0,1.15);}

\providecommand{\rulecol}{lmugreen}
\newcommand{\ruleDiagram}[1]{%
  \begin{tikzpicture}[rule]
    \ruleSetup
    \ifcase#1\or
      \def\cint{\rulecol}\def\cext{black}\def\cvtx{black}\def\cdr{}
    \or
      \def\cint{rulegrey}\def\cext{rulegrey}\def\cvtx{rulegrey}\def\cdr{\rulecol}
    \or
      \def\cint{rulegrey}\def\cext{\rulecol}\def\cvtx{\rulecol}\def\cdr{rulegrey}
    \else
      \def\cint{black}\def\cext{black}\def\cvtx{black}\def\cdr{black}
    \fi
    \ifnum#1>1
      \ruleDress{v1}{m}{\cdr}  \ruleDress{m}{v2}{\cdr}
      \ifnum#1<4
        \path[rule flow=\cdr] (v1) -- (m);  \path[rule flow=\cdr] (v2) -- (m);
      \fi
      \ifnum#1=4
        \path[rule flow=\rulecol] (v1) -- (m);  \path[rule flow=\rulecol] (v2) -- (m);
      \fi
      \node[rule meet=\cdr] at (m) {};
    \fi
    \draw[semithick, \cext] (e1) -- (v1) -- (e2)  (e3) -- (v2) -- (e4);
    \draw[semithick, \cint, rule flow=\cint] (v1) -- (v2);
    \node[rule vtx=\cvtx] at (v1) {};  \node[rule vtx=\cvtx] at (v2) {};
    \ifcase#1\or
      \node[rule lab, \rulecol, below=1pt] at (0,0) {$Q=(p,\vec q)$};
    \or
      \node[rule lab, \rulecol, left=3pt]  at ($(m)!0.45!(v1)$) {$P_1$};
      \node[rule lab, \rulecol, right=3pt] at ($(m)!0.45!(v2)$) {$P_2$};
    \or
      \foreach \e/\a/\i in {e1/left/1,e2/left/2,e3/right/3,e4/right/4}
        \node[rule lab, \rulecol, \a] at (\e) {$k_\i$};
      \node[rule lab, \rulecol, below=2pt] at (v1) {$\lambda$};
      \node[rule lab, \rulecol, below=2pt] at (v2) {$\lambda$};
    \else
      \foreach \e/\a/\i in {e1/left/1,e2/left/2,e3/right/3,e4/right/4}
        \node[rule lab, \a] at (\e) {$\vec k_\i$};
      \node[rule lab, below=1pt] at (0,0) {$Q=(p,\vec q)$};
      \node[rule lab, \rulecol, left=3pt]  at ($(m)!0.3!(v1)$) {$P_1=-p$};
      \node[rule lab, \rulecol, right=3pt] at ($(m)!0.3!(v2)$) {$P_2=p$};
      \node[rule lab, \rulecol, below] at (0,-0.78)
        {$\vec q=\vec k_1+\vec k_2=-\vec k_3-\vec k_4$};
      \foreach \v in {v1,v2}
        \draw[\rulecol, thick] (\v) circle (3.3pt);
    \fi
  \end{tikzpicture}%
}

\newcommand{\ruleLegs}[1]{\draw[semithick, #1] (e1) -- (v1) -- (e2)  (e3) -- (v2) -- (e4);}
\newcommand{\ruleVtcs}[1]{\node[rule vtx=#1] at (v1) {};  \node[rule vtx=#1] at (v2) {};}
\newcommand{\ruleProp}[1]{\draw[semithick, #1, rule flow=#1] (v1) -- (v2);}

\newcommand{\ruleCut}[2]{%
  \draw[semithick, #1] (v1) -- (v2);
  \path[rule flow=#2] (0,0) -- (v1);  \path[rule flow=#2] (0,0) -- (v2);
  \draw[line width=1.1pt, #1] (0,-0.2) -- (0,0.2);}

\newcommand{\ruleMomArrow}[1]{%
  \draw[#1, -{Stealth[length=3.4pt,width=3.4pt]}] (-0.25,-0.3) -- (0.25,-0.3)
    node[rule lab, midway, below=1pt] {$\vec q$};}

\newcommand{\ruleDressing}[2]{%
  \ruleDress{v1}{m}{#1}  \ruleDress{m}{v2}{#1}
  \path[rule flow=#2] (v1) -- (m);  \path[rule flow=#2] (v2) -- (m);
  \node[rule meet=#1] at (m) {};}

\newcommand{\ruleGenDiagram}[1]{%
  \ifcase#1\or
  
  \begin{tikzpicture}[rule, x=1cm, y=1cm]
    \foreach \y in {0.75,-0.75}{\begin{scope}[yshift=\y cm]
      \ruleSetup \ruleLegs{black}
      \draw[semithick, rule flow=black] (v1) -- (v2);
      \node[rule lab, below=1pt] at (0,0) {$\vec q$};
      \ruleVtcs{black}
    \end{scope}}
    \node at (0,0) {$+$};
    \node[rule lab, \rulecol] at (0,-1.5) {$2^{L}=2$ copies};
  \end{tikzpicture}%
  \or
  
  \begin{tikzpicture}[rule, x=1.3cm, y=1.3cm]
    \begin{scope}[yshift=0.9cm]
      \ruleSetup \ruleLegs{rulegrey} \ruleProp{\rulecol} \ruleVtcs{rulegrey}
      \node[rule lab, \rulecol, below=1pt] at (0,0) {$Q=(p,\vec q)$};
    \end{scope}
    \node at (0,0) {$+$};
    \begin{scope}[yshift=-0.9cm]
      \ruleSetup \ruleLegs{rulegrey} \ruleCut{\rulecol}{\rulecol} \ruleVtcs{rulegrey}
      \node[rule lab, \rulecol, above=1pt] at (-0.3,0) {$p$};
      \node[rule lab, \rulecol, above=1pt] at ( 0.3,0) {$p$};
      \ruleMomArrow{\rulecol}
    \end{scope}
  \end{tikzpicture}%
  \or
  
  \begin{tikzpicture}[rule, x=1.1cm, y=1.1cm]
    \begin{scope}[yshift=1.25cm]
      \ruleSetup
      \foreach \v in {v1,v2}
        \filldraw[fill=\rulecol!15, draw=\rulecol, thin] (\v) circle (0.17);
      \ruleDressing{\rulecol}{\rulecol}
      \ruleLegs{rulegrey} \ruleProp{rulegrey} \ruleVtcs{rulegrey}
      \node[rule lab, \rulecol] at (-0.6,-0.33) {$c_1$};
      \node[rule lab, \rulecol] at ( 0.6,-0.33) {$c_2$};
      \node[rule lab, \rulecol] at (0,-0.72) {$\mathcal D_{c_1}\mathcal D_{c_2}$};
    \end{scope}
    \node at (0,-0.07) {$+$};
    \begin{scope}[yshift=-1.65cm]
      \ruleSetup
      \filldraw[fill=\rulecol!15, draw=\rulecol, thin, rounded corners=1.7mm]
        (-0.78,-0.17) rectangle (0.78,0.17);
      \ruleDressing{\rulecol}{\rulecol}
      \ruleLegs{rulegrey} \ruleCut{rulegrey}{rulegrey} \ruleVtcs{rulegrey}
      \node[rule lab, \rulecol] at (0,-0.33) {$c$};
      \node[rule lab, \rulecol] at (0,-0.72) {$\mathcal D_{c}$};
    \end{scope}
  \end{tikzpicture}%
  \or
  
  \begin{tikzpicture}[rule]
    \ruleSetup
    \ruleDressing{rulegrey}{rulegrey}
    \ruleLegs{\rulecol} \ruleCut{rulegrey}{rulegrey} \ruleVtcs{\rulecol}
    \foreach \e/\a/\i in {e1/left/1,e2/left/2,e3/right/3,e4/right/4}
      \node[rule lab, \rulecol, \a] at (\e) {$k_\i$};
    \node[rule lab, \rulecol, below=2pt] at (v1) {$\lambda$};
    \node[rule lab, \rulecol, below=2pt] at (v2) {$\lambda$};
  \end{tikzpicture}%
  \else
 
  \begin{tikzpicture}[rule]
    \ruleSetup
    \ruleDressing{black}{\rulecol}
    \ruleLegs{black} \ruleCut{black}{black} \ruleVtcs{black}
    \foreach \e/\a/\i in {e1/left/1,e2/left/2,e3/right/3,e4/right/4}
      \node[rule lab, \a] at (\e) {$\vec k_\i$};
    \node[rule lab, above=1pt] at (-0.3,0) {$p$};
    \node[rule lab, above=1pt] at ( 0.3,0) {$p$};
    \ruleMomArrow{black}
    \node[rule lab, \rulecol, left=3pt]  at ($(m)!0.3!(v1)$) {$P_1=p$};
    \node[rule lab, \rulecol, right=3pt] at ($(m)!0.3!(v2)$) {$P_2=p$};
    \node[rule lab, \rulecol, above=2pt] at (m) {$P_1+P_2=2p$};
    \foreach \v in {v1,v2}
      \draw[\rulecol, thick] (\v) circle (3.3pt);
    \node[rule lab, \rulecol, below] at (0,-0.78)
      {$\vec q=\vec k_1+\vec k_2=-\vec k_3-\vec k_4$};
  \end{tikzpicture}%
  \fi}

\tikzset{amp lab/.style={font=\small, inner sep=1.5pt}}
\newcommand{\ampDiagram}[1]{%
  \begin{tikzpicture}[rule, x=1.9cm, y=1.9cm, baseline={(0,0)},
      rule vtx/.append style={minimum size=3.4pt}]
    \def\ruleDd{1.25pt}
    \ruleSetup
    \ruleDressing{black}{black}
    \ruleLegs{black}
    \ifnum#1=1
      \ruleProp{black}
      \node[amp lab, above=1pt] at (0,0) {$p$};
      \node[amp lab, left=3pt]  at ($(m)!0.35!(v1)$) {$-p$};
      \node[amp lab, right=3pt] at ($(m)!0.35!(v2)$) {$p$};
    \else
      \ruleCut{black}{black}
      \node[amp lab, above=1pt] at (-0.3,0) {$p$};
      \node[amp lab, above=1pt] at ( 0.3,0) {$p$};
      \node[amp lab, left=3pt]  at ($(m)!0.35!(v1)$) {$p$};
      \node[amp lab, right=3pt] at ($(m)!0.35!(v2)$) {$p$};
    \fi
    
    \draw[-{Stealth[length=4pt,width=4pt]}] (-0.2,-0.3) -- (0.2,-0.3)
      node[amp lab, midway, below=1pt] {$\vec q$};
    \foreach \e/\a/\i in {e1/left/1,e2/left/2,e3/right/3,e4/right/4}
      \node[amp lab, \a] at (\e) {$\vec k_\i$};
    \ruleVtcs{black}
  \end{tikzpicture}}

\newcommand{\ruleShadowDiagram}[1]{%
  \ifcase#1\or
  
  \begin{tikzpicture}[rule, x=1cm, y=1cm]
    \foreach \y/\s in {0.75/+1,-0.75/-1}{\begin{scope}[yshift=\y cm]
      \ruleSetup \ruleLegs{black}
      \draw[semithick, \rulecol] (v1) -- (v2);
      \node[rule lab, \rulecol, below=1pt] at (0,0) {$\sigma=\s$};
      \ruleVtcs{black}
    \end{scope}}
    \node at (0,0) {$+$};
    \node[rule lab, \rulecol] at (0,-1.5) {$2^{L}=2$ assignments};
  \end{tikzpicture}%
  \or
  
  \begin{tikzpicture}[rule]
    \ruleSetup \ruleLegs{rulegrey}
    \draw[semithick, \rulecol] (v1) -- (v2);
    \ruleVtcs{rulegrey}
    \node[rule lab, \rulecol, above=1pt] at (0,0) {$p,\,\sigma$};
    \ruleMomArrow{\rulecol}
  \end{tikzpicture}%
  \or
 
  \begin{tikzpicture}[rule]
    \ruleSetup
    \foreach \v in {v1,v2}
      \filldraw[fill=\rulecol!15, draw=\rulecol, thin] (\v) circle (0.17);
    \ruleLegs{rulegrey} \draw[semithick, rulegrey] (v1) -- (v2); \ruleVtcs{\rulecol}
    \node[rule lab, \rulecol] at (-0.6,-0.36) {$\Theta_1$};
    \node[rule lab, \rulecol] at ( 0.6,-0.36) {$\Theta_2$};
    \node[rule lab, \rulecol] at (0,-0.72) {$\mathcal V_1\,\mathcal V_2$};
  \end{tikzpicture}%
  \or
 
  \begin{tikzpicture}[rule]
    \ruleSetup
    \ruleLegs{\rulecol} \draw[semithick, rulegrey] (v1) -- (v2); \ruleVtcs{rulegrey}
    \foreach \e/\a/\i in {e1/left/1,e2/left/2,e3/right/3,e4/right/4}
      \node[rule lab, \rulecol, \a] at (\e) {$k_\i$};
  \end{tikzpicture}%
  \else
  
  \begin{tikzpicture}[rule]
    \ruleSetup
    \ruleLegs{black} \draw[semithick] (v1) -- (v2); \ruleVtcs{black}
    \foreach \e/\a/\i in {e1/left/1,e2/left/2,e3/right/3,e4/right/4}
      \node[rule lab, \a] at (\e) {$\vec k_\i$};
    \node[rule lab, above=1pt] at (0,0) {$p,\,\sigma$};
    \ruleMomArrow{black}
    \foreach \v in {v1,v2}
      \draw[\rulecol, thick] (\v) circle (3.3pt);
    \node[rule lab, \rulecol, above] at (0,0.42)
      {$\sum_{\sigma}\int_0^\infty\!\d p\,p$};
    \node[rule lab, \rulecol, below] at (0,-0.78)
      {$\vec q=\vec k_1+\vec k_2=-\vec k_3-\vec k_4$};
  \end{tikzpicture}%
  \fi}

\newcounter{dressingruleset}
\newcounter{dressingrule}[dressingruleset]

\newcommand{\newdressingruleset}{\stepcounter{dressingruleset}}
\newcommand{\rulekw}[1]{\textcolor{\rulecol}{#1}}  
\newlength{\rulediagw}  
\newcommand{\ruleBadge}[1]{\tikz[baseline=(b.base)]\node[circle, fill=#1,
  text=white, font=\scriptsize\bfseries, inner sep=1.2pt, minimum size=3.6mm] (b)
  {\thedressingrule};}
\newenvironment{dressingrule}[3]{%
  \refstepcounter{dressingrule}%
  \def\rulecol{#1}\def\rulediag{#3}%
  \begin{tcolorbox}[
      skin=bicolor, sidebyside, sidebyside align=center seam,
      righthand width=\rulediagw, sidebyside gap=4mm,
      colframe=#1, colback=white, colbacklower=#1!6,
      boxrule=0.8pt, arc=4pt, left=3mm, right=2mm, top=4.5mm, bottom=3mm,
      segmentation style={draw=none},
      title={\ruleBadge{#1}\enspace #2},
      attach boxed title to top center={yshift=-\tcboxedtitleheight/2},
      boxed title style={colback=white, colframe=#1, boxrule=0.8pt, arc=5pt,
        left=1.5mm, right=2mm, top=0.6mm, bottom=0.6mm},
      coltitle=black, fonttitle=\footnotesize,
      before skip=6mm, after skip=2mm,
      phantom={\ifdefined\phantomsection
          \let\ruleRaiseSaved\HyperRaiseLinkDefault
          \def\HyperRaiseLinkDefault{\dimexpr\baselineskip+11mm\relax}%
          \phantomsection
          \let\HyperRaiseLinkDefault\ruleRaiseSaved\fi},
    ]}%
  {\tcblower\centering\rulediag\end{tcolorbox}}

\title{Cosmological Dressing Rules for Massive Scalars}

\author[a,b]{Jonathan Gräfe\,\orcidlink{0009-0005-0663-9408},}
\author[c]{Matthias Nowinski\,\orcidlink{0009-0006-4736-6976},}
\author[c]{Ivo Sachs\,\orcidlink{0000-0003-2526-8655}}
\author[c]{and Siva Pawan Venna\,\orcidlink{0009-0008-7881-9549}}

\affiliation[a]{Max-Planck-Institut f\"ur Physik, Werner-Heisenberg-Institut,
Boltzmannstr. 8, 85748 Garching, Germany}
\affiliation[b]{Max Planck-IAS-NTU Center for Particle Physics, Cosmology and Geometry}
\affiliation[c]{Arnold-Sommerfeld-Center for Theoretical Physics, Ludwig-Maximilians-Universit\"at M\"unchen, 
Theresienstr. 37, 80333 Munich, Germany}

\emailAdd{graefe@mpp.mpg.de}
\emailAdd{m.nowinski@lmu.de}
\emailAdd{ivo.sachs@lmu.de}
\emailAdd{siva.venna@lmu.de}

\abstract{The Schwinger-Keldysh formulation of de Sitter correlation functions comes in different guises including harmonic analysis and dressed flat space correlators. We observe that these are equivalent spectral representations of the same Sturm-Liouville problem. Starting from the latter we derive dressing rules for scalar fields of arbitrary mass with auxiliary propagators in the form of Euler integrals that are accessible to flat space diagrammatic tools. 

\par\vskip 30pt
\noindent\textsc{Note:} This work was instigated by the master thesis of the last author.
}

\begin{document} 
\maketitle

\flushbottom

\section{Introduction}

Cosmological correlators are traces of interactions in the inflationary phase of the early universe that presently still hide in cosmological observations \cite{Mukhanov:1981xt, Maldacena:2002vr, Weinberg:2005vy, Cheung:2007st, Chen:2009zp}. In parallel to constant efforts to identify such signals in large amounts of observational data \cite{Planck:2018vyg, Cabass:2024wob, Chudaykin:2025vdh}, there is currently a steady push on the theoretical side to understand these objects better \cite{Arkani-Hamed:2015bza, Lee:2016vti, Cohen:2020php, Baumann:2022jpr, Li:2026xhw, Beneke:2026nkd}. As such, the cosmological collider program \cite{Arkani-Hamed:2015bza, Lee:2016vti} aims at predicting the signals of massive fields during inflation, complemented by research in flat space amplitudes to advance the state-of-the-art  when it comes to diagrammatic computations \cite{Benincasa:2022gtd, Albayrak:2026auu}. An important tool for mapping correlators to amplitudes is provided by cosmological dressing rules \cite{Chowdhury:2025ohm, Chowdhury:2025nnk}, which allow us to compute any Schwinger-Keldysh (SK) Feynman diagram in de Sitter QFT using integrals over the corresponding flat space scattering process.  Applications of dressing rules include the derivation of cutting rules analogous to flat space scattering amplitudes \cite{Ansari:2026xkm, Das:2026vfv, Chowdhury:2026upp}, the symbol analysis of cosmological correlators \cite{Chowdhury:2026dwm, Henn:2026lfz}, as well as methods based on differential equations \cite{Sachs:2026gcr}. 

The aim of this paper is to broaden the bridge between these two efforts by extending  cosmological dressing rules to scalar fields of generic  masses.  In the process of constructing these rules we furthermore provide a simple relation between dressing rules \cite{Chowdhury:2023arc, Chowdhury:2025ohm, Nowinski:2025cvw, Chowdhury:2025nnk, Chowdhury:2026upp} and harmonic analysis in de Sitter space (e.g. \cite{Bros:1995js,Hogervorst:2021uvp,Loparco:2023rug,Werth:2024mjg,Lee:2025kgs,Grafe:2026qsm,Belrhali:2026rkn, Grafe:2026avi}). The latter is particularly well suited to formulate perturbation theory for arbitrary masses while the former is motivated by its proximity to flat space Feynman integrals.  
We show that both formalisms are two equivalent interpretations of the same Sturm-Liouville problem, whose solution features two spectral parameters: The mass and the momentum. Choosing one over the other to base the construction on lands us in either of the above formulations. This makes it clear that both of the formulations are indeed spectral methods.

We then use this insight to build a direct translation of results derived in either formalism to the other. This also gives rise  to an alternative derivation of cosmological dressing rules for conformally coupled scalar fields \cite{Chowdhury:2025ohm} that we find logically rather appealing.  
Using this understanding we are then able to generalize this construction to general masses, using the very same steps. 
The result is a formulation of generalized cosmological dressing rules that inherits most of the structure from the conformally coupled case, with one important difference. In general, the propagator of scalar fields in the complementary series for fields with masses expressed through $\nu =\sqrt{\frac{d^2}{4}-\frac{m^2}{H^2}}\in \R$ contains a term with a coefficient $\cos(\pi\nu)$. This term vanishes for conformally coupled and massless fields in 3 space dimensions (or more generally speaking the exceptional series), but is non-zero for essentially all others. This term gives rise to a second kind of lines in the dressed Feynman diagrams that we dub an \textit{oppagator} since it does not propagate energy as in flat space and therefore has no direct interpretation in a flat space diagram.

In a separate analysis we then reconsider the  original derivation of the dressing rules and its generalization to generic masses. We find that the simplicity present for massless and conformal coupling does not quite persist for massive fields. Indeed, in the last step in the original derivation, a cancellation across multiple diagrams, does not go through generally. The resulting dressed formulation, that we dub as \textit{shadow dressing rules}, is, in essence, a formulation of the shadow formalism in de Sitter QFT \cite{DiPietro:2021sjt, Sleight:2021plv, Chowdhury:2023arc, Chowdhury:2025ohm}, where we write the conformal time integrals as propagators that carry (non-directed) energy. These dressing rules give a nontrivial reorganization of the previous generalized dressing rules formulation above that turns out to be quite powerful as we will exemplify for the massive  tree-level exchange diagram.

An alternative set of dressing rules, obtained by representing individual SK-propagators as a Laplace transform of flat space auxiliary propagators with a dressing kernel, has been described in \cite{Belrhali:2026jqe}. There are several differences between that and the present approach, in particular the sum over Schwinger-Keldysh indices is carried out implicitly in our formulation, which is important for the simplicity of cosmological correlators, compared to wave-function coefficients \cite{Chowdhury:2023arc}.

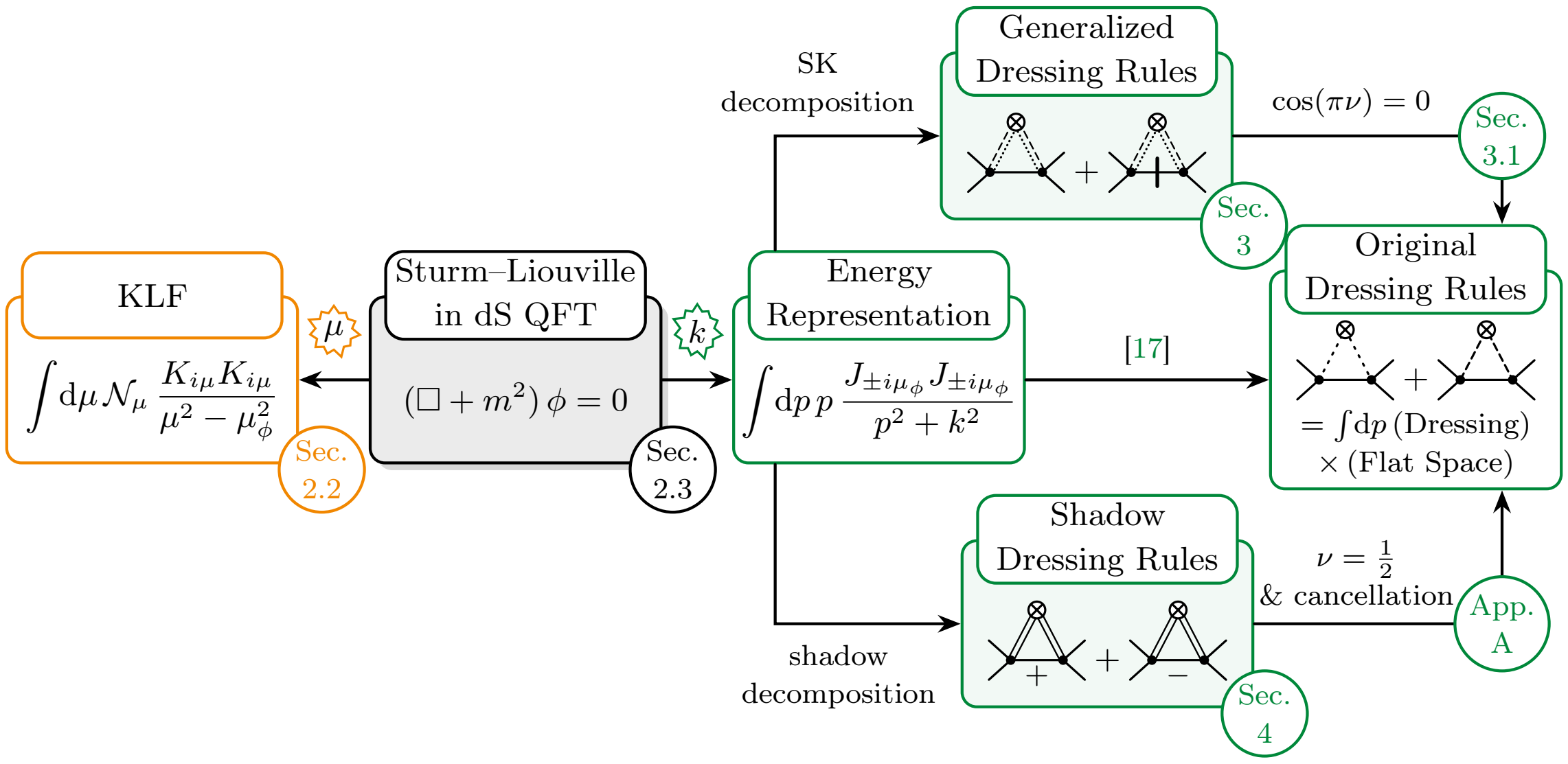
\begin{figure}[h]
\centering
\newcommand{\rmSecKLF}{\ref{sec: KLF in dS}}
\newcommand{\rmSecSL}{\ref{sec:sturm_liouville}}
\newcommand{\rmSecPC}{\ref{sec:SK_dressing_rules}}
\newcommand{\rmSecCC}{\ref{sec:SK_dressing_rules_conformal}}
\newcommand{\rmSecSH}{\ref{sec:shadow_dressing_rules}}
\newcommand{\rmAppSH}{\ref{app:shadow-conformal-graph-proof}}
\newcommand{\rmRefDR}{\cite{Chowdhury:2025ohm}}

\providecolor{lmugreen}{RGB}{0, 136, 58}   
\providecolor{lmuorange}{RGB}{241, 135, 0}   
\newcommand{\rmBox}[7][]{%
  \node[box=#5, #1, minimum width=2.8cm, minimum height=#4] (#2) at (#3) {};
  \node[tab=#5] (#2-tab) at (#2.north) {#6};
  \node[font=\footnotesize, align=center] at ([yshift=-2mm]#2.center) {#7};}
\newcommand{\rmSingle}[1]{%
  \draw[semithick, #1] (-0.6,0) -- (0,1.15) -- (0.6,0);
  \node[meet] at (0,1.15) {};}
\newcommand{\rmDouble}[2]{%
  \def\dd{0.065}
  \draw[line width=0.45pt, #1]
    ({-0.6-0.8866*\dd},{0.4626*\dd}) -- ({-0.8866*\dd},{1.15+0.4626*\dd})
    ({0.8866*\dd},{1.15+0.4626*\dd}) -- ({0.6+0.8866*\dd},{0.4626*\dd});
  \draw[line width=0.45pt, #2]
    ({-0.6+0.8866*\dd},{-0.4626*\dd}) -- ({0.8866*\dd},{1.15-0.4626*\dd})
    ({-0.8866*\dd},{1.15-0.4626*\dd}) -- ({0.6-0.8866*\dd},{-0.4626*\dd});
  \node[meet] at (0,1.15) {};}
\newcommand{\rmDiag}[2]{%
  #1
  \draw[semithick] (-1.1,0.45) -- (-0.6,0) -- (-1.1,-0.45)
                   (1.1,0.45) -- (0.6,0) -- (1.1,-0.45) (-0.6,0) -- (0.6,0);
  #2
  \node[vtx] at (-0.6,0) {}; \node[vtx] at (0.6,0) {};}
\begin{tikzpicture}[x=1cm, y=1cm, line cap=round,
  box/.style={draw=#1, fill=white, thick, rounded corners=4pt},
  tab/.style={draw=#1, fill=white, thick, rounded corners=5pt, font=\footnotesize,
              align=center, inner sep=2pt, text width=2.36cm, minimum height=0.82cm},
  badge/.style={circle, draw=#1, fill=white, thick, text=#1, font=\scriptsize,
                align=center, inner sep=0.6pt,
                execute at begin node={\ifdefined\hypersetup\hypersetup{linkcolor=#1}\fi}},
  route/.style={-{Stealth[length=2.2mm,width=1.7mm]}, thick, draw=#1},
  lab/.style={font=\scriptsize, align=center},
  eigen/.style={starburst, starburst points=11, starburst point height=0.9mm,
                draw=#1, fill=white, semithick, inner sep=0.8pt, font=\footnotesize},
  vtx/.style={circle, fill=black, inner sep=0pt, minimum size=2.6pt},
  meet/.style={circle, draw, fill=white, semithick, inner sep=0pt, minimum size=4.6pt,
    path picture={\draw
      (path picture bounding box.north west) -- (path picture bounding box.south east)
      (path picture bounding box.south west) -- (path picture bounding box.north east);}},
  dressing/.style={dash pattern=on 2pt off 1.3pt},
  dots/.style={line width=0.6pt, dash pattern=on 0pt off 1.15pt},
  main/.style={fill=black!8, drop shadow={opacity=0.3, shadow xshift=0.7mm, shadow yshift=-0.7mm}},
]
  
  \rmBox{klf}{1.4,0}{1.6cm}{lmuorange}{KLF}
    {\raisebox{1mm}{$\displaystyle\int\!\mathrm{d}\mu\,\mathcal{N}_\mu\,\frac{K_{i\mu}K_{i\mu}}{\mu^2-\mu_\phi^2}$}}
  \rmBox[main]{sl}{4.9,0}{1.6cm}{black}{Sturm--Liouville\\ in dS QFT}
    {$(\Box+m^2)\,\phi=0$}
  \rmBox{en}{8.4,0}{1.6cm}{lmugreen}{Energy\\ Representation}
    {$\displaystyle\int\!\mathrm{d}p\,p\,\frac{J_{\pm i\mu_\phi}J_{\pm i\mu_\phi}}{p^2+k^2}$}

  \draw[route=black] (sl.west) -- (klf.east);
  \draw[route=black] (sl.east) -- (en.west);
  \node[eigen=lmuorange] at (3.15,0.47) {$\mu$};
  \node[eigen=lmugreen]  at (6.65,0.47) {$k$};

  \rmBox[fill=lmugreen!5]{pc}{10.4,2.35}{1.6cm}{lmugreen}{Generalized\\ Dressing Rules}{}
  \begin{scope}[shift={(9.72,2.0)}, scale=0.42] \rmDiag{\rmDouble{dressing}{dots}}{} \end{scope}
  \node[font=\footnotesize] at (10.4,2.0) {$+$};
  \begin{scope}[shift={(11.08,2.0)}, scale=0.42]
    \rmDiag{\rmDouble{dressing}{dots}}{\draw[line width=1pt] (0,-0.32) -- (0,0.32);} \end{scope}

  \rmBox[fill=lmugreen!5]{sh}{10.6,-2.35}{1.6cm}{lmugreen}{Shadow\\ Dressing Rules}{}
  \begin{scope}[shift={(9.92,-2.7)}, scale=0.42]
    \rmDiag{\rmDouble{}{}}{\node[font=\scriptsize] at (0,-0.36) {$+$};} \end{scope}
  \node[font=\footnotesize] at (10.6,-2.7) {$+$};
  \begin{scope}[shift={(11.28,-2.7)}, scale=0.42]
    \rmDiag{\rmDouble{}{}}{\node[font=\scriptsize] at (0,-0.36) {$-$};} \end{scope}

  \rmBox{orig}{13.57,0}{2.1cm}{lmugreen}{Original\\ Dressing Rules}{}
  \begin{scope}[shift={(12.89,0)}, scale=0.42] \rmDiag{\rmSingle{dotted}}{} \end{scope}
  \node[font=\footnotesize] at (13.57,0) {$+$};
  \begin{scope}[shift={(14.25,0)}, scale=0.42] \rmDiag{\rmSingle{dressing}}{} \end{scope}
  \node[font=\scriptsize, align=center] at (13.57,-0.62)
    {$=\int\!\mathrm{d}p\,(\text{Dressing})$\\ $\times\,(\text{Flat Space})$};

  \draw[route=black] (7.4,0 |- en-tab.north) |- (pc.west);
  \node[lab, anchor=south east] at (8.88,2.4) {SK\\ decomposition};
  \draw[route=black] (7.4,0 |- en.south) |- (sh.west);
  \node[lab, anchor=north east] at (9.08,-2.42) {shadow\\ decomposition};
  \draw[route=black] (en.east) -- node[lab, above] {\rmRefDR} (orig.west);

  \draw[route=black] (pc.east) -| (14.4,0 |- orig-tab.north);
  \node[lab, anchor=south] at (12.95,2.4) {$\cos(\pi\nu)=0$};
  \draw[route=black] (sh.east) -| (14.4,0 |- orig.south);
  \node[lab, anchor=south] at (13.0,-2.3) {$\nu=\frac12$\\ \& cancellation};

  \node[badge=lmuorange] at ([xshift=2.2mm,yshift=-0.5mm]klf.south east) {Sec.\\ \rmSecKLF};
  \node[badge=black]     at ([xshift=1mm,yshift=-0.5mm]sl.south east)  {Sec.\\ \rmSecSL};
  \node[badge=lmugreen]  at ([xshift=1mm,yshift=-0.5mm]pc.south east)  {Sec.\\ \rmSecPC};
  \node[badge=lmugreen]  at ([xshift=1mm,yshift=-0.5mm]sh.south east)  {Sec.\\ \rmSecSH};
  \node[badge=lmugreen]  at (14.4,2.35)  {Sec.\\ \rmSecCC};
  \node[badge=lmugreen]  at (14.4,-2.35) {App.\\ \rmAppSH};
\end{tikzpicture}
\caption{Roadmap of the paper. The scalar equation of motion in de Sitter space
  poses a Sturm--Liouville problem (Sec.~\rmSecSL) that can be read in two
  ways, depending on which quantity (starred) is taken as the eigenvalue: the
  mass parameter $\mu$, which gives the KLF representation (Sec.~\rmSecKLF), or
  the momentum $k$, which gives an energy representation with spectral variable
  $p$. Factorizing the propagator in its Schwinger-Keldysh structure
  yields generalized dressing rules using flat space propagators and oppagators
  (Sec.~\rmSecPC); decomposing into Bessel orders yields the shadow
  dressing rules (Sec.~\rmSecSH). For $\nu=n+\frac12$ both reduce to the
  original dressing rules of \rmRefDR: the former because the oppagator is
  proportional to $\cos(\pi\nu)=0$ (Sec.~\rmSecCC), the latter after
  cancellations between diagrams (App.~\rmAppSH).}
\label{fig:roadmap}
\end{figure}

We organize this paper as shown in the roadmap in figure \ref{fig:roadmap}: In section \ref{sec:theory}, we present the aforementioned formulation of de Sitter QFT via spectral methods. Afterwards, we present our thoughts on the Sturm-Liouville problem and how $\mu$-integrals are closely connected to $p$-integrals. Next, in section \ref{sec:SK_dressing_rules}, we show how we can use these considerations to derive conformally coupled dressing rules from the spectral methods. Then, we present how this construction can be generalized to arbitrary masses. As a demonstration of the utility of this tool, we show that we can recover the tree-level massive exchange diagram for general internal masses. Additionally, in section \ref{sec:shadow_dressing_rules}, we also perform the same construction -- general mass dressing rules, including an application -- for the shadow dressing rules. Finally, in section \ref{sec:conclusions}, we point out what we can use the results of this paper for and where to look next.

\subsection*{Conventions}

We mostly use the conventions laid out in \cite{Belrhali:2026rkn}. This means that the mass parameter is $\mu$, which is real in the principal series. When we talk about complementary series, we will use a real $\nu \in (0,d/2)$, which is connected to $\mu$ via $\mu = i \nu$. In particular, conformally coupled fields have $\nu = \frac{1}{2}$ and $\mu = \frac{i}{2}$. When we talk about physical mass parameters of fields that are not to be integrated over, we use an index and write $\mu_\phi, \nu_\phi$. These external parameters are always taken to be positive and real, unless otherwise stated.

We work in $d$ spatial dimensions in de Sitter space with Hubble parameter $H$ and conformal time $\eta \in (-\infty, 0)$, which we will mostly Wick-rotate to Euclidean Anti-de Sitter space with conformal time $z \geq 0$:
\begin{align}
    \d s^2 = \frac{\d \eta^2 - (\d \vec x)^2}{(\eta H)^2} \qquad \xlongleftrightarrow{\mathrm{Wick \ rotation}} \qquad \d s^2 = \frac{\d z^2 + (\d \vec x)^2}{(z H)^2}.
\end{align}
For diagrams, our conventions are: When we talk about a single vertex, internal energies connected to it are called $p_l$ and internal momenta $\vec q_l$, where $l$ runs from $1$ to $I$. In the same setting, external momenta are $\vec k_s$, where $s$ runs from $1$ to $E$. The total number of legs attached to a vertex is $N$. When we talk about a full diagram, the total number of external legs attached to it is $M$. Vertices are enumerated by $v$, running up to $V$. Propagators are enumerated by $j$, running up to $L$. In the dressing rules, the energies for dressing propagators are labeled $P_r$ where $r$ runs up to $R$.

\section{Spectral Methods in de Sitter} \label{sec:theory}
In de Sitter QFT computations, the words \textit{spectral methods} are used in many ways, so it is easy to be confused about what they actually mean. Let us therefore begin by reviewing harmonic analysis in de Sitter space \cite{Bros:1995js,DiPietro:2021sjt,Hogervorst:2021uvp,Loparco:2023rug,Werth:2024mjg,Grafe:2026qsm,Belrhali:2026rkn}. We want to recall that spectral methods use eigenvalues of specific operators as integration variables, just like the momentum $p$ in flat space QFT. Because it is a good summary that shares our main focus for this section, we will mainly follow the reference \cite{Belrhali:2026rkn}, which formulates  harmonic analysis, via  Kontorovich-Lebedev-Fourier integrals (KLF) over the mass parameter $\mu$ and the Fourier mode $\vec k$. 

\subsection{Flat Space Warm-Up}
In this subsection we begin with a somewhat pedantic review of the standard Feynman rules for scattering amplitudes. The purpose of this exercise is to motivate our language for formulating the dressing rules in the rest of this paper, and to stress our interpretation of why this construction works. 

In position space the Feynman rules specify propagators $\mathcal{G}(x_1, x_2)$, and vertices at positions $x$. These carry some factor $\mathcal{V}(x)$, containing coupling constants, symmetry factors or derivatives that act on the propagators. In general, vertex positions are to be integrated over, while propagators (that may include time ordering) connect these positions non-trivially and make the integrals complicated. A general diagram can then be written as
\begin{align}
    \mathcal{M} = \int \d^{d+1} x_1 \ldots \d^{d+1} x_V \left(\prod\limits_{\mathrm{vertices} \ v} \mathcal{V}_v(x_v) \right) \left(\prod\limits_{\mathrm{lines} \ j} \mathcal{G}(x_{1,j}, x_{2,j})\right).
\end{align}

In momentum space, the same expression $\mathcal{M}$ is much simpler to compute. We make use of the spectral properties of the equation of motion (EoM) operator $\Box + m^2$ and expand the propagator as an integral over pairs of \textit{eigenfunctions} $e^{i p x_1}$ and $ e^{-i p x_2}$. These are orthonormal with respect to the \textit{spectral measure} $\d^{d+1} p / (2\pi)^{d+1}$. The defining property of the propagator is that the action of the EoM operator produces a delta function, so we need to include a term that cancels its action, the \textit{denominator} $p^2 - m^2$ of the Feynman propagator. Ignoring $i\epsilon$ prescriptions for now, we write
\begin{align}\label{eq:pf}
    \mathcal{G}(x_1, x_2) = \int \frac{\d^{d+1} p}{(2\pi)^{d+1}} \frac{e^{ipx_1} e^{-ipx_2}}{p^2 - m^2}.
\end{align}
As a result, we have complicated the propagator from a closed expression to an integral, making the $x$-dependence manifest so that the integrals over the vertex positions are now easy to perform. All dependence on the vertex positions is contained within the eigenfunctions $e^{ipx}$ and the $\mathcal{V}_v(x)$. However, in flat space, $\mathcal{V}_v(x)$ usually only contains derivatives, which simply pull momenta out of the eigenfunctions and then no longer depend on $x$. We may define a new vertex factor $\mathcal{V}_v(\{p_j\})$ via the entire integral
\begin{align}
    \mathcal{V}_v(\{p_j\}) \coloneq \int \d^{d+1} x \, \mathcal{V}(x) \Big( e^{ip_1 x} \ldots e^{ip_N x} \Big) = \mathcal{V}^\prime_v(\{p_j\}) (2\pi)^{d+1} \delta^{(d+1)}(p_1 + \ldots + p_N).
\end{align}
$\mathcal{V}^\prime(\{p_j\})$ contains all additional factors and $p$-dependence other than the delta function. The momenta $p_1, \ldots, p_N$ are the $N$ momenta of the lines connected to the vertex $v$. At the level of the diagram, this trades the vertex position integrals for spectral integrals over the label $p$ representing the eigenvalues of the EoM operator, and so is a \textit{spectral method}. Our Feynman diagram is now
\begin{align}
    \mathcal{M} = \int \frac{\d^{d+1} p_1}{(2\pi)^{d+1}} \ldots \frac{\d^{d+1} p_{L}}{(2\pi)^{d+1}} \left(\prod\limits_{\mathrm{vertices} \ v} \mathcal{V}_v(\{p_j\}) \right) \left(\prod\limits_{\mathrm{lines} \ j} \frac{1}{p_j^2 - m^2} \right).
\end{align}
From this, we can read off the momentum space Feynman rules familiar from standard textbooks on QFT, in particular the momentum space Feynman propagator $\mathcal{G}(p) = 1 / (p^2 - m^2 + i \epsilon)$ after adding the $i\epsilon$ prescription. Poincaré invariance is clearly manifest in this flat space construction. 

\subsection{Harmonic Analysis in de Sitter}\label{sec: KLF in dS}
De Sitter space is maximally symmetric, like flat space, but the isometry group is different. So, while it is possible to repeat the spectral representation of the propagators in a manifestly de Sitter invariant way, the eigenfunctions are now expressed in terms of Hankel functions, while the vertex factors $\mathcal{V}(\eta)$ now contain an explicit dependence on conformal time $\eta$ through the metric determinant $\sqrt{g}$.

Furthermore, in de Sitter space one naturally computes \textit{in-in correlators} through the Schwinger-Keldysh (SK) formalism\footnote{Of course, the SK formalism can also be applied to flat space correlation functions. In fact, dressing rules can be derived in flat space as well, see appendix \ref{app:flat_space_dressing}.}, so each propagator and vertex now carry an SK index $\aa = \pm$. This also affects the Wick rotation from de Sitter space to Euclidean anti-de Sitter space (EAdS) with metric
\begin{align}
    \d s^2 = \frac{\d z^2 + (\d \vec x)^2}{(z H)^2}.
\end{align}
The SK formalism prescribes a contour to integrate along and we have to be careful during Wick rotation to not cross a branch cut \cite{DiPietro:2021sjt}. Therefore, we rotate the two branches of the contour in opposite directions, which introduces additional phase factors that we can absorb into the vertex factor $\mathcal{V}_\aa$, which now depends on the SK index.

To be more explicit, the de Sitter scalar eigenfunctions of the EoM operator, after Wick rotation, are given by
\begin{align}
    \Phi_{\vec k}^\mu (z, \vec x) = \frac{H^\frac{d+1}{2}}{\sqrt{\pi}} e^{-i \vec k \cdot \vec x} z^\frac{d}{2} K_{i\mu}(k z). \label{eq:KLF_eigenfunctions}
\end{align}
Here, the set of $(\mu, \vec k)$ represents the spectral parameter, while the set $(z, \vec x)$ are the position-space coordinates. These are orthonormal with respect to the measure
\begin{align}
    \d \mu \, \frac{\mu}{\pi} \sinh(\pi \mu)=:  \d \mu \, \mathcal{N}_\mu\,.
\end{align}
The analogue of the propagator \eqref{eq:pf} in de Sitter, including the SK phase factor $\mathcal{G}(\aa,\bb)$, therefore can be expressed as \cite{Sleight:2020obc,Loparco:2023rug,Cacciatori:2024zrv}
\begin{align}
    \mathcal{G}_{\aa\aa}(z_1, \vec x_1; z_2, \vec x_2) &= \mathcal{G}(\aa,\aa) \int\limits_{-\infty}^\infty \d \mu \, \mathcal{N}_\mu \int \frac{\d^d k}{(2\pi)^d} \frac{\Phi^\mu_{\vec k}(z_1, \vec x_1) \Phi^\mu_{-\vec k}(z_2, \vec x_2)}{\mu^2 - \mu_\phi^2} \notag\\
    &\eqcolon \int\limits_{-\infty}^\infty \d \mu \, \mathcal{N}_\mu \int \frac{\d^d k}{(2\pi)^d} \Phi^\mu_{\vec k}(z_1, \vec x_1) \Phi^\mu_{-\vec k}(z_2, \vec x_2) \mathcal{G}^{\mu_\phi}_{\aa\aa}(\mu) \label{eq:KLF_propagator}\,.
\end{align}
This is named a Kontorovich-Lebedev-Fourier (KLF) transformation in \cite{Belrhali:2026rkn}. This still needs to be completed with an appropriate $i\epsilon$ prescription and the precise form depends on the de Sitter representation of the field. In particular, there is a difference between fields in the principal series, $\mu_\phi \in \R$, and the complementary series, where $\nu_\phi \in (0, d/2)$:
\begin{subequations} \label{eq:KLF_propagators}
\begin{align}
    \mathcal{G}^{\mu_\phi}_{\aa\aa}(\mu) &= e^{-i\aa \pi \frac{d-1}{2}} \frac{1}{(\mu^2-\mu_\phi^2)_{i\aa\epsilon}} \quad \mathrm{for \ principal \ series}, \label{eq:KLF_propagators_principal} \\
    \mathcal{G}^{i\nu_\phi}_{\aa\aa}(\mu) &= e^{-i\aa \pi \frac{d-1}{2}} \left( \frac{1}{\mu^2 + \nu_\phi^2} - i \aa \frac{e^{i\aa \pi \nu_\phi}}{\mathcal{N}_{i\nu_\phi}} \hat\delta_{i\nu_\phi}(\mu) \right) \quad \mathrm{for \ complementary \ series}. \label{eq:KLF_propagators_complementary}
    \end{align}
\end{subequations}
The denominator in \eqref{eq:KLF_propagators_principal} is defined as
\begin{align}
    \frac{1}{(\mu^2 - \mu_\phi^2)_{i\aa \epsilon}} &\coloneq \frac{1}{2 \sinh(\pi \mu_\phi)} \left( \frac{e^{+\pi\mu_\phi}}{\mu^2 - \mu_\phi^2 + i\aa \epsilon} - \frac{e^{-\pi\mu_\phi}}{\mu^2 - \mu_\phi^2 - i\aa\epsilon} \right).
\end{align}
One might object that this is more than just an $i\epsilon$ prescription. However, the $i\epsilon$ in flat space is just there to steer the contour to include or exclude specific poles. The same is done here, but in a more complicated way. Still, the expression collapses in the expected way for $\epsilon \rightarrow 0$. Additionally, in \eqref{eq:KLF_propagators_complementary}, the $\hat\delta_{\mu_\phi}(\mu)$ is defined through
\begin{align}
    \int\limits_\gamma \d \mu \, \hat\delta_\alpha(\mu) f(\mu) \coloneq \frac{1}{2}\Big( f(\alpha) + f(-\alpha) \Big).
\end{align}
This definition is given for any function $f$ that is analytic in a small neighborhood of $\alpha$ and any contour $\gamma$, also when $\gamma$ does not include $\alpha$ itself. It is therefore also to be understood as a contour modification that adds a small circle around $\alpha$ to the $\mu$ contour.

For the SK formalism, we also need the distribution $\mathcal{G}^{i\nu_\phi}_{\aa,(-\aa)}$ which represents the 2-point function  where the SK indices of the two ends of a line are opposite. This term is special because it is actually not a propagator, but a Wightman function. Therefore, there is no denominator and the contour is just picking up two poles
\begin{align}
    \mathcal{G}^{i\nu_\phi}_{\aa,(-\aa)}(\mu) &= \frac{\hat\delta_{i\nu_\phi}(\mu)}{\mathcal{N}_{\mu}} \quad \mathrm{for \ any \ representation}. \label{eq:KLF_propagators_wightman}
\end{align}
This leaves us with the vertex integral. Just as in flat space, we simply take $N$ eigenfunctions of the EoM operator and plug them into the integral over the position of the vertex. However, additionally, we have to include the metric measure $\sqrt{g} = (z H)^{-d-1}$ and define the factor $\mathcal{V}(\aa)$ that contains the symmetry factor  $\mathcal{C}$, coupling constant $\lambda$, and now also an SK index. We assume, for simplicity that the vertex does not contain any derivatives. Putting everything together, the general vertex factor is
\begin{align}
    \mathcal{V}^{\mu_1, \ldots, \mu_N}_{\vec k_1, \ldots, \vec k_N}(\aa) \coloneq \mathcal{V}(\aa) \int\limits_0^\infty \frac{\d z}{(zH)^{d+1}} \int \d^d x \ \Phi_{\vec k_1}^{\mu_1} (z, \vec x) \ldots \Phi_{\vec k_N}^{\mu_N} (z, \vec x).
\end{align}
The spatial integrals here are trivial, as they just give a momentum-conserving delta function like in flat space. The problematic part is the $z$ integral, which does not have a closed form in elementary functions. We factorize the spatial and temporal components into
\begin{align}
    \mathcal{V}^{\mu_1, \ldots, \mu_N}_{\vec k_1, \ldots, \vec k_N}(\aa) &= -i \aa e^{i \pi \aa \frac{d}{2}} (2\pi)^d \delta^{(d)}\left( \sum\limits_{j=1}^N\vec k_j \right) \frac{\mathcal{C}\lambda H^\frac{(N-2)(d+1)}{2}}{\pi^\frac{N}{2}} \mathcal{I}^{\mu_1, \ldots, \mu_N}_{\vec k_1, \ldots, \vec k_N}, \label{eq:KLF_vertex_factor}
\end{align}
where we introduced the vertex function $\mathcal{I}$ as
\begin{align}
    \mathcal{I}^{\mu_1, \ldots, \mu_N}_{\vec k_1, \ldots, \vec k_N} = \int\limits_0^\infty \d z \, z^{\frac{d(N-2)}{2}-1} K_{i\mu_1}(k_1 z) \ldots K_{i\mu_N}(k_N z). \label{eq:KLF_vertex_function}
\end{align}
In particular, this latter object does not carry any additional phase and does not depend on $\aa$. There are now two types of integrals in our diagram, $\mu$ and $z$ integrals. 

As a last step, we consider the external legs of the diagram. They carry information about the observable we want to compute, which includes the masses of the external fields $\mu_{\phi,s}$, their momenta $\vec k_s$, and also the conformal time at which the observable is evaluated, which we call $\eta_0$. Note that we do not Wick rotate $\eta_0$, so here the SK phase factor is a bit different from the internal vertices. To avoid unnecessary clutter, we also do not KLF-transform the external legs. They can be treated as Wightman functions. Altogether, every diagram then gets multiplied by the following factor for every external leg
\begin{subequations} \label{eq:KLF_external}
\begin{align}
    \mathcal{E}_{\aa}(\eta_0) = \ &\frac{H^{\frac{d-3}{2}}}{\sqrt{\pi}} (-\eta_0)^\frac{d}{2} e^{-\frac{i \aa \pi d}{4}} K_{i\mu_\phi}\Big( (- \eta_0 ) k (- i \aa) \Big) \notag\\
    \xrightarrow{\eta_0 \rightarrow 0} \, &H^\frac{d-3}{2} \frac{(-\eta_0)^\frac{d-1}{2}}{\sqrt{2k}} e^{-i \aa \pi \frac{d-1}{4}} \quad (\mathrm{conformally \ coupled}) \\
    \xrightarrow{\eta_0 \rightarrow 0} \, &\frac{H^{\frac{d-3}{2}}}{\sqrt{\pi}} \frac{\Gamma(\nu_\phi)}{2} (-\eta_0)^{\frac{d}{2} - \nu_\phi} \Big( \frac{2}{k} \Big)^{\nu_\phi}  e^{-i \aa \pi \frac{d - 2\nu_\phi}{4}} \quad (\mathrm{complementary}) \\
     \xrightarrow{\eta_0 \rightarrow 0} \, &\frac{H^{\frac{d-3}{2}}}{\sqrt{\pi}} \frac{\Gamma(i\mu_\phi)}{2} (-\eta_0)^{\frac{d}{2} - i\mu_\phi} \Big( \frac{2}{k} \Big)^{i\mu_\phi}  e^{-i \aa \pi \frac{d}{4} - \aa \frac{\pi}{2} \mu_\phi}  + \Big[ \mu_\phi \rightarrow -\mu_\phi \Big] \quad (\mathrm{principal}) \label{eq:KLF_external_principal}
\end{align}
\end{subequations}
We call the product of all of these terms $\mathcal{E}_{\{\aa_v\}}$. Note that there are two terms for the principal series, while there is one for the complementary series. This is because the complementary series terms have extra suppression in $\eta_0$ due to $\nu$ being real, while in the principal series both terms with $\nu = \pm i \mu$ are of the same order in late times. In the rest of this paper, we will only consider the first term in \eqref{eq:KLF_external_principal} and assume $\mu > 0$, then symmetrize in $\mu$ at the very end. This is possible, because the rest of the diagram is symmetric in $\mu$.

Further, each external leg connects to some vertex in the diagram. We thus set $\mu$ of the corresponding vertex function $\mathcal{I}$ to the value $\mu_\phi$ of the mass parameter of the external field. Above, we explicitly singled out the conformally coupled $\mu = i/2$, because we will treat this case below. In addition to being the simplest to compute, the conformally coupled case can act as a seed integral from which one can reach the correlators relevant for inflation with weight-shifting operators \cite{Baumann:2019oyu}.

For later reference, we recapitulate that in total, there are three sources of phase factors: The propagators \eqref{eq:KLF_propagators}, the vertices \eqref{eq:KLF_vertex_factor}, and external legs \eqref{eq:KLF_external}. The structure of the complete diagram $\mathcal{M}$, that  contains a sum over all SK indices, is then
\begin{align}
\begin{split}
    \mathcal{M} &= \sum\limits_{\{\aa_v = \pm\}} \mathcal{E}_{\{\aa_v\}}(\eta_0) \int\limits_{-\infty}^\infty \d \mu_1 \mathcal{N}_{\mu_1} \ldots \d \mu_{L} \mathcal{N}_{\mu_L} \int \frac{\d^{d} k_1}{(2\pi)^{d}} \ldots \frac{\d^{d} k_{L}}{(2\pi)^{d}} \\
    &\qquad\qquad \times \left(\prod\limits_{\mathrm{vertices} \ v} \mathcal{V}^{\mu_{1,v}, \ldots, \mu_{N,v}}_{\vec k_{1,v}, \ldots, \vec k_{N,v}}(\aa_v) \right) \left(\prod\limits_{\mathrm{lines} \ j} \mathcal{G}^{\mu_{\phi,j}}_{\aa_{1,j}\aa_{2,j}}(\mu_j) \right). \label{eq:KLF_diagram_final}
\end{split}
\end{align}
A vertex factor depends on the $\mu$ and $\vec k$ of the adjacent lines, and its own $\aa$, while a propagator depends on its own $\mu$ and the $\aa_1$ and $\aa_2$ of its adjacent vertices.

\subsection{Sturm-Liouville Relation} \label{sec:sturm_liouville} 

\begin{figure}
\centering
\providecolor{lmugreen}{RGB}{0, 136, 58}    
\providecolor{lmuorange}{RGB}{241, 135, 0}   
\newcommand{\rmBox}[7][]{%
  \node[box=#5, minimum width=#4, minimum height=1.6cm, #1] (#2) at (#3) {};
  \node[tab=#5] (#2-tab) at (#2.north) {#6};
  \node[font=\footnotesize, align=center] at ([yshift=-2mm]#2.center) {#7};}

\newcommand{\rmDiag}[2]{%
  \def\dd{0.065}%
  \draw[line width=0.45pt, #1]
    ({-0.6-0.8866*\dd},{0.4626*\dd}) -- ({-0.8866*\dd},{1.15+0.4626*\dd})
    ({0.8866*\dd},{1.15+0.4626*\dd}) -- ({0.6+0.8866*\dd},{0.4626*\dd});
  \draw[line width=0.45pt, #2]
    ({-0.6+0.8866*\dd},{-0.4626*\dd}) -- ({0.8866*\dd},{1.15-0.4626*\dd})
    ({-0.8866*\dd},{1.15-0.4626*\dd}) -- ({0.6-0.8866*\dd},{-0.4626*\dd});
  \node[meet] at (0,1.15) {};
  \draw[semithick] (-1.1,0.45) -- (-0.6,0) -- (-1.1,-0.45)
                   (1.1,0.45) -- (0.6,0) -- (1.1,-0.45) (-0.6,0) -- (0.6,0);
  \node[vtx] at (-0.6,0) {}; \node[vtx] at (0.6,0) {};}
\begin{tikzpicture}[x=1cm, y=1cm, line cap=round,
  box/.style={draw=#1, fill=white, thick, rounded corners=4pt},
  tab/.style={draw=#1, fill=white, thick, rounded corners=5pt, font=\footnotesize,
              align=center, inner sep=2pt, text width=2.36cm, minimum height=0.82cm},
  badge/.style={circle, draw=#1, fill=white, thick, text=#1, font=\scriptsize,
                align=center, inner sep=0.6pt,
                execute at begin node={\ifdefined\hypersetup\hypersetup{linkcolor=#1}\fi}},
  route/.style={-{Stealth[length=2.2mm,width=1.7mm]}, thick, draw=#1},
  lab/.style={font=\scriptsize, align=center},
  eigen/.style={starburst, starburst points=11, starburst point height=0.9mm,
                draw=#1, fill=white, semithick, inner sep=0.8pt, font=\footnotesize},
  vtx/.style={circle, fill=black, inner sep=0pt, minimum size=2.6pt},
  meet/.style={circle, draw, fill=white, semithick, inner sep=0pt, minimum size=4.6pt,
    path picture={\draw
      (path picture bounding box.north west) -- (path picture bounding box.south east)
      (path picture bounding box.south west) -- (path picture bounding box.north east);}},
  dressing/.style={dash pattern=on 2pt off 1.3pt},
  dots/.style={line width=0.6pt, dash pattern=on 0pt off 1.15pt},
  main/.style={fill=black!8, drop shadow={opacity=0.3, shadow xshift=0.7mm, shadow yshift=-0.7mm}},
]

  \rmBox[main]{sl}{4.3,3.0}{2.8cm}{black}{Sturm--Liouville\\ in dS QFT}
    {$(\Box+m^2)\,\phi=0$}

  \rmBox{klf}{1.7,0}{3.4cm}{lmuorange}{KLF}
    {$\displaystyle\int\!\d\mu\,\mathcal N_\mu\,\frac{\Phi_k(\mu)\Phi_{-k}(\mu)}{\mu^2-\mu_\phi^2}$}
  \rmBox{en}{6.9,0}{3.4cm}{lmugreen}{Energy\\ Representation}
    {$\displaystyle\int\!\d p\,\mathcal N_p\,\frac{\Psi_\mu(p)\Psi^*_\mu(p)}{p^2+k^2}$}

  \draw[route=black] ([xshift=-6mm]sl.south) |- (1.7,1.7) -- (klf-tab.north);
  \draw[route=black] ([xshift=6mm]sl.south)  |- (6.9,1.7) -- (en-tab.north);
  \node[eigen=lmuorange] at (2.7,1.7) {$\mu$};
  \node[eigen=lmugreen]  at (5.9,1.7) {$k$};

  \rmBox[fill=lmugreen!5]{dr}{11.0,0}{2.8cm}{lmugreen}{Dressing Rules}{}
  \begin{scope}[shift={(11.0,-0.25)}, scale=0.42] \rmDiag{dressing}{dots} \end{scope}
  \draw[route=black] (en.east) -- (dr.west);
  \node[badge=lmugreen] at ([xshift=1mm,yshift=-0.5mm]dr.south east)
    {Sec.\\ \ref{sec:SK_dressing_rules}/\ref{sec:shadow_dressing_rules}};

  \draw[{Stealth[length=2.2mm,width=1.7mm]}-{Stealth[length=2.2mm,width=1.7mm]},
        thick, dash pattern=on 4pt off 2.5pt]
    (klf.south) to[out=-40, in=-140]
    node[lab, fill=white, inner sep=1.5pt] {transition identities}
    (en.south);
\end{tikzpicture}
\caption{Schematic flow chart of the central statements of this section,
  repeating the left part of the roadmap in figure~\ref{fig:roadmap} together
  with the transition identities between the two representations. Shown are
  mainly the propagators from both the KLF and energy representation
  formulations. The functions $\Phi$ are given by \eqref{eq:KLF_eigenfunctions},
  while the functions $\Psi$ on the right hand side are still to be determined.
  The missing $i\epsilon$ prescriptions on the KLF side are given by
  \eqref{eq:KLF_propagators} and \eqref{eq:KLF_propagators_wightman}. On the
  right hand side, we do not need $i\epsilon$ if we interpret it as an
  Euclidean propagator. The transition identities are derived below. In
  sections \ref{sec:SK_dressing_rules} and \ref{sec:shadow_dressing_rules}, we
  show how to derive dressing rules from the energy representation.}
\label{fig:sturm_liouville_schematic}
\end{figure}
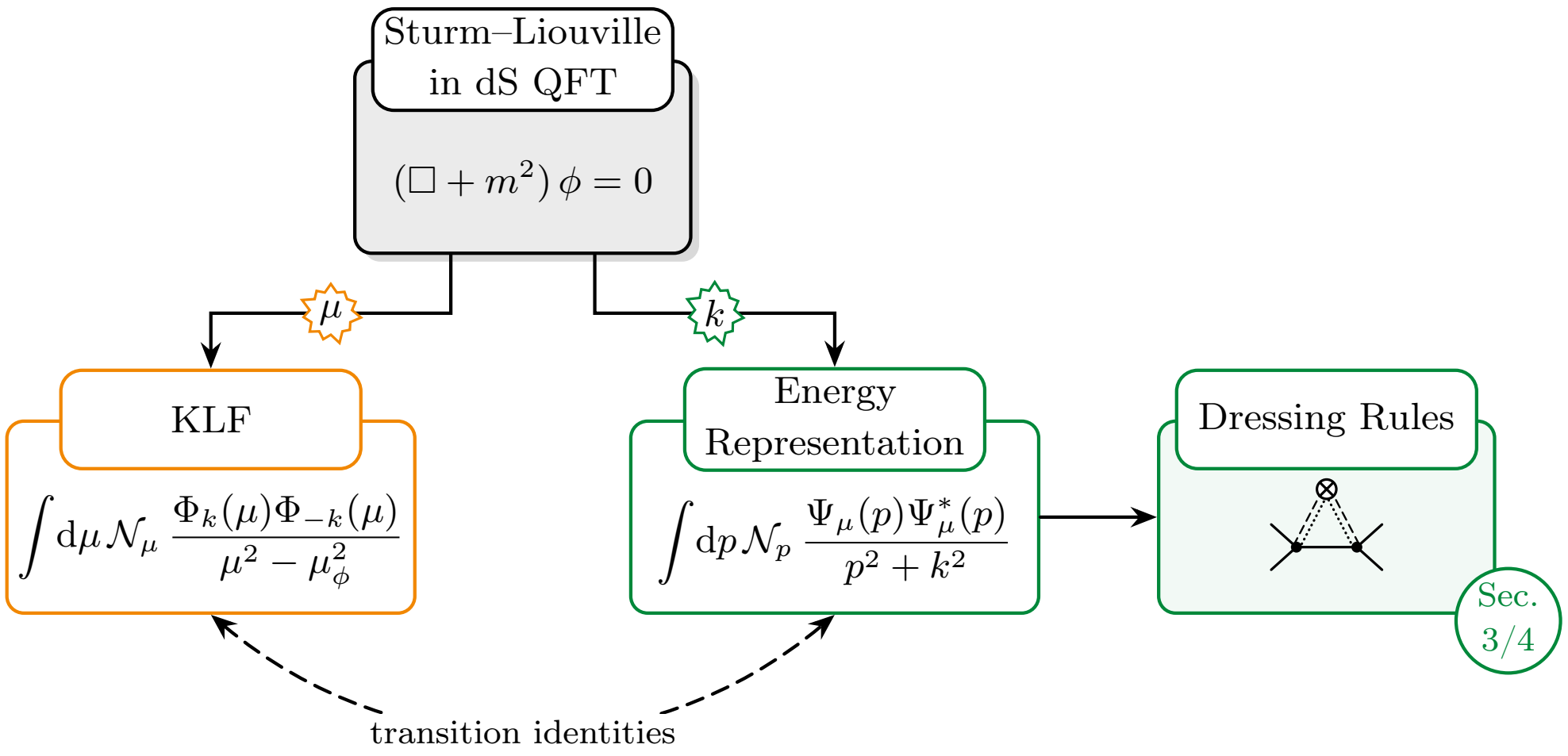

In this section, we come to the reason why we introduced harmonic analysis in a paper on dressing rules, by choosing an alternative interpretation of the eigenvalue problem induced by the scalar EoM operator in Wick-rotated de Sitter space (EAdS)
\begin{align}
    \Big( z^2 \partial_z^2 - (d-1) z \partial_z + z^2 \partial_j^2 + \mu^2 + \frac{d^2}{4} \Big) \phi = 0\,, \label{eq:dS_EoM}
\end{align}
acting on any  scalar field $\phi$. Here we have parametrized the mass of the field as
\begin{align}
    \frac{m^2}{H^2} = \mu^2 + \frac{d^2}{4}.
\end{align}
Further, we can parametrize the spatial dependence via plane waves and transition to \textit{partial momentum space} in the variables $(z, \vec k)$, where we replace $\partial_j^2 \longrightarrow - k^2$. Also, we split off some of the $z$-dependence and use $\phi = z^{d/2} g$. Then, the EoM \eqref{eq:dS_EoM} can also be written as
\begin{align}
    \partial_z\big(z\, \partial_z g_k\big) - k^2 z\, g_k = -\frac{\mu^2}{z}\, g_k. \label{eq:SL_eigenvalue_mu}
\end{align}
This formulation has the well-known shape of a Sturm-Liouville problem, which in general is
\begin{align}
    \partial_z \Big( p(z) \partial_z f \Big) + q(z) f = -\lambda w(z) f,
\end{align}
with a \textit{coefficient} function $p(z)$, a \textit{potential} function $q(z)$, a \textit{weight} or \textit{measure} function $w(z)$ and an eigenvalue $\lambda$. In our case, the eigenvalue is $\lambda=\mu^2$ and the potential carries a \textit{parameter} $k^2$. As long as we consider $k$ to be fixed, it is just a label to be carried through to the solution. The differential equation is the well-known Bessel differential equation with Bessel functions as the regular solutions. Following the construction of spectral propagators reviewed in the earlier sections, we are naturally led to propagators featuring $\mu$ integrals and denominators depending on $\mu$. This is the KLF construction and schematically depicted on the left side of figure \ref{fig:sturm_liouville_schematic}. In de Sitter, this route might be the most natural, since the eigenfunctions of this interpretation also diagonalize the symmetries of de Sitter space, as shown for example in \cite{Belrhali:2026rkn}.

But, the interpretation that $\mu$ is the eigenvalue is not unique. Nothing prevents us from rearranging the Sturm-Liouville equation \eqref{eq:SL_eigenvalue_mu} as
\begin{align}
    \partial_z\big(z\,\partial_z g_\mu\big) + \frac{\mu^2}{z}\, g_\mu = k^2 z\, g_\mu. \label{eq:SL_eigenvalue_k}
\end{align}
Now, the eigenvalue is $\lambda=-k^2$ and the parameter is $\mu^2$. The eigenvalue equation has not changed, only the names. Additionally, the potential and weight functions have swapped places. However, this difference has an important implication on how we interpret the results that we get from solving the Sturm-Liouville problem. Following the procedure to construct spectral propagators, we write down integrals over $k$ and denominators including $k^2$. We call this construction the \textit{energy representation}\footnote{It will be shown in section \ref{sec:SK_dressing_rules_general} that one can write the Bessel $J$ functions via Hankel functions. The energy representation therefore could also be called Hankel transform in analog to the name Kontorovich-Lebedev transform on the $\mu$ side.}. Conceptually, we are now at the level of the paper \cite{Chowdhury:2023arc}, which can be understood as a precursor of the dressing rules \cite{Chowdhury:2025ohm}. This form should immediately remind us of the flat-space construction, where the structure is very similar. We will show in section \ref{sec:SK_dressing_rules} that one can re-shuffle terms in the energy representation to derive the dressing rules, which schematically generate expressions like
\begin{align}
    \mathcal{M} = \int\limits_{-\infty}^\infty \d p_1 \ldots \d p_R \Big( \mathrm{dressing} \Big) \Big( \mathrm{flat \ space \ diagram \ with \ injected \ energies} \ \{p_j\} \Big).
\end{align}
In the remainder of this section, we find integral identities that link the two propagator shapes. These identities are called \textit{transition identities} in figure \ref{fig:sturm_liouville_schematic}.

\subsubsection*{Transition Identities}

From now on, we drop the subscripts $\mu$ and $k$ and define $k^2=-p^2$ with $p^2>0$ to have positive eigenvalues $\lambda = -k^2$ and separate the eigenvalue from the momentum in our notation. This is an important step, as we have to remember that the eigenvalue is integrated over at fixed external parameters. This is the same as saying that the $p^2$ appearing in the flat space Feynman propagator is not set to $m^2$, in other words we are \textit{off-shell}. The transition identities have to be derived for the two cases of complementary and principal series scalars, since the $i\epsilon$ structure in $\mu$ space is very different between them, see \eqref{eq:KLF_propagators}.
For the complementary series $\nu_\phi > 0$, we use the integral identities\footnote{These integrals can be found in \cite[eq. 2.16.52.11/12]{prudnikov} and \cite[eq. 6.577.1/2]{gradshteyn}, respectively.}
\begin{align}
  \frac{1}{\pi} \int\limits_{-\infty}^{+\infty} \d\mu \,\mathcal{N}_\mu \frac{K_{i\mu}(kz_1)K_{i\mu}(kz_2)}{\mu^2+\nu_\phi^2} = I_{\nu_\phi}(kz_<) K_{\nu_\phi}(kz_>) = \int\limits_{0}^{\infty} \d p\, p \frac{J_{\nu_\phi}(pz_1)J_{\nu_\phi}(pz_2)}{p^2+k^2}\,, \label{eq:spectral_transition}    
\end{align}
where $z_< \coloneq \min(z_1,z_2)$ and $z_>\coloneq \max(z_1,z_2)$. We can read off that $\mathcal{N}_p = p$ and $\Psi_\mu \sim J_{\pm \nu}(p z)$ in figure \ref{fig:sturm_liouville_schematic}. We also need to treat the $\hat\delta$ term in \eqref{eq:KLF_propagators_complementary}, for which we need
\begin{align}
    K_\nu(k z) = \frac{\pi}{2\sin(\pi\nu)} \Big( I_{-\nu}(k z) - I_{\nu}(kz) \Big),
\end{align} 
which we can use to write
\begin{equation}
    \frac{1}{\pi}K_\nu(kz_1)K_\nu(kz_2)
    =\frac{1}{2\sin(\pi\nu)}\int_0^\infty\d p\,p
    \frac{J_{-\nu}(pz_1)J_{-\nu}(pz_2)-J_\nu(pz_1)J_\nu(pz_2)}{p^2+k^2}.
    \label{eq:homogeneous_connection}
\end{equation}
The last integral form holds directly for $|\Re(\nu)|<1$, $\nu\neq 0$, so that it can also be used for the Wightman function that appears in the principal series SK formalism as well, see \eqref{eq:KLF_propagators_wightman}. Outside this range it is understood by analytic continuation, with integer orders taken by a limit.

For principal-series fields, $\mu_\phi>0$, so we formally continue
$\nu_\phi\to\epsilon\pm i\mu_\phi$ in \eqref{eq:spectral_transition}, keeping the
pole prescription. This gives
\begin{align}
    \frac{1}{\pi}\int_{-\infty}^{\infty}\d\mu\,\mathcal{N}_\mu
    \frac{K_{i\mu}(kz_1)K_{i\mu}(kz_2)}{\mu^2-\mu_\phi^2\pm i\epsilon}
    &=I_{\pm i\mu_\phi}(kz_<)K_{i\mu_\phi}(kz_>)\notag\\
    &=\int_0^\infty\d p\,p
    \frac{J_{\pm i\mu_\phi}(pz_1)J_{\pm i\mu_\phi}(pz_2)}{p^2+k^2}.
    \label{eq:spectral_transition_principal}
\end{align}
We have to use both cases $\pm i \epsilon$ for the KLF propagator, as written in \eqref{eq:KLF_propagators_principal}. 

We can use the transition identities to go from $\mu$ space to $p$ space, but we can also stop in the middle in partial position space in the variables $(z, \vec k)$. There, the propagator is
\begin{subequations} \label{eq:propagator_z_space}
\begin{align}
    \mathcal{G}^{i\nu_\phi}_{\aa\aa}(kz_1, kz_2) &\coloneq \int\limits_{-\infty}^\infty \d \mu \, \mathcal{N}_\mu K_{i\mu}(k z_1) K_{i\mu}(k z_2) \mathcal{G}_{\aa \aa}^{i\nu_\phi}(\mu) \notag\\
    = e^{-i\aa\pi\frac{d-1}{2}} &\left[ \frac{\pi}{2} \Big( I_{\nu_\phi}(kz_<) + I_{-\nu_\phi}(kz_<) \Big) K_{\nu_\phi}(kz_>) - i \aa \cos(\pi\nu_\phi) K_{\nu_\phi}(kz_1) K_{\nu_\phi}(kz_2) \right], \label{eq:propagator_z_space_aa}\\
    \mathcal{G}^{i\nu_\phi}_{\aa,(-\aa)}(kz_1, kz_2) &\coloneq \int\limits_{-\infty}^\infty \d \mu \, \mathcal{N}_\mu K_{i\mu}(k z_1) K_{i\mu}(k z_2) \mathcal{G}_{\aa,(-\aa)}^{i\nu_\phi}(\mu) = K_{\nu_\phi}(kz_1) K_{\nu_\phi}(kz_2)\,,
\end{align}
\end{subequations}
In $p$ space, the propagator again factorizes and is
\begin{subequations} \label{eq:propagator_as_Jintegral_intermediate}
\begin{align}
    \mathcal{G}^{i\nu_\phi}_{\aa\aa}(kz_1, kz_2) &= \frac{\pi e^{-i\aa \pi \frac{d}{2}}}{2\sin(\pi\nu_\phi)} \int\limits_0^\infty \d p \, p \, \frac{e^{i\aa\pi \nu_\phi} J_{-\nu_\phi}(p z_1) J_{-\nu_\phi}(p z_2) - e^{-i\aa\pi\nu_\phi} J_{\nu_\phi}(p z_1) J_{\nu_\phi}(p z_2)}{p^2 + k^2}, \\
    \mathcal{G}^{i\nu_\phi}_{\aa,(-\aa)}(kz_1, kz_2) &= \frac{\pi}{2\sin(\pi\nu_\phi)} \int\limits_0^\infty \d p \, p \, \frac{J_{-\nu_\phi}(p z_1) J_{-\nu_\phi}(p z_2) - J_{\nu_\phi}(p z_1) J_{\nu_\phi}(p z_2)}{p^2+k^2}.
\end{align}
\end{subequations}
Both of these representations hold for both complementary and principal series fields, where we replace $\nu_\phi \rightarrow i\mu_\phi$. For terms where $\sin(\pi\nu_\phi) \rightarrow 0$, we understand them by a limit argument. Splitting the above propagators in two parts with either $\pm \nu_\phi$ results in the aforementioned shadow formalism. We also want to point out that some of the previous arguments were already found in the appendix of the original dressing rules paper \cite{Chowdhury:2025ohm}, but their significance was not realized there.

\section{Generalized Dressing Rules} \label{sec:SK_dressing_rules}

We are now ready to generalize dressing to general masses. To avoid troubles with divergent time integrals, we focus on masses that are at least as large as the mass of a conformally coupled field, which translates to
\begin{align}
    | \Re(\nu) | \leq \frac{1}{2}.
\end{align}
From now on, there will no longer be any $\mu$ integrals, so we drop the subscript $\phi$ everywhere. To illustrate how the argument works, we first re-derive the dressing rules for conformally coupled fields. We then point out the single point of failure of this argument for general masses, and how to push around it. The result is a generalized version of dressing rules, which is more complicated than the conformally coupled case, but still contains the original spirit of writing curved space QFT diagrams as integrals over flat space ones.

\subsection{Conformal Coupling} \label{sec:SK_dressing_rules_conformal}

Once we have the energy representation introduced in section \ref{sec:sturm_liouville}, deriving the dressing rules is as easy as performing the $z$-integrals that are still present in the overall diagram structure. In flat space, the time integrals become delta functions enforcing energy conservation. Instead, in de Sitter, they become expressions that depend on the energies of the connecting propagators. We interpret these expressions as new propagators and draw additional lines that indicate them diagrammatically. One obstacle in this construction is the SK index sum, but we will show below that this can be treated in full generality without jeopardizing the elegance of the construction.

Despite the conceptual simplicity explained above, the many pieces and factors of a given diagram can make the notation quite cloudy. It is very hard to see the big picture when writing the derivation down step by step. To circumvent this, we present a very schematic computation and fill in the details in the rest of this section:
\begin{align}
    \mathcal{M} &\overset{(0)}{=} \sum\limits_{\mathrm{SK}} \Big(\mathrm{external} \Big)_\aa \int_k \int_p \Big( \frac{1}{p^2+k^2} \ \mathrm{- \ propagators} \Big) \int_z \Big( e^{i p z}, K \ \mathrm{- \ functions}\Big)_\aa \notag\\
    &\overset{(1)}{=} \sum\limits_{\mathrm{SK}} \Big(\mathrm{ext. \ no \ } \aa \Big) \Big( \mathrm{flat \ space \ diagram} \Big) \prod_v Z_v(\aa_v) e^{i \aa_v \psi_v} \notag\\
    &\overset{(2)}{=} \Big(\mathrm{ext. \ no \ } \aa \Big) \Big( \mathrm{flat \ space \ diagram} \Big) \prod\limits_v \Big( \mathrm{per \ vertex \ dressing} \Big) \notag\\
    &\overset{(3)}{=} \Big(\mathrm{ext. \ no \ } \aa \Big) \int_p \prod\limits_v \Big( \mathrm{dressing} \Big) \Big( \mathrm{flat \ space \ diagram \ with \ energy \ conservation} \Big) \notag
\end{align}
Components that depend on the SK labels are denoted with the subscript $\aa$.

\subsubsection*{Step (0) - Energy Representation}

We pick up where we ended in the Sturm-Liouville section \ref{sec:sturm_liouville} with the energy representation. It does not matter if we take the KLF or the partial momentum space formulation as the starting point, both can be translated to $p$ space via the transition identities \eqref{eq:spectral_transition}:

If we start from \textit{KLF} with the expression \eqref{eq:KLF_diagram_final}, we take apart the vertex functions $\mathcal{V}$ and write them as $z$ integrals. This way, we can perform the $\mu$ integrals and transition to partial momentum space. The propagators are now Bessel functions of products of momenta $k$ and times $z$, as written in \eqref{eq:spectral_transition}.

Alternatively, if we start from the \textit{partial momentum space} formulation, we already have Bessel functions of the momenta and times, so the first step is not needed. Regardless of where we started, we can introduce the $p$ integrals and write the propagators according to \eqref{eq:propagator_as_Jintegral_intermediate}, which for conformally coupled fields $\mu = i/2$ becomes
\begin{align}
    \mathcal{G}_{\aa\aa}^{i/2}(k z_1, k z_2) &= \frac{e^{-i\aa\pi \frac{d-1}{2}}}{\sqrt{z_1 z_2}} \int\limits_0^\infty \d p \, \frac{\cos\Big( p (z_1 - z_2)\Big)}{p^2 + k^2} =  \frac{e^{-i\aa\pi \frac{d-1}{2}}}{2\sqrt{z_1 z_2}} \int\limits_{-\infty}^\infty \d p \, \frac{e^{ipz_1} e^{-ipz_2}}{p^2 + k^2} \\
    \mathcal{G}_{\aa,(-\aa)}^{i/2}(k z_1, k z_2) &= \frac{1}{\sqrt{z_1 z_2}} \int\limits_{0}^\infty \d p \, \frac{\cos\Big( p (z_1 + z_2)\Big)}{p^2 + k^2} = \frac{1}{2\sqrt{z_1 z_2}} \int\limits_{-\infty}^\infty \d p \, \frac{e^{ipz_1} e^{ipz_2}}{p^2 + k^2}
\end{align}
Both of these cases, $\aa \aa$ and $\aa, (-\aa)$ can be combined in
\begin{align}
    \mathcal{G}^\frac{i}{2}_{\aa\bb}(k z_1, k z_2) = e^{-i \pi \frac{d-1}{4} (\aa + \bb)} \frac{\pi}{\sqrt{z_1 z_2}} \int\limits_{-\infty}^\infty \frac{\d p}{2\pi} \frac{e^{i \aa p z_1} e^{ - i \bb p z_2}}{p^2+k^2}. \label{eq:conformal_propagator_p_space}
\end{align}
This form factorizes under the $p$ integral, both in  $z_{1/2}$ and $\aa / \bb$. By combining the $p$-integral with the spatial $\vec k$ integrals, we can interpret the denominator of the above propagator as a Euclidean flat space propagator. Momentum conservation for $\vec k$ already holds, so up to energy conservation in $p$, we have now assembled the flat space block of the diagram, as indicated in the schematic procedure. From now on, we use the name $\vec q$ for momenta carried by internal lines to distinguish them from the external momenta $\vec k$. The composite momenta are defined as $Q = (p, \vec q)$.

What is left of the propagators after these steps are the phase factors and the exponentials $e^{i\aa p z}$ that together with the $K_{1/2}(k z)$ from the external legs form the $z$ integrals that are discussed in the next step. Note the opposite signs in the two exponential factors under the $p$ integral in \eqref{eq:conformal_propagator_p_space}. We can interpret them as a direction of energy flow through the propagator (at least for $\aa = \bb$). These signs play an important role in the global energy structure of the diagram, as we discuss below. The transformation $\bb \rightarrow - \bb$ seems to redirect this energy flow. However, it also changes the phase factor in front of the propagator, so the direction is reflected either in the phase or the sign of $p$. Where does not matter, so we decide to keep it with the $p$ in the exponent.

\subsubsection*{Step (1) - Performing the $z$-Integral}

Now, we consider all contributions in a diagram that are connected to a given vertex $v$, meaning they contain either a $z_v$ or an $\aa_v$. According to \eqref{eq:KLF_vertex_function}, every $z$ integral comes with the measure $\d z \, z^{\frac{d(N-2)}{2} - 1}$, where $N$ is the number of legs -- internal and external -- connected to this vertex. The propagators \eqref{eq:conformal_propagator_p_space} each produce their own factor of $1/\sqrt{z}$ at each end. Further, external legs connecting to a vertex leave a Bessel $K$ function
\begin{align}
    K_{\frac{1}{2}}(k_\mathrm{ext} z) = \sqrt{\frac{\pi}{2}} \frac{1}{\sqrt{k_\mathrm{ext} z}} e^{-k_\mathrm{ext} z}. \label{eq:conformal_bessel_function}
\end{align}
Therefore, all legs contribute $1/\sqrt{z}$ each.

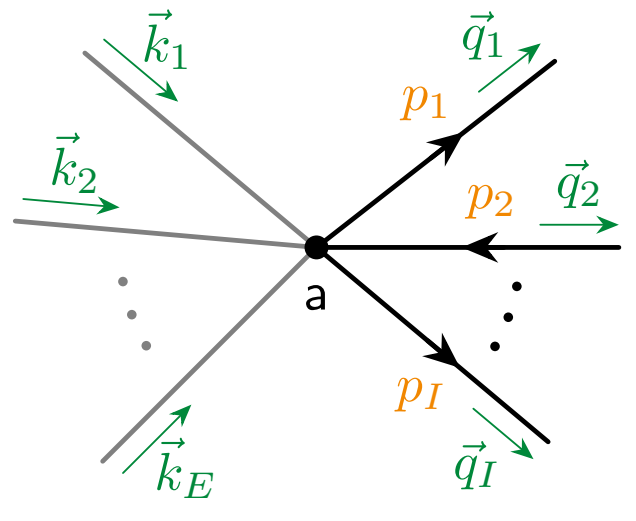
\begin{figure}[h]
\centering
\providecolor{lmugreen}{RGB}{0, 136, 58}    
\providecolor{lmuorange}{RGB}{241, 135, 0}  
\begin{tikzpicture}[x=1cm, y=1cm, line cap=round,
  vtx/.style={circle, fill=black, inner sep=0pt, minimum size=4.5pt},
  leg/.style={line width=0.9pt},
  eflow/.style={leg, -{Stealth[length=2.4mm,width=2mm]}},                
  flow/.style={-{Stealth[length=1.5mm,width=1.2mm]}, line width=0.35pt},  
  lbl/.style={font=\small},
]
  \def\L{2.0}      
  \def\off{0.15}  
  \def\loff{0.42}  
  \foreach \a/\k/\sd in {140/{\vec k_1}/-1, 175/{\vec k_2}/-1, 225/{\vec k_E}/1}{
    \draw[leg, gray] (0,0) -- (\a:\L);
    \draw[flow, lmugreen] ($(\a:0.98*\L)+({\a+90*(\sd)}:\off)$) -- ($(\a:0.66*\L)+({\a+90*(\sd)}:\off)$);
    \node[lbl, text=lmugreen] at ($(\a:0.82*\L)+({\a+90*(\sd)}:\loff)$) {$\k$};
  }
  \foreach \t in {190,200,210} \fill[gray] (\t:1.3) circle (0.9pt);

  \foreach \a/\p/\q/\s/\sd in {38/{p_1}/{\vec q_1}/-1/1, 0/{p_2}/{\vec q_2}/1/1, -40/{p_I}/{\vec q_I}/-1/-1}{
    \pgfmathsetmacro\ra{(0.55-0.07*\s)*\L}
    \pgfmathsetmacro\rb{(0.55+0.07*\s)*\L}
    \draw[leg] (0,0) -- (\a:\L);
    \draw[eflow] (\a:\rb) -- (\a:\ra);
    \node[lbl, text=lmuorange] at ($(\a:0.58*\L)+({\a+90*(\sd)}:0.3)$) {$\p$};
    \draw[flow, lmugreen] ($(\a:0.74*\L)+({\a+90*(\sd)}:\off)$) -- ($(\a:1.0*\L)+({\a+90*(\sd)}:\off)$);
    \node[lbl, text=lmugreen] at ($(\a:0.87*\L)+({\a+90*(\sd)}:\loff)$) {$\q$};
  }
  \foreach \t in {-11,-20,-29} \fill (\t:1.35) circle (0.9pt);

  \node[vtx] at (0,0) {};
  \node[lbl, anchor=north] at (0,-0.12) {$\mathsf{a}$};   
  \node[font=\small] at (0,-2.2)
    {$\delta^{(d)}\bigl(\textcolor{lmugreen}{\vec k_1}+\dots+\textcolor{lmugreen}{\vec k_E}
      -\textcolor{lmugreen}{\vec q_1}-\dots-\textcolor{lmugreen}{\vec q_I}\bigr)$};
\end{tikzpicture}
\caption{Notation at a single vertex with SK index $\mathsf{a}=\pm$. External legs (grey) carry the spatial momenta $\vec k_s$, which flow into the vertex, but they carry no energy $p$. Internal legs (black) carry the energies $p_l$ in the direction of the arrows, as well as the momenta $\vec q_l$. Writing $\sigma_l=+1$ ($-1$) for energy flowing into (out of) the vertex, the net energy entering it is $P=\sum_l\sigma_l p_l$. Analogously for external legs, we also use $K=\sum_s|\vec k_s|$. For conformally coupled fields the vertex integral depends only on these two combinations.}
\label{fig:vertex}
\end{figure}

Next, we focus on the $e^{\pm i p \aa z}$ exponential factors from internal lines \eqref{eq:conformal_propagator_p_space}, where we absorb the $\pm$ into a new variable $\sigma$, and the $e^{-k_\mathrm{ext} z}$ factors from external lines and perform the $z$ integration. The general structure of the $z$-integral is then
\begin{align}
   Z_\aa(P, K) \coloneq& \int\limits_0^\infty \d z \, z^{\frac{d(N-2)}{2} - 1 - \frac{N}{2}} \prod\limits_{\mathrm{internal} \ l} e^{i \aa \sigma_l p_l z} \prod\limits_{\mathrm{external} \ s} e^{- k_s z}\nonumber\\
    =& \int\limits_0^\infty \d z \, z^{\frac{d(N-2) - N}{2} - 1} e^{-(K - i \aa P) z} = (K - i \aa P)^{d - \frac{N}{2}(d-1)} \Gamma\left( \frac{N}{2}(d-1) - d \right) \label{eq:conformal_z_integral}\,,
\end{align}
where we assumed $\Re\big(d(N-2)-N\big) > 0$, and like in figure \ref{fig:vertex}, defined 
\begin{align}
    P = \sum\limits_{\mathrm{internal} \ l} \sigma_l p_l, \qquad K = \sum\limits_{\mathrm{external} \ s} k_s\,.
\end{align}

If there are no external legs, we understand the above expression in the limit $K \rightarrow 0$. Note that in the definition of $Z_\aa(P, K)$, the complex unit $i$ always appears together with $\aa$ and $P$. The important structural property of this object therefore is the decomposition
\begin{align}
    Z_\aa(P, K) = C(P, K) + i \aa S(P, K), \qquad C, S \ \mathrm{real}; \quad C \ \mathrm{even \ in} \ P \longrightarrow -P, \ S \ \mathrm{odd}. \label{eq:Z_split_definition}
\end{align}
In particular, $C(P,K) = \Re(Z_+(P,K))$ and $S(P,K) = \Im(Z_+(P,K))$. The even and odd structures $C$ and $S$ will later form the two possible dressings via dashed and dotted lines.

\medskip

In addition to the $z$-dependence of the vertex, we have a phase structure associated with the SK index $\aa$. As discussed before, there are three sources of phase factors: the propagators \eqref{eq:KLF_propagators}, the vertices \eqref{eq:KLF_vertex_factor}, and external legs \eqref{eq:KLF_external}. To treat the latter, we have to extract the $\aa$ dependent terms from $\mathcal{E}_{\{\aa_v\}}(\eta_0)$, which explains the difference between $(\mathrm{external})_\aa$ and $(\mathrm{ext. \ no } \ \aa)$ in the schematic derivation.

All of the phases factorize such that we can attribute them to their vertex of origin. Altogether, we find an overall factor for each vertex
\begin{align}
    \underbrace{-i \aa e^{i \aa \pi \frac{d}{2}}}_{\mathrm{vertex}} \times \prod\limits_{\mathrm{internal} \ l} e^{-i \aa \pi \frac{d-1}{4}} \times \prod\limits_{\mathrm{external} \ s} e^{i \aa \pi \frac{1 - d}{4}} \eqcolon e^{i \aa \psi} = \cos\psi + i \aa \sin\psi, \label{eq:psi_angle_definition}
\end{align}
where we defined $\psi$ as the sum over the exponents
\begin{align}
    \psi = \pi \left( \frac{d-1}{2} - \frac{d-1}{4} N \right). \label{eq:conformal_psi}
\end{align}
In dimensional regularization $d = 3 + 2\epsilon$, we infer from \eqref{eq:conformal_z_integral}
\begin{subequations} \label{eq:example_dressings_conformal}
\begin{align}
    N = 3: \qquad Z_\aa(P,K) &= (K - i \aa P)^{-\epsilon} \ \Gamma(\epsilon), \quad \psi = - \frac{\pi (1 + \epsilon)}{2}, \\
    N = 4: \qquad Z_\aa(P,K) &= (K - i \aa P)^{-1-2\epsilon} \ \Gamma(1+2\epsilon), \quad \psi = - \pi \left( 1 + \epsilon \right).
\end{align}
\end{subequations}
We can read off that for vertices with 3 legs, dimensional regularization is needed in order to deal with the IR divergent $z$ integral. However, we will see soon that the problems cancel out at the level of the final diagram, as expected. This coincides with the findings in the basic SK-formalism for these interactions.

\subsubsection*{Step (2) - The Schwinger-Keldysh Sum}

To continue we just have to rearrange. Our schematic expression at the moment looks like
\begin{align}
    \mathcal{M} = \sum\limits_{\mathrm{SK}} \Big(\mathrm{ext. \ no \ } \aa \Big) \Big( \mathrm{flat \ space \ diagram} \Big) \prod\limits_v Z_{\aa_v}(P_v, K_v) e^{i \aa_v \psi_v}.
\end{align}
We then use that the sum over SK indices $\aa_v$, at fixed $v$, of a product of localized terms simply is the product over the sum $\aa_v = \pm$ of these terms:
\begin{align}
    \mathcal{M} = \Big(\mathrm{ext. \ no \ } \aa \Big) \Big( \mathrm{flat \ space \ diagram} \Big) \prod\limits_v \left(\sum\limits_{\aa_v = \pm} Z_{\aa_v}(P_v, K_v) e^{i \aa_v \psi_v} \right). \label{eq:SK_sum_factorization}
\end{align}
This sum is easy to evaluate as we have split forms for $Z_\aa(P,K)$ in \eqref{eq:Z_split_definition} and $\psi$ in \eqref{eq:psi_angle_definition}:
\begin{align}
    \sum\limits_{\aa = \pm} Z_\aa(P,K) e^{i \aa \psi} = 2\cos\psi \cdot C(P,K) - 2\sin\psi \cdot S(P,K), \label{eq:conformal_z_integral_SK_sum}
\end{align}
while the other terms cancel out because they are proportional to $\aa$ under the sum. The combinations $2\cos\psi \cdot C(P,K)$ and $-2\sin\psi \cdot S(P,K)$ are the same as the original dressing propagators defined in \cite{Chowdhury:2025ohm}. We can read this off of table \ref{tab:conformally_coupled_dressings}. Since both dressings always show up together in the sum \eqref{eq:conformal_z_integral_SK_sum}, it seems to be a more invariant definition to call the sum the dressing, even if it is not yet apparent that it has a propagator-like form in general. We indicate this combination by drawing dressings with both dotted and dashed lines, as can be seen for example in the explanation of dressing Feynman rule \ref{rule:dressing}.

\begin{table}[h]
    \centering
    \begin{tabular}{c||c|c||c|c}
        $N$ & $2\cos\psi \cdot C(P,K)$ & equivalent & $-2\sin\psi \cdot S(P,K)$ & equivalent \\\hline
        $3\vphantom{\Big(}$ & $-\pi$ & dotted $\phi^3$ & $2\arctan\left( \frac{P}{K} \right)$ & dashed $\phi^3$ \\
        $4\vphantom{\Big(}$ & $-\frac{2K}{K^2 + P^2}$ & dashed $\phi^4$ & $0$ & does not exist
    \end{tabular}
    \caption{Dressing propagators obtained from our derivation for $N = 3, 4$ which corresponds to the conformally coupled scalar with $\phi^3$ and $\phi^4$ interactions, respectively. The dashed $\phi^3$ from \cite{Chowdhury:2025ohm} has an extra $i$ relative to ours. The $i$ is there in the original dressing rules to generate an additional factor of $(-1)$ for every pair of dashed propagators. In our construction, this sign appears automatically as we are tracking the direction of the energies $p$. One dressing propagator will have the opposite sign of $P$, and as $S$ is odd in $P$, this sign will show up in the overall diagram, also refer to appendix \ref{app:shadow-conformal-graph-proof}. Note that the roles of dotted and dashed propagators are exchanged between different $N$, which is simply a mismatch of labeling between this paper and \cite{Chowdhury:2025ohm}.}
    \label{tab:conformally_coupled_dressings}
\end{table}
Note that the pole in $\epsilon$ for $N=3$ has disappeared because it canceled with $\sin(\epsilon)$. Further, the $\sin\psi \cdot S$ term is $0$ for $\epsilon \rightarrow 0$ for any even $N \geq 4$, while for odd $N > 4$ the cosine term drops out, which means that there are not two, but only one dressing. This has already been observed in the original paper \cite{Chowdhury:2025ohm}. The reason is that the $z$ integral \eqref{eq:conformal_z_integral} is IR finite in these cases, which means that we can set $\epsilon = 0$, which in turn eliminates $\sin\psi = 0$.

One more feature of the dressing rules is that for conformally coupled $\phi^3$, the dashed propagators may only appear in pairs. In our derivation, this is an immediate consequence of the fact that $S$ is odd for $P \rightarrow -P$. Since all $p_j$ are integrated over, we can flip the direction of integration of all $p_j \rightarrow -p_j$. Terms with an odd number of $S$ are odd under this transformation and therefore have to integrate to $0$.

\subsubsection*{Step (3) - Dressing Propagators}

If we can count the energy flowing into the vertex, it is simply $P = \sum_l \sigma_l p_l$. We can then draw a new line -- the \textit{dressing propagator} -- at every vertex and define its outflowing energy as $P$. Now, by construction, we have local energy conservation at the vertex in the sense that the signed energies of all connecting lines add to zero if we include the new dressing propagator. In this sum, external lines are defined to not carry any energy. 

The sum of absolute values of the momenta of external lines is defined as $K$ and kept as a label to the dressing propagator. Note that we do not assign momentum to the dressing propagator since momentum is already conserved in the flat space part of the diagram. 

So far, these lines do not go anywhere, they simply carry away energy. However, we can count the total energy carried by all these lines:
\begin{align}
    \sum\limits_{\mathrm{vertices} \ v} P_v = \sum\limits_{\mathrm{vertices} \ v} \left(\sum\limits_{\substack{\mathrm{internal}\\\mathrm{legs} \ l \ \mathrm{at} \ v}} \sigma_l \, p_l \right) = \sum\limits_{\substack{\mathrm{all \ internal}\\\mathrm{propagators} \ j}} (p_j - p_j) = 0.
\end{align}
This is a simple consequence of the fact that all the $p$ propagators \eqref{eq:conformal_propagator_p_space} are directed and their energy flows into one vertex with a $+$ and into one vertex with a $-$. Summing over all vertices necessarily equals 0. We can introduce another bookkeeping device, the common point where all dressing lines meet, called \textit{dressing point}, as a diagrammatic representation of this fact.

To finally arrive at the original formulation of the dressing rules, we can now integrate over $P_r$ for all dressing propagators. In the same step, we add a delta function that enforces energy conservation at every vertex of the flat space diagram, in order to kill the new integrals while setting their respective $P_r$ to the correct value. We do not need to add another delta function to enforce energy conservation at the point where all dressing propagators meet, as energy conservation there holds automatically, as argued above.

This concludes our derivation of dressing rules for conformally coupled fields. For future reference, we will now simply state the effective Feynman rules resulting from the above procedure.

\subsubsection*{Feynman Rules for Conformally Coupled Dressing Rules}

To compute a given Feynman diagram for conformally coupled scalars in de Sitter space with any monomial interaction $\lambda \phi^N$, follow steps \ref{rule:propagators}--\ref{rule:conservation} below. The diagrams on the right illustrate each step for a tree-level exchange with $N=3$ and $M=4$ external legs.
\begin{dressingrule}{lmugreen}{Propagators}{\ruleDiagram{1}}\label{rule:propagators}
    Draw the flat space diagram and assign momentum and energy flow directions as usual. For \rulekw{every propagator}, write
    \begin{align*}
        \pi \int \frac{\d^{d+1} Q}{(2\pi)^{d+1}}\frac{1}{Q^2},
    \end{align*}
    where $Q^2$ is the Euclidean $(d+1)$-dimensional scalar product. The additional factor of $\pi$ follows from \eqref{eq:conformal_propagator_p_space}.
\end{dressingrule}
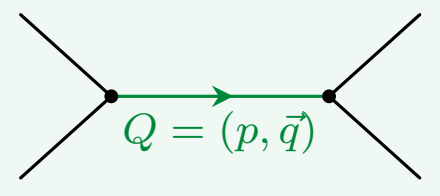
\begin{dressingrule}{lmuorange}{Dressing Propagators}{\ruleDiagram{2}}\label{rule:dressing}
    Attach \rulekw{dressing propagators} to every vertex of the diagram, such that all of them meet at one point, the \rulekw{dressing point}. For every such dressing propagator, write
    \begin{align*}
        \int\limits_{-\infty}^\infty \frac{\d P}{2\pi} \left( \sum\limits_{\aa = \pm} Z_\aa(P,K) e^{i\aa \psi} \right)
    \end{align*}
    where $Z_\aa(P,K)$ is defined in \eqref{eq:conformal_z_integral} and $\psi$ is defined in \eqref{eq:conformal_psi}. The argument $K$ is the sum of the magnitudes of the external momenta flowing into a vertex and $P$ is the propagators energy.
\end{dressingrule}
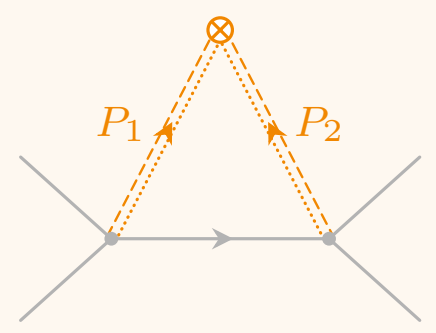
\begin{dressingrule}{lmugreen}{External Legs and Vertices}{\ruleDiagram{3}}\label{rule:prefactors}
    For every \rulekw{external leg} with momentum $\vec k_s$, multiply by
    \begin{align*}
        \frac{\sqrt\pi H^\frac{d-3}{2}}{2} \frac{(-\eta_0)^\frac{d-1}{2}}{k_s},
    \end{align*}
    which originates from \eqref{eq:KLF_external} (without the $\aa$ dependent structure), together with the $\sqrt{\pi/(2k_s)}$ leftover from the Bessel function \eqref{eq:conformal_bessel_function}. Then, multiply by an extra factor of
    \begin{align*}
        \mathcal{C} \lambda \frac{H^\frac{(N-2)(d+1)}{2}}{\pi^\frac{N}{2}}
    \end{align*}
    for every \rulekw{vertex}, representing the left-over piece of \eqref{eq:KLF_vertex_factor} that contains the usual symmetry factors $\mathcal{C}$ of the flat space Feynman diagram, coupling constants $\lambda$ and some powers of $\pi$ and $H$.
\end{dressingrule}
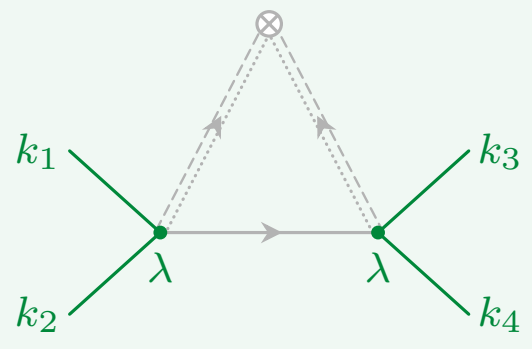
\begin{dressingrule}{lmugreen}{Energy and Momentum Conservation}{\ruleDiagram{4}}\label{rule:conservation}
    Finally, enforce \rulekw{$d$-momentum conservation} by inserting a delta function $(2\pi)^d \, \delta^{(d)}(\vec k_s + \ldots - \vec q_l + \ldots)$ at every flat space vertex. Then, implement \rulekw{energy conservation} at every flat space vertex by inserting the delta function $(2\pi)\,\delta(P_r+\ldots - p_j + \ldots)$. Note that energies and momenta are directed and that external lines carry no energy, while dressings carry no momentum. Energy is automatically conserved at the dressing point.
\end{dressingrule}
\vspace{5mm}
Note that the above rules can be generalized to multiple different polynomial interaction terms in a very straight-forward way.
This formulation of dressing rules differs slightly from the original phrasing in \cite{Chowdhury:2025ohm} so that they generalize more easily to the general mass case. Mainly, we do not differentiate between dotted and dashed dressings. Parity of the energy integrals automatically kills the contributions which would not have been present under the original formulation. One may wonder if the original derivation via the shadow formalism allows for a similar generalization to massive fields. This will be addressed in section \ref{sec:shadow_dressing_rules}.

\subsection{General Masses} \label{sec:SK_dressing_rules_general}

The construction for dressing rules from the previous section can be extended to general masses in a very similar way although there is one important difference, which is the existence of the second term proportional to $\cos(\pi\nu)$ in \eqref{eq:propagator_z_space_aa}. Below we follow the same scheme as for the conformally coupled scalars, but  highlight the relevant generalizations.

\subsubsection*{Modification to Step (0) - Energy Representation}

The energy representation for the propagator in \eqref{eq:propagator_as_Jintegral_intermediate} holds for general masses. However, it is significantly more complicated to find a factorized representation that replaces \eqref{eq:conformal_propagator_p_space}. To find one, we define the (Tricomi) function 
\begin{align}
    \Pi_\nu(\omega) \coloneq \frac{1}{\Gamma\big( \nu + \frac{1}{2} \big)} \int\limits_0^\infty \d s \ e^{-s} s^{\nu - \frac{1}{2}} \Big( 1 + \frac{\omega s}{2} \Big)^{\nu - \frac{1}{2}} = \sqrt{\frac{2}{\pi \omega}} e^\frac{1}{\omega} K_\nu\Big( \frac{1}{\omega} \Big). \label{eq:Pi_definition}
\end{align}
This function has a branch cut along the negative real line in $\omega$ and the integral is not well-defined for $\Re(\nu) \leq - \frac{1}{2}$, which is excluded. Using the shorthand notation $H_\nu^{(+)} \coloneq H_\nu^{(1)}$ and $H_\nu^{(-)} \coloneq H_\nu^{(2)}$ and the identity
\begin{align}
    H_\nu^{(\kappa)}(x) = \sqrt{\frac{2}{\pi x}} e^{i \kappa \big( x - \frac{\pi\nu}{2} - \frac{\pi}{4} \big)} \Pi_\nu\Big( \frac{i\kappa}{x} \Big), \label{eq:hankel_Pi_identity}
\end{align}
we can split the Bessel $J$ functions appearing in \eqref{eq:propagator_as_Jintegral_intermediate} as
\begin{align}
    J_{\pm \nu}(x) = \frac{1}{2} \sum\limits_{\kappa = 1,-1} e^{i\pi \kappa \nu \frac{1 \mp 1}{2}} H_\nu^{(\kappa)}(x) = \frac{1}{2} \sum\limits_{\kappa = 1,-1} e^{-i\pi \kappa \frac{1 \pm 2 \nu}{4}} \sqrt{\frac{2}{\pi x}} e^{i \kappa x} \Pi_\nu\Big( \frac{i\kappa}{x} \Big). \label{eq:J_split_identity}
\end{align}
The $\Pi$ function will be convenient to generalize the $z$-integral from  conformal coupling to the general mass case in a natural way. However, for that it is also useful to define the auxiliary function
\begin{align}
    X_\aa^\nu(\omega, z) \coloneq \frac{1}{\sqrt{z}} e^{i\aa \omega z} \Pi_\nu\Big( \frac{i \aa}{\omega z} \Big). \label{eq:X_definition}
\end{align}
We insert \eqref{eq:J_split_identity} in terms of $X$ into all the Bessel $J$ functions appearing in the propagator \eqref{eq:propagator_as_Jintegral_intermediate}. Note that each line is a sum of two terms under the integral, where the sum essentially goes over the sign of $\nu$. After inserting the expansion for $J$, we therefore have three different sums: The sign of $\nu$ and two independent $\kappa$ sums. There are many nontrivial cancellations in these sums, but we can work out the result
\begin{align}
\begin{split}
    \mathcal{G}^{i\nu}_{\aa\bb}(qz_1, qz_2) &=\pi e^{-i\pi \frac{d-1}{4} (\aa + \bb)} \int\limits_{-\infty}^\infty \frac{\d p}{2\pi} \frac{X_\aa^\nu(p, z_1) X_\bb^\nu(-p, z_2)}{p^2 + q^2} \\
    &\quad + 2\pi\delta_{\aa\bb} \cos(\pi\nu)e^{-i\aa\pi \frac{d}{2}} \int\limits_0^\infty \frac{\d p}{2\pi} \frac{X_\aa^\nu(p, z_1) X_\aa^\nu(p,z_2)}{p^2+q^2}. \label{eq:general_nu_factorized_propagator}
\end{split}
\end{align}
In a way, we have now gone full circle from the Hankel basis in the Schwinger-Keldysh formulation. However, the time ordering has been removed through the energy representation and was traded for the non-trivial dependence on $\aa$ and $\bb$. Again, the imaginary unit only ever appears together with $\aa$ and $p$, so
\begin{align}
    X_{\mp}^\nu(p, z) = \overline{X_\pm^\nu(p,z)} = X_\pm^\nu(-p, z).
\end{align}
The combined form \eqref{eq:general_nu_factorized_propagator} shows the same extra $\aa = \bb$ piece in the second line as \eqref{eq:propagator_z_space_aa}, also in combination with a $\cos(\pi\nu)$. It is this piece that adds additional structures to the original dressing rules, while the first term above plays the same role as \eqref{eq:conformal_propagator_p_space} in the next steps. The additional term cannot be interpreted as a propagator, since the energy direction is in some sense opposite. For this reason, we call the first piece the \textit{propagator} and the second piece the \textit{oppagator} (opposite-propagator).

Note that the domain of integration of the oppagator cannot be extended to all real numbers in general,  without spoiling the SK indices of the $X$ functions (see \eqref{eq:F(u,v) flat space form}). In order to keep the SK structure simple, we have to accept this as a true modification of the conformally coupled dressing rules: In the presence of oppagators, the energy integrals have a boundary. In addition, as we will see, the flat space momentum integration is not Lorentz invariant, in contrast to conformal coupling. These points will be discussed in detail later. 

\subsubsection*{Modification to Step (1) - Performing the $z$ Integral}

The terms we have to assemble to build the $z$ integral are the same as before, but lead to more complicated expressions. Essentially, this amounts to the replacements 
\begin{subequations}
\begin{align}
    e^{-i\pi \aa \frac{d-1}{4}} \frac{1}{\sqrt{z}} e^{\pm i p \aa z} &\xto{\mathrm{propagator}} e^{-i\pi \aa \frac{d-1}{4}} \frac{1}{\sqrt{z}} e^{\pm i p \aa z} \Pi_\nu\Big( \frac{i \aa}{\pm p z} \Big), \\
    &\xto{\mathrm{oppagator}} e^{-i\pi \aa \frac{d}{4}} \frac{1}{\sqrt{z}} e^{+ i p \aa z} \Pi_\nu\Big( \frac{i \aa}{+ p z} \Big), \\
    \frac{1}{\sqrt{z}} e^{- k z} &\xto{\mathrm{external}} \frac{1}{\sqrt{z}} e^{- k z} \Pi_\nu\Big( \frac{1}{k z} \Big)\,.
\end{align}
\end{subequations}
Just as for the conformally coupled case, we have moved a factor of $\sqrt{\pi/(2k)}$ to the external factor for the external legs. Note that the oppagators do not have a directed energy flow, so both ends show up with a $+p$. In other words, their $\sigma$ is always $+$, which is the only difference as far as the $z$ integral is concerned. Put together, the general $z$ integral takes the form
\begin{align}
    Z_\aa(\{\sigma_l p_l\}, \{k_s\}) \coloneq \int\limits_0^\infty \d z \, z^{\frac{d(N-2) - N}{2} - 1} e^{-(K - i \aa P) z} \hspace{-3mm}\prod\limits_{\mathrm{external} \ s} \hspace{-1.5mm} \Pi_{\nu_s} \Big( \frac{1}{k_s z} \Big) \hspace{-1.5mm} \prod\limits_{\mathrm{internal} \ l} \hspace{-1mm} \Pi_{\nu_l}\Big( \frac{i \aa}{\sigma_l p_l z} \Big). \label{eq:general_z_integral}
\end{align}
This is the conformally coupled vertex integrand times a collection of $\Pi$-functions, which collapses in the conformal limit as $\Pi_{1/2}(x) \equiv 1$. In general, $Z$ depends on the individual energies and external momenta flowing into this vertex, which is similar to a normal Feynman vertex with derivatives attached. This feature has also been observed with the massless dressings in \cite{Chowdhury:2025ohm}. As can be seen in \eqref{eq:general_z_integral}, the vertex factors in the energy representation are multiple-$K$ integrals, just like in the KLF formalism. Thus, the discussions in \cite{Grafe:2026avi, Grafe:2026hum} regarding their explicit evaluation and analytic continuation to the physical kinematic region can be applied directly. Equation \eqref{eq:general_z_integral} holds for $d(N-2)/2 > \sum_j |\Re(\nu_j)|$, where the $z$ integral converges in the IR. For other values, it is to be understood as an analytic continuation. As for conformal coupling, $Z$ decomposes into $Z = C + i\aa S$.

Importantly, we can rewrite \eqref{eq:general_z_integral} by using the integral representation of $\Pi_\nu$ from \eqref{eq:Pi_definition} so that it manifests itself as a propagator. For this, we first define (and remember)
\begin{align}
\begin{split}
    \beta_s &= k_s \quad \mathrm{for \ external \ legs}, \quad \beta_l = -i\aa \sigma_l p_l \quad \mathrm{for \ internal \ legs},\\
    \Lambda_v &= \frac{d(N-2)}{2} + \sum\limits_{\mathrm{all \ legs} \ j} \nu_j \quad \mathrm{for \ vertices}, \quad K = \sum\limits_s k_s, \quad P = \sum\limits_l \sigma_l p_l.
\end{split}
\end{align}
Then, we can write
\begin{align}
\begin{split}
    Z_\aa(\{\sigma_l p_l\}, \{k_s\}) &= \prod\limits_{\mathrm{legs} \ j} \left( \frac{(2\beta_j)^{\frac{1}{2} - \nu_j}}{\Gamma(\nu_j + \frac{1}{2})}  \int\limits_0^\infty \d s_j \Big( s_j (2\beta_j + s_j) \Big)^{\nu_j -\frac{1}{2}} \right) \frac{\Gamma\Big( \Lambda_v \Big)}{\Big(K - i\aa P + \sum\limits_{\mathrm{legs} \ j^\prime} s_{j^\prime} \Big)^{\Lambda_v}} \label{eq:general_Z_Euler_form}
\end{split}
\end{align}
This form is already much closer to parametric integrals known from flat-space Feynman diagrams. However, we can bring it into the even more familiar form
\begin{equation}\label{eq:general-Z-flatspace-form}
    Z_\aa(\{\sigma_lp_l\},\{k_s\}) = e^{\frac{i\pi}{2}\Lambda_v}\Gamma(\Lambda_v) \prod_{\mathrm{legs} \ j} \frac{\Gamma(\frac12-\nu_j)(2\beta_j)^{\frac12-\nu_j}}{2\pi} \int\limits_{\mathbb{R}^N} \d^N t\, \frac{\prod\limits_{\mathrm{legs} \ j^\prime} (t_{j^\prime}^2+\beta_{j^\prime}^2)^{\nu_{j^\prime}-\frac12}}{\Big(\sum\limits_{\mathrm{legs} \ j^\prime} t_{j^\prime} +i\epsilon\Big)^{\Lambda_v}}\,.
\end{equation}
We will provide the derivation in appendix \ref{app:vertex_functions}. 
Note that the denominators inside the two above integrals have the structure of a generalized propagator in the variables $\beta_j$ and $s_j$, or $t_j$, respectively. These representations of the vertex integral hold for any mix of complementary series and principal series fields, where one would simply analytically continue $\nu \rightarrow i \mu$. For this, it is useful to keep in mind that the Bessel $K$ function (and by extension $\Pi$) is even in its parameter, so the entire expression is invariant under $\mu \rightarrow -\mu$, even if this is not manifest. We also note that the vertex integral expression simplifies significantly if the external (or internal) legs are conformally coupled.

\medskip

The phase $\psi$ does also depend on masses. Here, we further have to track the phase difference between propagators and oppagators. In total, we find
\begin{align}
    \psi = \frac{\pi (d-1)}{2} - \frac{\pi}{2} \sum\limits_{\mathrm{external} \ s} \Big( \frac{d}{2} - \nu_s \Big) - \frac{\pi(d-1)}{4} n_\mathrm{prop} - \frac{\pi d}{4} n_\mathrm{opp}. \label{eq:general_psi_phase}
\end{align}
Interestingly, this only depends on the masses of the external fields, not the internal ones. If we assume the external masses to be conformal, the general $\psi$ angle is the same as in the overall conformally coupled case \eqref{eq:conformal_psi} if there are no oppagators, and is modified by $\pi/4$ per oppagator connected to a vertex.

Finding a closed form expression for the $Z$ term is quite involved for general $\nu$. For this reason, we do not state examples here. Instead, we will present the full computation of the massive tree level exchange, including $Z$, in the next section \ref{sec:SK_dressing_massive_exchange}.

\subsubsection*{Modification to Step (2) - The SK Sum}

As long as a diagram does not involve oppagator lines in the sense of \eqref{eq:general_nu_factorized_propagator}, step (2) is actually not modified for general masses. If there is an oppagator, however, there is one important difference: An oppagator involves $\delta_{\aa\bb}$, which forces the SK indices of the two connected vertices to be the same. Their SK indices are no longer independent and the SK sum does not factorize.

To track this efficiently, we define \textit{clusters}, which are sets of vertices that are connected via oppagator lines, and therefore share the same SK index. Instead of \eqref{eq:SK_sum_factorization}, we have
\begin{align}
    \mathcal{M} = \Big(\mathrm{ext. \ no \ } \aa \Big) \Big( \mathrm{flat \ space \ diagram} \Big) \prod\limits_{\mathrm{clusters} \ c} \left(\sum\limits_{\aa_v = \pm} \prod\limits_{v \in c} Z_{\aa_v}(\{\sigma_l p_l\}, \{k_s\}) e^{i \aa_v \psi_v} \right). \label{eq:cluster_SK_sum_explanation}
\end{align}
We use the shorthand notation $v \in c$ for "vertices $v$ that are part of cluster $c$". For clusters of only one vertex, i.e.\ no oppagator lines, this reduces to the previous case. For larger clusters, it corresponds to a mixing of what are called dotted and dashed propagators ($C$ and $S$) in \cite{Chowdhury:2025ohm}. For example, for a cluster of two vertices with the same $\psi$, we have
\begin{align}
\begin{split}
    \sum\limits_{\aa = \pm} Z_\aa(1) Z_\aa(2) e^{2 i \aa \psi} = \, &+2\cos(2\psi) \Big( C(1) C(2) - S(1) S(2) \Big) \\
    &- 2\sin(2\psi) \Big(S(1) C(2) + C(1) S(2) \Big). \label{eq:cluster_SK_sum_example}
\end{split}
\end{align}
We have simplified the arguments of $Z, C$ and $S$ for readability. Note that the terms with only one $S$ propagator no longer necessarily integrate to $0$, since oppagator lines are only integrated over the positive real axis and therefore we are not free to flip the sign of the argument. For propagator lines, we would have instead expected
\begin{align}
    \sum\limits_{\aa,\bb = \pm} Z_\aa(1) Z_\bb(2) e^{i (\aa+\bb) \psi} = 4\cos^2(\psi) C(1) C(2) + 4\sin^2(\psi) S(1) S(2), \label{eq:product_SK_sum_example}
\end{align}
where we have canceled the odd terms containing $S \cdot C$. Comparing \eqref{eq:cluster_SK_sum_example} and \eqref{eq:product_SK_sum_example}, we can see that oppagators do not follow the parity rule for dressing propagators and therefore have more possible combinations of dressings. Furthermore, the combinations $\cos(\psi) \times S$ and $\sin(\psi) \times C$ are possible, which means that the terms shown in table \ref{tab:conformally_coupled_dressings} are not the only possible ones in the presence of oppagators. It is better to think of the vertex integral $Z$ as the dressing propagator. The split that worked for conformally coupled scalars was a special case.

Because $i$ always appears in combination with the SK index $\aa$ in all involved objects, we can still write the product of all dressing propagators in a given cluster, i.e.\ the content of the last set of parentheses in \eqref{eq:cluster_SK_sum_explanation}, as
\begin{align}
    \mathcal{D}_c = 2 \,\Re\Big( \prod\limits_{v \in c} e^{i \psi_v} Z_+(\{\sigma_l p_l\}, \{k_s\}) \Big). \label{eq:general_dressing_propagator_cluster}
\end{align}
This way, our final result does not contain any reference to the SK labels, making it easier to package in the final formulation of the generalized dressing rules. Note that the real part does not introduce non-analytic dependence in the $p_l$ because we can understand it as a sum of $Z_+(\{\sigma_l p_l\}, \{k_s\})$ and $Z_+(\{-\sigma_l p_l\}, \{k_s\})$.

\subsubsection*{Modification to Step (3) - Dressing Propagators}

For vertices that are not connected to oppagators, there is no modification to the conformally coupled discussion. If they do contain oppagators, we can still draw dressing propagators, assign them the energy $P$, and rewrite the book-keeping in a way that energy is conserved at that vertex, but there is a global mismatch: The sum of the energies of all dressing propagators is now no longer $0$, but $2$ times the sum over all energies carried by oppagator lines, since they do not appear once with a $+$ and once with a $-$, but twice with a $+$.

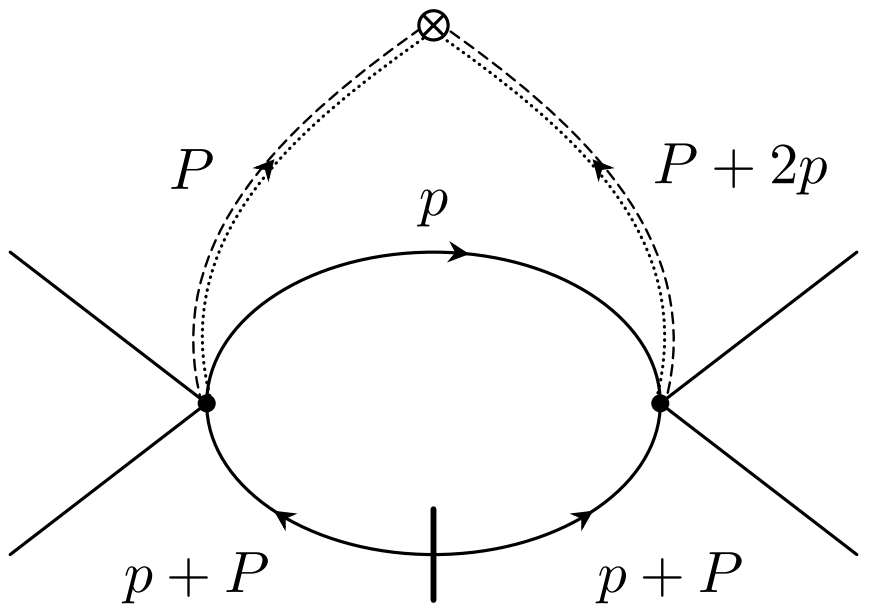
\begin{figure}
\centering
\begin{tikzpicture}[line cap=round, x=1cm, y=1cm]
  \coordinate (a) at (0,0);  \coordinate (b) at (3,0);  \coordinate (t) at (1.5,2.5);
  
  \ruleDressCurve{a}{t}{105}{215}{black}
  \ruleDressCurve{t}{b}{-35}{75}{black}
  \path[rule flow=black] (a) to[out=105, in=215] (t);
  \path[rule flow=black] (b) to[out=75, in=-35] (t);
  \node[left=1pt]  at (0.22,1.55) {$P$};
  \node[right=1pt] at (2.78,1.55) {$P+2p$};
  \node[rule meet, minimum size=5.5pt] at (t) {};
 
  \draw[semithick] (-1.3,1) -- (a) -- (-1.3,-1)  (4.3,1) -- (b) -- (4.3,-1);
  
  \draw[semithick, rule flow=black] (a) arc[start angle=180, end angle=0, x radius=1.5, y radius=1];
  \node[above=1pt] at (1.5,1) {$p$};
  
  \draw[semithick] (a) arc[start angle=180, end angle=360, x radius=1.5, y radius=1];
  \path[rule flow=black] (1.5,-1) arc[start angle=270, end angle=180, x radius=1.5, y radius=1];
  \path[rule flow=black] (1.5,-1) arc[start angle=270, end angle=360, x radius=1.5, y radius=1];
  \draw[line width=1.1pt] (1.5,-1.3) -- (1.5,-0.7);
  \node[below left=0pt]  at (0.55,-0.85) {$p+P$};
  \node[below right=0pt] at (2.45,-0.85) {$p+P$};
  
  \node[rule vtx, minimum size=3.4pt] at (a) {};  \node[rule vtx, minimum size=3.4pt] at (b) {};
\end{tikzpicture}
\caption{Misdirected energy flow of a dressed loop diagram. Usually, one would expect the momentum $p$ to only flow within the loop, but here it can be seen to "leak out" because the oppagator "mirrors" its contained energy. This could be read as a non-Lorentz covariant energy flow. Instead, it would be better to define $p + P = p^\prime$. This way, it is clearer that the lowest point of the diagram, the oppagator, is a source of energy (here $p^\prime$). Note how the total energy $2p^\prime$ is lost at the top point where the dressing propagators meet.}
\label{fig:misdirected_energy_loop}
\end{figure}

It is better to think of the middle of oppagators as sources of energy and the dressing point as a sink of that energy. In this sense, having an oppagator line in a loop breaks the usual structure. The loop energy now "leaks out" of the loop, as demonstrated in figure \ref{fig:misdirected_energy_loop}. Unfortunately, this also somewhat removes the interpretation of the flat space part of the expression as such, because oppagators carry momentum, but source energy. This concludes the construction. 

\subsubsection*{Feynman Rules for General Mass Scalar Dressing Rules}

Here we will formulate a set of Feynman rules for the generalized dressing rules that are agnostic of their derivation and can be used as a standalone. The diagrams on the right illustrate each step for the same exchange diagram, which has $L=1$ flat space propagators and hence two copies; from step~\ref{rule:gen:external} on, only the copy with the oppagator is shown.
\newdressingruleset
\begin{dressingrule}{lmugreen}{Copies of the Diagram}{\ruleGenDiagram{1}}\label{rule:gen:copies}
    Draw the flat space diagram \rulekw{$2^{L}$ times}, where $L$ is the number of propagators, and assign momentum flow as usual. Sum over all copies of the diagram.
\end{dressingrule}
 
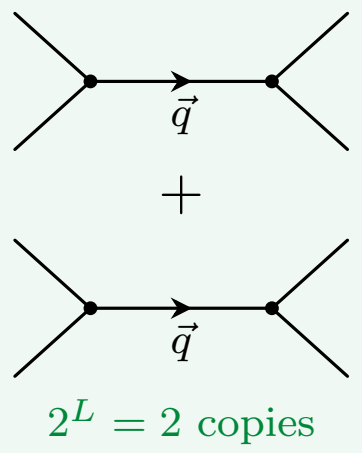
\begin{dressingrule}{lmuorange}{Propagators and Oppagators}{\ruleGenDiagram{2}}\label{rule:gen:cuts}
    Assign which lines are \rulekw{propagators} and which are \rulekw{oppagators}. For propagators, energy follows the momentum flow. Write
    \begin{align*}
        \pi \int \frac{\d^{d+1} Q}{(2\pi)^{d+1}}\frac{1}{Q^2}.
    \end{align*}
    For oppagators, energy flows outwards to both sides. Write
    \begin{align*}
        2\pi \cos(\pi\nu) \int\limits_{-\infty}^\infty \frac{\d p}{2\pi} \theta(p) \int\frac{\d^{d} q}{(2\pi)^{d}}\frac{1}{Q^2}.
    \end{align*}
\end{dressingrule}
 
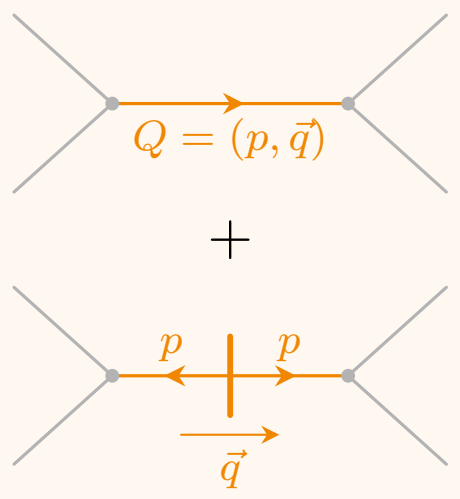
\begin{dressingrule}{lmuorange}{Dressing Propagators and Clusters}{\ruleGenDiagram{3}}\label{rule:gen:dressing}
    Attach \rulekw{dressing propagators} to every vertex of the diagram, connecting to the \rulekw{dressing point}. Determine \rulekw{clusters} of vertices that are connected by oppagators. For any cluster $c$, write
    \begin{align*}
        2 \int\limits_{-\infty}^\infty \frac{\d P_1 \ldots \d P_c}{(2\pi)^c}  \Re\Big( \prod\limits_{v \in c} e^{i \psi_v} Z_+(\{\sigma_l p_l\}, \{k_s\}) \Big)
    \end{align*}
    from \eqref{eq:general_dressing_propagator_cluster}. $Z_+$ is defined in \eqref{eq:general_Z_Euler_form} and $\psi$ in \eqref{eq:general_psi_phase}. Here, $c$ is used as the number of vertices in cluster $c$.
\end{dressingrule}
 
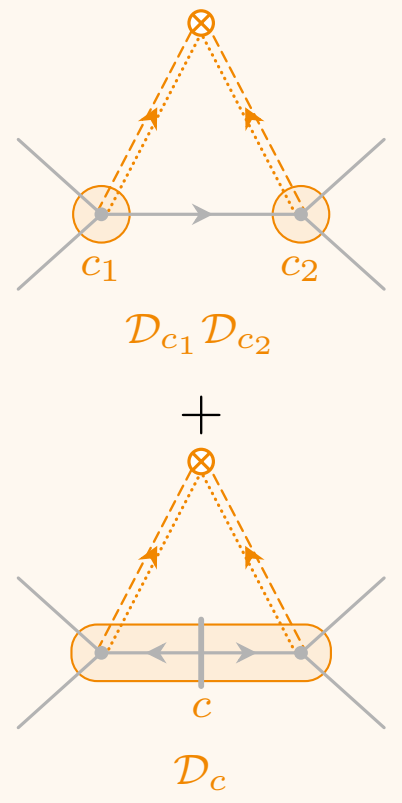
\begin{dressingrule}{lmugreen}{External Legs and Vertices}{\ruleGenDiagram{4}}\label{rule:gen:external}
    For every \rulekw{external leg} with momentum $\vec k_s$, multiply by
    \begin{align*}
        H^{\frac{d-3}{2}} \frac{\Gamma(\nu_s)}{2^{\frac{3}{2}-\nu_s}} (-\eta_0)^{\frac{d}{2} - \nu_s} k_s^{-\frac{1}{2}-\nu_s}
    \end{align*}
    coming from \eqref{eq:KLF_external} (without the SK structure) and parts of \eqref{eq:Pi_definition}. For principal series fields, replace $\nu_s \rightarrow i\mu_s$ and symmetrize in $\mu_s \rightarrow -\mu_s$ at the very end. Then, multiply by
    \begin{align*}
        \mathcal{C} \lambda \frac{H^\frac{(N-2)(d+1)}{2}}{\pi^\frac{N}{2}}
    \end{align*}
    for every \rulekw{vertex}, coming from \eqref{eq:KLF_vertex_factor} that contains symmetry factors $\mathcal{C}$, coupling constants $\lambda$ and some powers of $\pi$ and $H$.
\end{dressingrule}
 
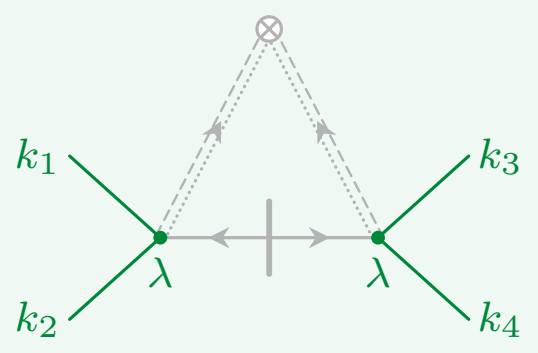
\begin{dressingrule}{lmugreen}{Energy and Momentum Conservation}{\ruleGenDiagram{5}}\label{rule:gen:conservation}
    Finally, enforce \rulekw{$d$-momentum conservation} by inserting a delta function $(2\pi)^d \, \delta^{(d)}(\vec k_s + \ldots - \vec q_l + \ldots)$ at every flat space vertex. Then, implement \rulekw{energy conservation} at every flat space vertex by inserting the delta function $(2\pi) \, \delta(P_r + \ldots - p_j + \ldots)$. Note that energies and momenta are directed and that external lines carry no energy, while dressings carry no momentum. Note the difference in energy direction between propagators and oppagators. Energy is automatically conserved at the dressing point up to 2 times the energies sourced by oppagators.
\end{dressingrule}

\subsection{Example: Massive Exchange} \label{sec:SK_dressing_massive_exchange}

In this section, we demonstrate how to use the generalized dressing rules to compute an example diagram, the tree level massive exchange. 
To this end, we consider a four-point correlation function of conformally coupled scalars $\phi$ with a heavy scalar field $\chi$ with mass parameter $|\Re(\nu)| < 1/2$ that is exchanged in the $s$-channel. The interaction term in the Lagrangian shall have the form $\mathcal{L}\supset -\frac{\lambda}{2}\sqrt{-g} \, \phi^2\chi$. We will show that we recover the known literature expressions exactly \cite{Arkani-Hamed:2018kmz, Sleight:2019mgd, Sleight:2019hfp, Sleight:2020obc, Baumann:2019oyu, Aoki:2023wdc, Liu:2024xyi, Qin:2022fbv, Werth:2024mjg, Liu:2024str, Xianyu:2025lbk, Baumann:2026atn, Arundine:2026myr, Grafe:2026avi}.

\begin{figure}[h]
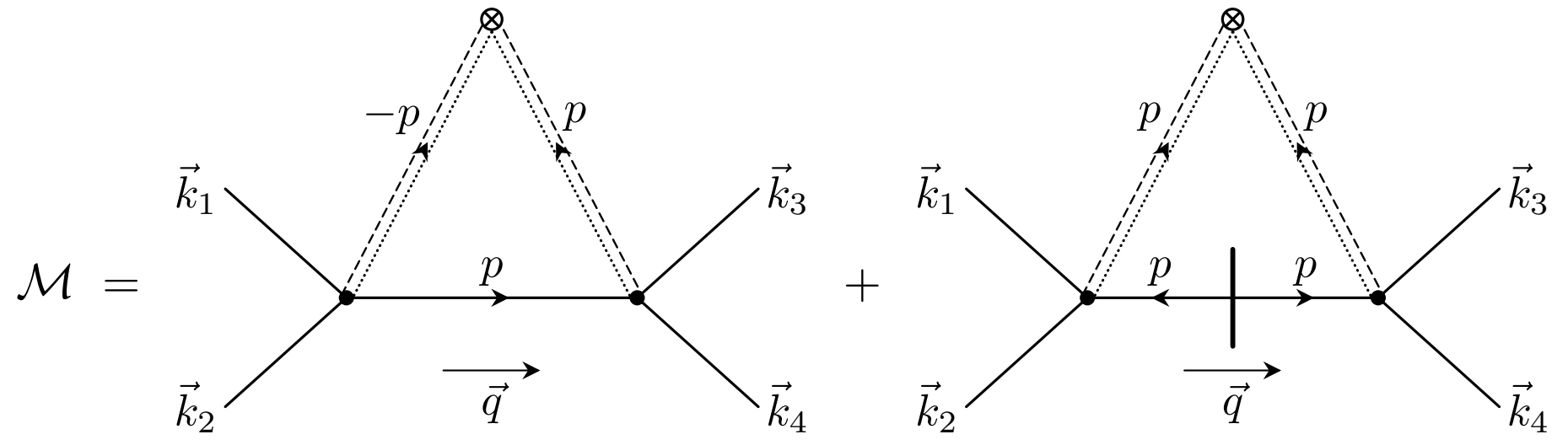

\centering
$\displaystyle\mathcal{M} \;=\;$\;\ampDiagram{1}\;\;$+$\;\;\ampDiagram{2}
\caption{The $2$ diagrams generated by the generalized dressing rules \ref{rule:gen:copies}-\ref{rule:gen:conservation} for the single massive exchange. The momentum flow through the internal line is always $\vec q = \vec k_1 + \vec k_2$, while the energy flow is as indicated.}
\label{fig:tree_level_exchange_dressing}
\end{figure}

The two dressing contributions are shown in figure \ref{fig:tree_level_exchange_dressing}. For both contributions, propagator and oppagator, we need the $z$ integral $Z_\aa(\{\sigma_j p_j\}, \{k_j\})$ for two external legs with momenta $\vec k_1, \vec k_2$ (using $k_j = |\vec k_j|$) and one internal leg with momentum $\vec q = \vec k_1 + \vec k_2$, energy $p$ and mass parameter $\nu$. We also define $k_{12} = k_1 + k_2$. For this example, we set the external legs to conformal coupling. We can use \eqref{eq:general_Z_Euler_form} to compute
\begin{align}\label{eq:eu1}
\begin{split}
    Z_\aa(\sigma p, k_{12}) &= \int\limits_0^\infty \d s_1\, \d s_2 \,\d s \frac{\Gamma\left( \frac{d+2}{2} + \nu\right) (-2i\aa\sigma p)^{\frac{1}{2}-\nu}\Big( s (-2i\aa\sigma p + s) \Big)^{\nu - \frac{1}{2}}}{\Gamma\left( \nu + \frac{1}{2} \right)\big( k_{12} - i \aa \sigma p + s_1 + s_2 + s \big)^{\frac{d+2}{2} + \nu}}\,.
\end{split}
\end{align}
The $s_1$ and $s_2$ integrals can be performed easily. This is always true for conformally coupled external legs. Using dimensional regularization $d = 3 + 2\epsilon$, we find
\begin{align}\label{eq:eu2}
    Z_\aa(\sigma p, k_{12}) &= (-2i\aa\sigma p)^{\frac{1}{2}-\nu}\frac{\Gamma\left(\frac12+\nu+\epsilon\right)}{\Gamma\big(\frac{1}{2}+\nu\big)}  \int\limits_0^\infty \d s \frac{\Big(s\big(-2i\aa\sigma p + s\big)\Big)^{\nu - \frac{1}{2}}}{\left( k_{12} - i\aa \sigma p + s\right)^{\frac{1}{2} + \epsilon + \nu}}\,.
\end{align}
For conformal coupling, $\nu = \frac{1}{2}$, this integral is divergent for $\epsilon = 0$, in agreement with our previous discussion in section \ref{sec:SK_dressing_rules_conformal}. It is straight-forward to recover the conformally coupled dressings \eqref{eq:example_dressings_conformal} from this limit. For all $\nu$ with $|\Re(\nu)| < \frac{1}{2}$, we can immediately set $\epsilon = 0$, which we shall do from now on. Then, the result is found to be
\begin{align}\label{eq: massive vertex function}
    Z_\aa(\sigma p, k_{12}) &= \frac{\pi}{\cos(\pi\nu)} \,P_{\nu - \frac{1}{2}}\left( \frac{i \aa k_{12}}{\sigma p} \right).
\end{align}
Here, $P_\lambda$ is the Legendre function of the first kind with a branch cut along $(-\infty, -1]$ (type 3, order 0). This form of the vertex function resembles the Laplace-transforms obtained in \cite{Belrhali:2026jqe,CCUP}. The phase angles $\psi$ are different between the propagator and the oppagator diagram at $d=3$ and are given by
\begin{align}
    \psi_\mathrm{prop} = - \frac{\pi}{2}, \qquad \psi_\mathrm{opp} = - \frac{3\pi}{4}.
\end{align}
From \eqref{eq:cluster_SK_sum_example} and \eqref{eq:product_SK_sum_example}, we know that for the first diagram with the propagating line, we need trigonometric functions of $\psi$, squared. For the oppagator diagram, we need trigonometric functions with argument $2\psi = - 3\pi/2$. We note
\begin{subequations}
\begin{alignat}{2}
    \cos^2(\psi_\mathrm{prop}) &= 0,  &\qquad \sin^2(\psi_\mathrm{prop}) &= 1, \\
    \sin(2\psi_\mathrm{opp})   &= 1, &\qquad \cos(2\psi_\mathrm{opp})   &= 0.
\end{alignat}
\end{subequations}
We can therefore immediately simplify the dressing propagators by dropping the respective vanishing combination. As a result, we can write down the full Feynman integral
\begin{align}
\begin{split}
    \mathcal{M}_\mathrm{prop} &+ \mathcal{M}_\mathrm{opp} = \frac{H^4 \lambda^2}{8\pi} \frac{(-\eta_0)^4}{k_1 k_2 k_3 k_4} \left[ \ \int\limits_{-\infty}^{+\infty} \frac{\d p}{p^2 + q^2} S(p, k_{12}) S(-p, k_{34}) \right.\\
    &- \left.\cos(\pi\nu)\int\limits_{0}^\infty \frac{\d p}{p^2 + q^2} \Big( S(p, k_{12}) C(p, k_{34}) + C(p, k_{12}) S(p, k_{34}) \Big) \right].
\end{split}\label{eq:massive-exchange-dressing-integral}
\end{align}
Note how $S$ and $C$ dressings are mixed for the oppagator diagram. From \eqref{eq: massive vertex function} it is apparent that $Z_\aa(-p) = \overline{Z_\aa(p)}$ and hence $S(-p) = -S(p)$. This parity of $S$ allows us to rewrite the first term as an integral on $(0,\infty)$ again. It turns out to be more convenient in the following, so we un-expand the combination in the second line to $S(1)C(2)+C(1)S(2) = \Im\left[Z_+(1)Z_+(2)\right]$. 
Due to conformal invariance the four-point function will only depend on the two conformal momentum ratios
\begin{equation}
    u  \equiv \frac{q}{k_{12}}\,,\,\,\,\, v\equiv \frac{q}{k_{34}}\,.
\end{equation}
Following the conventions in \cite{Arkani-Hamed:2018kmz}, the massive exchange correlator can then be expressed in terms of a dimensionless function $F(u,v)$ of these two momentum ratios:
\begin{equation}
    \M_\text{prop} +\M_\text{opp} = \frac{ H^4\lambda^2(-\eta_0)^4}{16 k_1k_2k_3k_4 q}\, F(u,v)\,.
\end{equation}
 This dimensionless function $F(u,v)$ has a representation in terms of the momentum integrals above. For later convenience we set  $Z(x) \coloneq \frac{\pi}{\cos\pi\nu} P_{\nu-\frac12}(x)$ because the vertex function $Z_+(p,k_{12})$ only depends on ratios of momenta and not on the individual arguments; and we will perform a substitution to inverse momenta and also render the integration variable dimensionless by setting $x \equiv q/p$. We obtain:

\begin{equation}\label{eq: massive exchange F(u,v)}
    F(u,v) = -\frac{2}{\pi}\int\limits_0^\infty \frac{\d x}{1+x^2} \left\{2\,\Im Z\left(\frac{ix}{u}\right) \,\Im Z\left(\frac{ix}{v}\right) +\cos\pi\nu\, \Im\left[Z\left(\frac{ix}{u}\right)Z\left(\frac{ix}{v}\right)\right]\right\}\,.
\end{equation}
In the following we will make use of the fact that $Z(-i\bar{z})=\overline{Z(iz)}$ for $\nu\in\mathbb{R}$ or $\nu\in i\mathbb{R}$. Moreover, for real $z$ we have $\overline{Z(iz)}=Z(-iz)$ since the Legendre function depends only on the square $\nu^2$ of the mass parameter. This also implies that $Z$ is real on $(-1,\infty)$. The statement does not extend to a statement about the whole real line because the Legendre function possesses a branch cut which is why $Z$ is analytic only on $\mathbb{C}\setminus (-\infty, -1]$.\\
With the previous observations, we can disentangle the imaginary parts in \eqref{eq: massive exchange F(u,v)},
\begin{align}
\begin{split}
    2\,\Im Z\left(\frac{ix}{u}\right)\Im Z\left(\frac{ix}{v}\right) = &-\frac12 \left[Z\left(\frac{ix}{u}\right)Z\left(\frac{ix}{v}\right) + (x\leftrightarrow -x)\right] \\
    &+ \frac12 \left[Z\left(\frac{ix}{u}\right)Z\left(-\frac{ix}{v}\right)
    + (x\leftrightarrow -x)\right]\,.
\end{split}
\end{align}
Each of the above brackets is symmetric under the exchange $x\leftrightarrow -x$ and hence we can write the first term in \eqref{eq: massive exchange F(u,v)} as an integral along the full real line. To be precise, we obtain
\begin{align}\label{eq: massive exchange decomposition}
\begin{split}
    F(u,v) =& \frac{1}{\pi} \int\limits_{-\infty}^{+\infty} \frac{\d x}{1+x^2} Z\left(\frac{ix}{u}\right)\left[Z\left(\frac{ix}{v}\right) - Z\left(-\frac{ix}{v}\right)\right] \\
    &-\frac{2}{\pi}\cos \pi\nu\, \Im \int\limits_0^\infty \frac{\d x}{1+x^2} Z\left(\frac{ix}{u}\right)Z\left(\frac{ix}{v}\right)\,.
\end{split}
\end{align}
For each of the three pieces we will close the integration contour appropriately. To do so, we need to fully understand the analytic structure of the integrand. Due to the branch cut of the Legendre function, the function $Z(\pm ix/u)$ is singular at $x\in \pm i u [1,\infty)$. Furthermore, for large arguments we have the behavior $|Z(x)| =\mathcal{O}(|x|^{|\Re \,\nu|-\frac12})$. Thus, the arc at infinity does not contribute when closing the contour. Additionally, the flat-space propagator $(1+x^2)^{-1}$ will contribute simple poles at $\pm i$. In the physical kinematic region we have $0<u,v <1$ and hence these poles lie on the cuts.

\vskip 4pt 
Let us start with the first term where we integrate the product $Z(ix/u)Z(ix/v)$. Both factors are analytic in the lower half plane, which is why we close the integration contour downwards and pick up the residue of the simple pole at $x=-i$. Therefore, the first term is equivalent to $ Z(u^{-1})Z(v^{-1})$.

\vskip 4pt
The second term contains the product $Z(ix/u)Z(-ix/v)$ which means that one factor has a branch cut in the upper and the other one a cut in the lower half plane. We can close the contour either way and here we will choose the upper half plane. Thus, the contour must be wrapped around the branch cut $\C_u$ of $Z(ix/u)$ ranging along $iu[1,\infty)$. We avoid the on-shell pole at $x=+i$ by moving around it along two small semi-circles. We will first account for the contribution of the wrapping along $\C_u$ which is given by the discontinuity of the integrand along the cut. The Legendre function $P_\lambda(z)$ obeys the connection formula
\begin{equation}
    P_\lambda (-z \pm i\epsilon) = e^{\pm i\pi\lambda }P_\lambda(z) -\frac{2}{\pi} \sin(\pi\lambda) \,Q_\lambda(z)\,,
\end{equation}
where $Q_\lambda(z)$ is the Legendre-$Q$ function. Then, 
\begin{equation}
    Z(-z+i\epsilon)-Z(-z-i\epsilon) = -2i\,\cos(\pi\nu)\, Z(z)\,,
\end{equation}
and integrating along the cut with $\d x= i\,\d t$ results in 
\begin{equation}
    F(u,v) \supset -\frac{2}{\pi} \cos(\pi\nu)\,\text{PV}\int_u^\infty \frac{\d t}{1-t^2} Z(u^{-1} t) Z(v^{-1} t)\,.
\end{equation}
The principal value $\text{PV}$ denotes the fact that we have not yet dealt with the pole at $t=1$ which is part of the integration contour. When going around this pole along the small counter-clockwise semi-circles, we pick up $i\pi$ times the boundary value $Z(-u^{-1} \pm i\epsilon)Z(v^{-1})/2i$ at each turn. Combining both pieces together with the overall pre-factor $-\pi^{-1}$ of the integral results in the contribution $F(u,v)\supset -Z(v^{-1}) \,\Re \,Z(e^{i\pi}u^{-1})$.

\vskip 4pt
Finally, we also need to evaluate the third term. There, we can simply rotate the integration contour to $\d x = -i\,\d t$ by crossing the fourth quadrant of the complex plane. There are no poles inside the enclosed area if we avoid the on-shell pole at $x=-i$. The latter is done by once again integrating along a small semi-circle around this pole which leads to the additional contribution $-i\pi$ times the residue of the integrand at the pole. altogether we find
\begin{equation}
    F(u,v) \supset -\frac{2}{\pi}\cos(\pi\nu) \,\Im\left\{\frac{\pi}{2}Z(u^{-1})Z(v^{-1}) -i\,\text{PV} \int_0^\infty \frac{\d t}{1-t^2}Z(u^{-1}t)Z(v^{-1}t)\right\}\,.
\end{equation}
The on-shell part is real and hence drops out completely from the expression while the second term is fully imaginary and survives in the final result. 

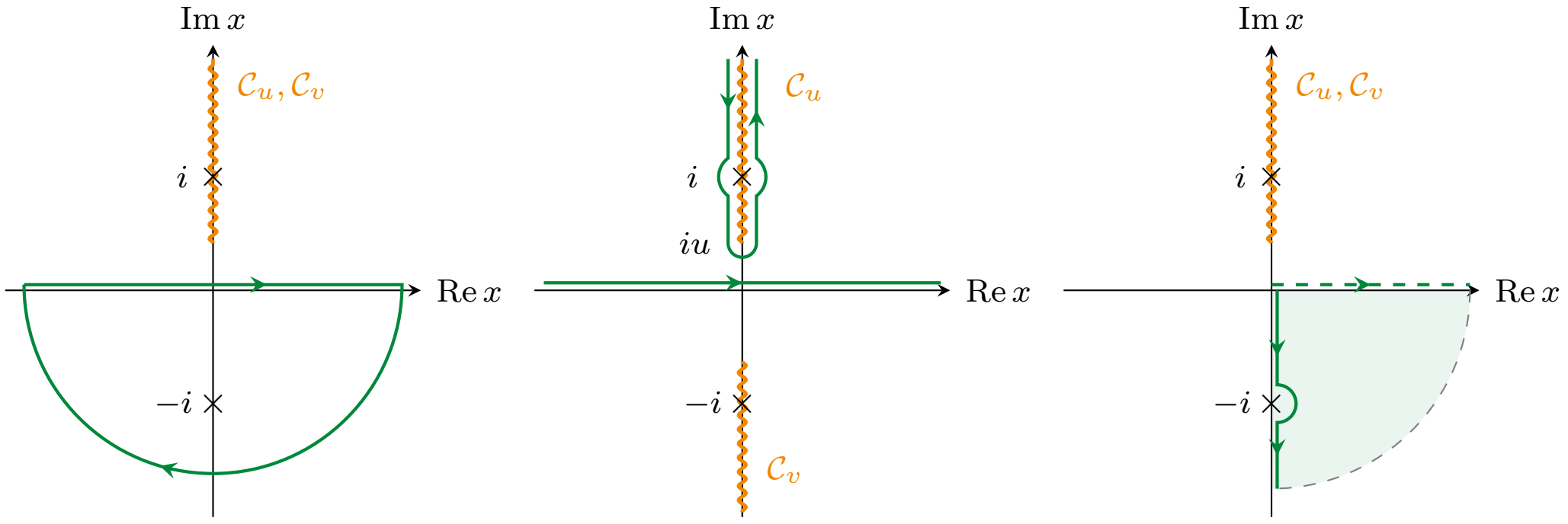
\begin{figure}[t]
\centering
\begin{tikzpicture}[scale=0.82,>=stealth,
  axis/.style={->,black},
  cut/.style={lmuorange,line width=0.9pt,decorate,decoration={snake,amplitude=0.9pt,segment length=3.5pt}},
  contour/.style={lmugreen,thick,postaction={decorate}},
  pole/.style={black}]

\begin{scope}
\draw[axis] (-2.2,0) -- (2.2,0) node[right]{\scriptsize $\operatorname{Re}x$};
\draw[axis] (0,-2.4) -- (0,2.6) node[above]{\scriptsize $\operatorname{Im}x$};
\draw[cut] (0,0.5) -- (0,2.45);
\node[lmuorange,right] at (0.1,2.15) {\scriptsize  $\C_u,\C_v$};
\draw[contour,decoration={markings,mark=at position 0.25 with {\arrow{>}},mark=at position 0.75 with {\arrow{>}}}]
  (-2.0,0.06) -- (2.0,0.06) arc[start angle=0,end angle=-180,radius=2.0];
\node[pole] at (0,-1.2) {\footnotesize $\times$}; \node[left] at (-0.05,-1.2) {\scriptsize $-i$};
\node[pole] at (0,1.2) {\footnotesize $\times$}; \node[left] at (-0.1,1.2) {\scriptsize $i$};
\node at (0,-3.0) {\small (a) first term: close downwards};
\end{scope}

\begin{scope}[xshift=5.6cm]
\draw[axis] (-2.2,0) -- (2.2,0) node[right]{\scriptsize $\operatorname{Re}x$};
\draw[axis] (0,-2.4) -- (0,2.6) node[above]{\scriptsize $\operatorname{Im}x$};
\draw[cut] (0,-0.75) -- (0,-2.35);
\node[lmuorange,right] at (0.1,-1.9) {\scriptsize  $\C_v$};
\draw[cut] (0,0.5) -- (0,2.45);
\node[lmuorange,right] at (0.3,2.15) {\scriptsize $\mathcal C_u$};
\draw[contour,decoration={markings,mark=at position 0.5 with {\arrow{>}}}] (-2.1,0.08) -- (2.1,0.08);
\draw[contour,decoration={markings,mark=at position 0.12 with {\arrow{>}},mark=at position 0.88 with {\arrow{>}}}]
  (-0.15,2.45) -- (-0.15,1.4) arc[start angle=126.87,end angle=233.13,radius=0.25] -- (-0.15,0.5)
  arc[start angle=180,end angle=360,radius=0.15] -- (0.15,1.0) arc[start angle=-53.13,end angle=53.13,radius=0.25] -- (0.15,2.45);
\node[pole] at (0,1.2) {\footnotesize $\times$}; \node[left] at (-0.3,1.2) {\scriptsize $i$};
\node[pole] at (0,-1.2) {\footnotesize $\times$}; \node[left] at (-0.05,-1.2) {\scriptsize $-i$};
\node[left] at (-0.17,0.5) {\scriptsize $iu$};
\node at (0,-3.0) {\small (b) second term: wrap $\mathcal C_u$};
\end{scope}

\begin{scope}[xshift=11.2cm]
\fill[lmugreen!8] (0,0) -- (2.1,0) arc[start angle=0,end angle=-90,radius=2.1] -- cycle;
\draw[axis] (-2.2,0) -- (2.2,0) node[right]{\scriptsize $\operatorname{Re}x$};
\draw[axis] (0,-2.4) -- (0,2.6) node[above]{\scriptsize $\operatorname{Im}x$};
\draw[cut] (0,0.5) -- (0,2.45);
\node[lmuorange,right] at (0.1,2.15) {\scriptsize $\C_u,\C_v$};
\draw[contour,dashed,decoration={markings,mark=at position 0.5 with {\arrow{>}}}] (0,0.06) -- (2.1,0.06);
\draw[gray,dashed] (2.1,0) arc[start angle=0,end angle=-90,radius=2.1];
\draw[contour,decoration={markings,mark=at position 0.3 with {\arrow{>}},mark=at position 0.85 with {\arrow{>}}}]
  (0.06,0) -- (0.06,-1.0) arc[start angle=90,end angle=-90,radius=0.2] -- (0.06,-2.1);
\node[pole] at (0,-1.2) {\footnotesize $\times$}; \node[left] at (-0.05,-1.2) {\scriptsize $-i$};
\node[pole] at (0,1.2) {\footnotesize $\times$}; \node[left] at (-0.1,1.2) {\scriptsize $i$};
\node at (0,-3.0) {\small (c) third term: rotate to $x=-it$};
\end{scope}
\end{tikzpicture}
\caption{The three integrals of \eqref{eq: massive exchange decomposition} in the inverse-energy plane $x=q/p$; branch cuts are drawn in orange.
(a) In the first term both factors are analytic below the real axis, so closing downwards picks up only the on-shell pole $x=-i$.
(b) In the second term the factors are analytic in opposite half-planes. The real line is pushed up and wraps the partial-energy cut $\mathcal C_u=i[u,\infty)$. The on-shell pole $x=i$ on that cut is avoided by a semicircle on each path.
(c) The half-line integral of the third term (dashed) is rotated through the fourth quadrant onto the negative imaginary axis. It passes $x=-i$ on its right.}
\label{fig:contours}
\end{figure}

Combining all three pieces together, we end up with the expression
\begin{equation}\label{eq: F(u,v) with integral left}
    F(u,v) = Z(v^{-1})\left[Z(u^{-1}) - \Re Z(e^{i\pi} u^{-1})\right] +\frac{2}{\pi}\cos \pi\nu \int_0^u \frac{\d t}{1-t^2} Z(u^{-1}t)Z(v^{-1}t)\,.
\end{equation}
The principal value integrals combine and no longer need the respective labeling $\text{PV}$ as the pole at $t=1$ lies outside the integration contour for physical kinematics $u<1$. The on-shell pieces in this final expression are exactly the local and non-local signal of the exchange diagram. Using identities between the Legendre functions it can be directly shown that these terms can be written equivalently in the form familiar from spectral representations \cite{Werth:2024mjg, Grafe:2026avi}. Let us make this last statement explicit. First, note that the Legendre function can be expressed in terms of Gauss' hypergeometric function ${}_2F_1$ as
\begin{equation}
    Z(x) = \Z_+(x)+\Z_-(x) =\frac{1}{\sqrt{\pi}}\sum\limits_{\aa=\pm} \Gamma(-\aa\nu)\Gamma(\tfrac12+\aa\nu)\left(2x\right)^{-\frac12-\aa\nu}\,{}_2F_1\left[\left.\begin{matrix}
        \frac{\frac12+\aa\nu}{2}, \frac{\frac32+\aa\nu}{2}\\ 1+\aa\nu
    \end{matrix}\right\vert x^{-2}\right]\,.
\end{equation}
This is exactly the representation of the vertex functions used in \cite{Grafe:2026avi} where they are decomposed into positive and negative frequency modes $\Z_+$ and $\Z_-$. With the help of the reflection formula of the Legendre function one can then show that the homogeneous part of $F(u,v)$ (all the on-shell contributions in \eqref{eq: F(u,v) with integral left} without the remaining integral) can be rearranged to
\begin{align}
\begin{split}
    F_\text{hom}(u,v) = \pi^{-1}\sum\limits_{\aa,\bb=\pm} (1+\aa \sin \pi\nu)\Gamma[-\aa\nu,-\bb\nu,\tfrac12+\aa\nu,\tfrac12+\bb\nu]\left(\frac{u}{2}\right)^{\frac12+\aa\nu}\left(\frac{v}{2}\right)^{\frac12+\bb\nu}  \\
    \times {}_2F_1\left[\left.\begin{matrix}
        \frac{\frac12+\aa\nu}{2}, \frac{\frac32+\aa\nu}{2}\\ 1+\aa\nu
    \end{matrix}\right\vert u^2\right]{}_2F_1\left[\left.\begin{matrix}
        \frac{\frac12+\bb\nu}{2}, \frac{\frac32+\bb\nu}{2}\\ 1+\bb\nu
    \end{matrix}\right\vert v^2\right]
\end{split}
\label{eq:massive-exchange-homogeneous}
\end{align}
We used the short-hand $\Gamma[a_1,\dots ,a_n]=\Gamma(a_1)\cdots\Gamma(a_n)$ above. The remaining integral on the other hand can be directly mapped to the analytic background of the correlator which was first computed using the cosmological bootstrap \cite{Arkani-Hamed:2018kmz}. We will show that we can directly recover the series representation from the bootstrap result by expanding the integrand. We start by making the substitution $t\to u t$ and define the background integral (following the notation in \cite{Arkani-Hamed:2018kmz} up to a factor $2$) as 
\begin{equation}
    I(u,v) \equiv \frac{2u}{\pi}\cos\pi\nu \int_0^1\frac{\d t}{1-u^2t^2} Z(t)Z(\tfrac{u}{v}t)\,.
\end{equation}
We assume $u<v$ and expand the second Legendre function as $Z(tu/v) = \sum_{n=0}^\infty a_n t^n\left(\frac{u}{v}\right)^n$ and the denominator as a geometric series $(1-u^2t^2)^{-1}=\sum_{m=0}^\infty u^{2m}t^{2m}$. We thus get a double sum
\begin{equation}
    I(u,v) =\sum\limits_{m,n=0}^\infty c_{mn} u^{2m+1}\left(\frac{u}{v}\right)^n\,,
\end{equation}
with exactly the structure from \cite{Arkani-Hamed:2018kmz}. The coefficients are given by 
\begin{equation}
    c_{mn} = \frac{2}{\pi}\cos(\pi\nu) a_n \int_0^1\d t\, t^{2m+n} Z(t)\,,
\end{equation}
where the coefficients $a_n$ from the Taylor expansion of the Legendre function are 
\begin{equation}
    a_n = \frac{(-1)^n 2^{n-1}}{\sqrt{\pi}\, n!} \Gamma\left(\tfrac{n+\nu+\frac12}{2}\right)\Gamma\left(\tfrac{n-\nu+\frac12}{2}\right)\,,
\end{equation}
and the moments of the Legendre function can be found in \cite[Eq.~7.126.1]{gradshteyn}. They evaluate to
\begin{equation}
    \int_0^1 \d t\, t^k\,P_\lambda(t) = \frac{\sqrt{\pi}\,\Gamma(k+1)}{2^{k+1}\Gamma(1+\frac{k-\lambda}{2})\Gamma(\frac32+\frac{k+\lambda}{2})}\,.
\end{equation}
Combining these expressions leads to the coefficients
\begin{equation}
    c_{mn} = \frac{(-1)^n (n+1)_{2m}}{2^{2m+1}\left(\frac{n-\nu+\frac12}{2}\right)_{m+1}\left(\frac{n+\nu+\frac12}{2}\right)_{m+1}} = \frac{2(-1)^n (n+1)_{2m}}{\prod_{j=0}^m\left[(n+2j+\frac12)^2-\nu^2\right]}\,.
\end{equation}
These are exactly the coefficients of the series ansatz that fulfills the bootstrap equation $(\Delta_u +\frac14-\nu^2)I(u,v)=2uv/(u+v)$ with the Ward identity operator $\Delta_u = u^2(1-u^2)\partial_u^2-2u^3\partial_u$.\\
Throughout this discussion we did not yet make explicit use of the flat-space form of the vertex function (dressing propagator). Here, we were mainly interested in a straightforward consistency check. Alternatively, we could have started with expression \eqref{eq:general-Z-flatspace-form} and (together with the discussion of conformally coupled limits for external legs in appendix \ref{app:vertex_functions}) used the alternative representation
\begin{align}
    \begin{split}
        F(u,v) = \frac{1}{8\pi\cos^2\pi\nu}& \sum\limits_{\aa,\bb=\pm}\kappa_{\aa\bb} \int\limits_{-\infty}^{+\infty}\d p\,\d t_1\,\d t_2 \frac{(4p^2)^{\frac12-\nu}}{1+p^2} \\
       &\times \frac{1}{(t_1^2+p^2)^{\frac12-\nu}(t_1+u^{-1}-i\aa\epsilon)^{\frac12+\nu}(t_2^2+p^2)^{\frac12-\nu}(t_2+v^{-1}-i\bb\epsilon)^{\frac12+\nu}}\,,
    \end{split}
    \label{eq:F(u,v) flat space form}
\end{align}
where the coefficients of the integrals are given by $\kappa_{\aa\aa} = (\cos\pi\nu-i\aa)e^{-i\aa\pi\nu}$ and $\kappa_{\aa,-\aa}=-1$. Here, all parametric and momentum integrals are dimensionless and run along the full real axis and the propagators have a form that looks familiar from flat-space Feynman integrals.

\subsection{What Is It Good For?}

Clearly, the massive dressing rules are not as simple to formulate as those for conformally coupled fields. Neither are the auxiliary propagators simple rational functions in "momentum" space. So, one may wonder whether it is worthwhile to reformulate the original SK Feynman rules in this way. The answer depends on what one wants to do with the Feynman integrals. If the purpose is to perform the integration explicitly it is not clear whether the dressing formulation presents an advantage. In the previous subsections we illustrated this for the massive tree-level exchange diagram which can be computed in several different ways. While we recover the expected result, it does not appear to be more efficient to use the dressing formulation. On the other hand, if one is more interested in the analytic properties rather than the explicit computation of the integrals, the dressing formulation may indeed be adequate. This becomes clear if we take the representation \eqref{eq:general_Z_Euler_form} for the dressing propagators which is indeed of the form of general Feynman diagrams \cite{Smirnov:2012gma}. It is thus amenable to the usual procedures, such as Schwinger-, Feynman- or Baikov parametrization, unitarity cuts etc. One complication, though, arises from the oppagator lines where the energy integral is over the positive real line only. This will give rise to extra contributions in the Baikov polynomial for example. On the other hand, unitarity cuts or, more generally, the Landau analysis is not obstructed by that restriction. 

 The form of the dressing integrals makes the dispersive structure of massive cosmological correlators as discussed in \cite{Liu:2024xyi, Liu:2026jzn} manifest as we can directly read off monodromies with regards to line energies $q^2$ at the integrand level. As an illustration, consider again the single-exchange diagram from section \ref{sec:SK_dressing_massive_exchange}. Consider the dressing integrals \eqref{eq:massive-exchange-dressing-integral} of the real and imaginary parts of the vertex function. We use the parity of the imaginary parts to write both integrals in \eqref{eq:massive-exchange-dressing-integral} as integrals along the positive real axis $[0,\infty)$. The vertex functions have the Euler representation \eqref{eq:eu2} from which we infer that the integrand of $Z_+(p,k_{12})$ must have branch points at $p=0$, $p=-is/2$ and $p=-i(k_{12}+s)$ as well as cuts along $(-is/2,0]$ and $(-i\infty, -i(k_{12}+s)]$ on the negative imaginary axis. The conjugate vertex function $Z_-(p,k_{12})$ has branch points and cuts in the upper half plane accordingly. After the $s$-integration, these branch points combine to a single branch cut of the Legendre function that runs along the imaginary $p$-axis at $[-ik_{12},0]$ for $Z_+(p,k_{12})$ and $[0,ik_{12}]$ for $Z_-(p,k_{12})$.\\
 For the dressing integrals we are interested in the real and imaginary parts, $C(p,k_{12})$ and $ S(p,k_{12})$, of the vertex function, which therefore have a branch cut along $[-ik_{12},ik_{12}]$. In the physical region, due to momentum conservation and the triangle inequality, we have $q <k_{12},k_{34}$ and hence the poles of the flat-space propagator at $p=\pm iq$ lie exactly on this cut. 
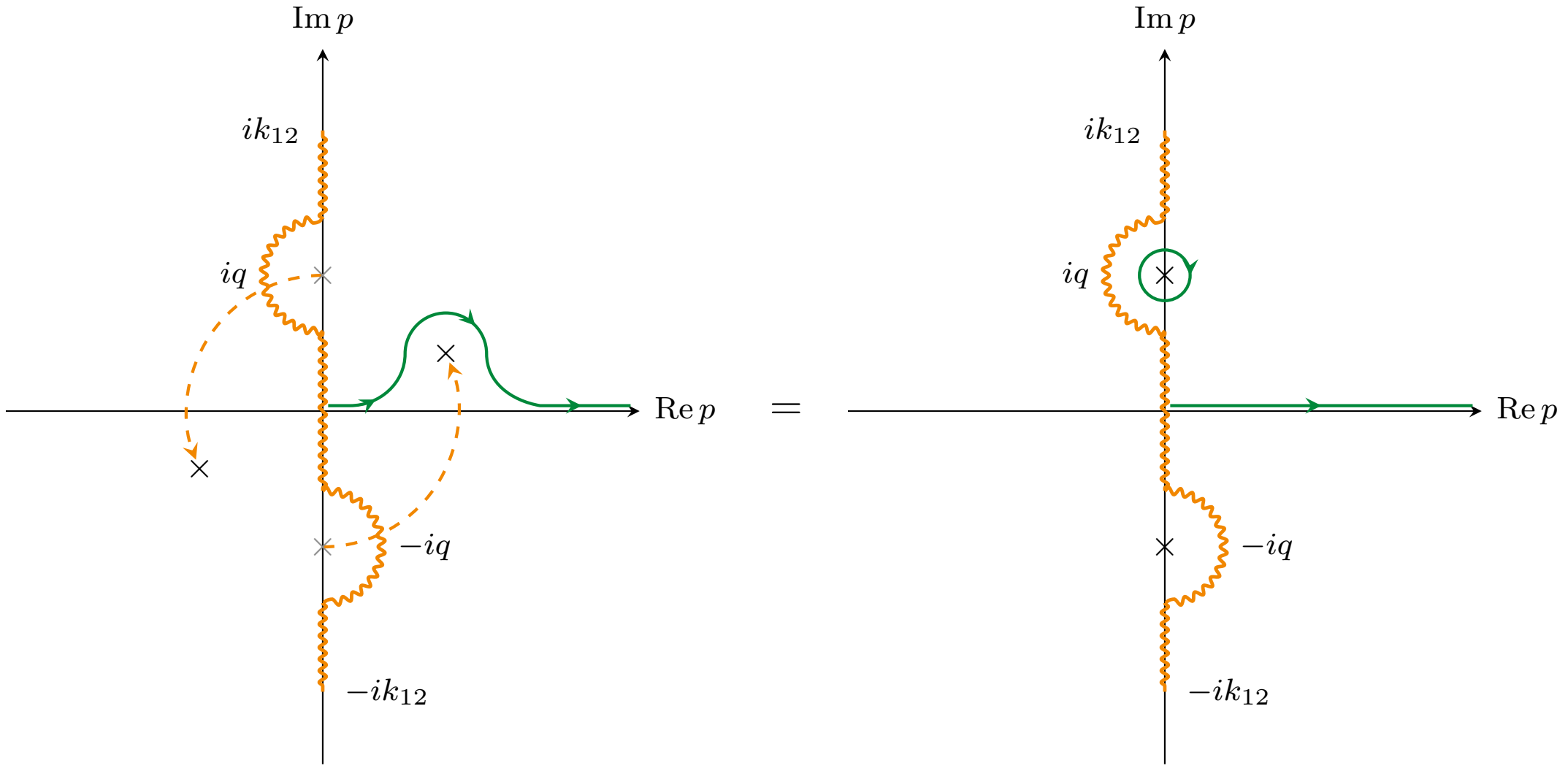
\begin{figure}
    \centering
    \begin{tikzpicture}[scale=0.82,>=stealth,
  axis/.style={->,black},
  cut/.style={lmuorange,line width=0.9pt,decorate,decoration={snake,amplitude=0.9pt,segment length=3pt}},
  contour/.style={lmugreen,thick,postaction={decorate}},
  move/.style={lmuorange,dashed,line width=0.8pt,->},
  pole/.style={black}]

\def\kk{1.5}     
\def\cka{3.1}    
\def\bump{0.62} 
\def\rr{0.45}   
\def\rc{0.28}    
\def\cuts{%
  \draw[cut] (0,0) -- (0,\kk-\bump);   \draw[cut] (0,\kk+\bump) -- (0,\cka);
  \draw[cut] (0,\kk-\bump) arc[start angle=270,end angle=90,radius=\bump];
  \draw[cut] (0,0) -- (0,-\kk+\bump);  \draw[cut] (0,-\kk-\bump) -- (0,-\cka);
  \draw[cut] (0,-\kk+\bump) arc[start angle=90,end angle=-90,radius=\bump];
  \node[left]  at (-0.08,\cka)  {\scriptsize $ik_{12}$};
  \node[right] at (0.08,-\cka)  {\scriptsize $-ik_{12}$};
  \node[left]  at (-\bump-0.05,\kk) {\scriptsize $iq$};
  \node[right] at (\bump+0.05,-\kk) {\scriptsize $-iq$};}
\begin{scope}
\coordinate (pA) at (205:\kk);   
\coordinate (pB) at (25:\kk);    
\draw[axis] (-3.5,0) -- (3.5,0) node[right]{\scriptsize $\operatorname{Re}p$};
\draw[axis] (0,-3.9) -- (0,4.0) node[above]{\scriptsize $\operatorname{Im}p$};
\cuts
\draw[contour,decoration={markings,mark=at position 0.12 with {\arrow{>}},
    mark=at position 0.5 with {\arrow{>}},mark=at position 0.88 with {\arrow{>}}}]
  (0.06,0.06) -- (0.3,0.06)
  to[out=0,in=270] ($(pB)+(180:\rr)$)
  arc[start angle=180,end angle=0,radius=\rr]
  to[out=270,in=170] (2.4,0.06) -- (3.4,0.06);
\draw[move] (0,\kk)  arc[start angle=90, end angle=201,radius=\kk];
\draw[move] (0,-\kk) arc[start angle=270,end angle=381,radius=\kk];
\node[pole] at (pA) {\footnotesize $\times$};
\node[pole] at (pB) {\footnotesize $\times$};
\node[pole, black!45] at (0,\kk)  {\footnotesize $\times$};
\node[pole, black!45] at (0,-\kk) {\footnotesize $\times$};
\node at (0,-4.4) {\small (a) $q\to e^{i\phi}q$, $\phi<\pi$: contour dragged along};
\end{scope}
\node at (4.65,0) {$\qquad=$};
\begin{scope}[xshift=9.3cm]
\draw[axis] (-3.5,0) -- (3.5,0) node[right]{\scriptsize $\operatorname{Re}p$};
\draw[axis] (0,-3.9) -- (0,4.0) node[above]{\scriptsize $\operatorname{Im}p$};
\cuts
\draw[contour,decoration={markings,mark=at position 0.5 with {\arrow{>}}}]
  (0.06,0.06) -- (3.4,0.06);

\draw[contour,decoration={markings,mark=at position 0.25 with {\arrow{>}}}]
  ($(0,\kk)+(90:\rc)$) arc[start angle=90,end angle=-270,radius=\rc];
\node[pole] at (0,\kk)  {\footnotesize $\times$};
\node[pole] at (0,-\kk) {\footnotesize $\times$};
\node at (0,-4.4) {\small (b) $\phi=\pi$: positive real axis and one residue};
\end{scope}
\end{tikzpicture}

    \caption{Contour dragging for the dressing integral \eqref{eq:massive-exchange-dressing-integral} of the tree-level exchange diagram under the monodromy $q\to e^{i\phi}q$. The energy pole at $p=-iq$ rotates to the positive imaginary axis and thereby crosses the integration contour, while the pole at $p=+iq$ does not interact with the contour. When taking the discontinuity, only the residue from the rotated pole will contribute. } 
    \label{fig:monodromy-massive-exchange}
\end{figure}
 We would now like to study the \textit{monodromy} of the $p$-integral. To this end, we continue the correlator $\M$ along the path $q \to e^{i\phi}q$ where $\phi\in[0,\pi]$. Let us take a look at the pole at $p=-iq$: Along this path, it starts moving through the fourth quadrant, crosses the positive real axis and then approaches $p=+iq$ for $\phi=\pi$ from the right half plane. When crossing the positive real axis, the integration contour is dragged with the pole and at $\phi=\pi$ the full contour is equivalent to the initial contour along $[0,\infty)$ plus a small clockwise circle around the pole at $p=+iq$. \\
 There is of course also the second pole at $p=+iq$ which starts to move through the second quadrant, crosses the negative real axis and then approaches $p=-iq$ for $\phi=\pi$ from the left half plane. However, this pole never crosses any integration contour and hence no contour dragging happens.\\
 Along the real axis the integrand is invariant under $q\to -q$, as $q$ only appears in the propagator $\frac{1}{p^2+q^2}$. Thus, when taking the discontinuity of the correlator with respect to $q^2$, the contribution from the integration along the positive real axis will cancel and everything that is left is the residue from the contour around the pole at $p=iq^+$ (the $+$ stresses that we approached from the right, i.e. with positive real part). Thus, the discontinuities for the two contributions are given by

 \begin{align}
 \begin{split}
         \Disc_{q^2} \,\M_\text{prop} &\equiv \M_\text{prop}(e^{i\pi}q)-\M_\text{prop}(q) = \frac{2\pi}{q} \frac{H^4\lambda^2(-\eta_0)^4}{8\pi k_1k_2k_3k_4} S(iq^+,k_{12})S(iq^+,k_{34})\,, \\
         \Disc_{q^2} \,\M_\text{opp} &=\frac{\pi}{q} \frac{H^4\lambda^2(-\eta_0)^4}{8\pi k_1k_2k_3k_4} \cos(\pi\nu)\Big[S(iq^+,k_{12})C(iq^+,k_{34})+C(iq^+,k_{12})S(iq^+,k_{34})\Big] \,.
\end{split}
\end{align}
Remember from section \ref{sec:SK_dressing_massive_exchange} that we can decompose the vertex functions into positive and negative frequency modes $\Z_+$ and $\Z_-$. At $p=iq$, the vertex function acquires real arguments and following the notation in the previous section, we have $Z_+(iq,k_{12}) = \sum_{\aa=\pm} \Z_{\aa}(u^{-1})$. However, there is some care required for the conjugate vertex function because the momentum $p=iq^+$ corresponds to the argument $-u^{-1}$ of the mode function  for which we have to consider the respective connection formula. Then, our prescription dictates $Z_-(iq^+,k_{12}) = Z(-u^{-1}-i\epsilon)$ and hence $Z_-(iq^+,k_{12})= i\sum_{\aa=\pm}e^{i\aa\pi\nu}\Z_\aa(u^{-1})$. Thus

\begin{align}
    \begin{split}
        S(iq^+,k_{12}) &= \frac{1}{2i}\sum\limits_{\aa=\pm} \left(1-i \,e^{i\aa\pi\nu}\right)\Z_\aa(u^{-1})\,,\\
        C(iq^+,k_{12}) &= \frac{1}{2}\sum\limits_{\aa=\pm} \left(1+i \,e^{i\aa\pi\nu}\right)\Z_\aa(u^{-1})\,.
    \end{split}
\end{align}
When looking at the discontinuity of the full correlator $\M=\M_\text{prop}+\M_\text{opp}$ all the mixed terms $\propto \Z_{\aa}(u^{-1})\Z_{-\aa}(v^{-1})$ cancel. Introducing the \textit{non-local signal} of the correlator, 
\begin{equation}
    \M_\text{NL}^{(\aa)} \equiv \frac{H^4\lambda^2(-\eta_0)^4}{16k_1k_2k_3k_4 q} \big(1+\aa\sin(\pi\nu)\big) \Z_\aa(u^{-1})\Z_{\aa}(v^{-1})\,,
\end{equation}
we can write the discontinuity of the correlator as
\begin{equation}
    \Disc_{q^2} \,\M = \sum\limits_{\aa=\pm} \left(e^{2\pi i\aa\nu}-1\right) \M_\text{NL}^{(\aa)}\,.
\end{equation}
This agrees with previous findings from dispersive bootstrap approaches \cite{Liu:2024xyi, Liu:2026jzn}. For more complicated diagrams, where the explicit computations of the vertex functions may not be possible one can still perform the above analysis at the $s$-integrand level. One can also  start from the integral expression for \eqref{eq:massive-exchange-dressing-integral} to derive sum rules \cite{Chowdhury:2026upp} for massive correlators. 

\section{Shadow Dressing Rules} \label{sec:shadow_dressing_rules}

In the previous section, we rewrote the Bessel functions $J_{\pm\nu}$ that appeared in the energy representation in terms of Hankel functions. This resulted in dressing rules that contained propagating as well as oppagator lines. The oppagators forced us to perform the SK sum over clusters of vertices instead of single vertices. Here, we take the other natural route and work with the Bessel functions $J_{\pm\nu}$ of the transition identity \eqref{eq:propagator_as_Jintegral_intermediate} as they are. Consequently, we end up with two different kinds of internal lines that correspond to the two conformal weights $\Delta_\pm =\frac{d}{2}\pm\nu$ as the respective order parameters of the Bessel functions. These conformal weights are the weights of the field and its shadow \cite{DiPietro:2021sjt,Chowdhury:2023arc,Chowdhury:2025ohm, Sleight:2021plv}. We therefore call the resulting rules \textit{shadow dressing rules}.

There is one immediate advantage of the shadow dressing rules compared to the generalized dressing rules in section \ref{sec:SK_dressing_rules}, namely that the SK sum can be carried out at every vertex separately. Moreover, we do not have to deal with the combinatorics of oppagators or clusters. In contrast to conformal coupling, where the two copies combined nicely into a single object, for massive fields we have to sum over the two copies of the fields for each propagator manually.  Still we can formulate dressing rules for general mass in section \ref{subsec:shadow-dressing-complementary}. In section \ref{sec:shadow_dressing_massive_exchange}, we will apply them to the massive exchange once again, for comparison.

\subsection{General Mass} \label{subsec:shadow-dressing-complementary}

The derivation of the shadow dressing rules works similarly as in subsection \ref{sec:SK_dressing_rules_conformal}. As before, we assume $|\Re(\nu)| \leq 1/2$, so that the integrals below converge. Beyond that, we employ dimensional regularization and analytic continuation. We work in the variable $\nu$ which is natural for the complementary series, but everything works the same for principal series fields.

\subsubsection*{Step (0) - Shadow Representation of the Propagator}

Our starting point is the propagator in the energy representation \eqref{eq:propagator_as_Jintegral_intermediate}. In section \ref{sec:SK_dressing_rules_general}, we split the Bessel functions $J_{\pm\nu}$ into Hankel functions via \eqref{eq:J_split_identity}. This moved the SK index into the vertex functions and produced the additional oppagator term in \eqref{eq:general_nu_factorized_propagator}. This time we keep the Bessel functions the way they are so that all SK propagators in \eqref{eq:propagator_as_Jintegral_intermediate} can be combined into the single expression
\begin{align}
\begin{split}
    \mathcal{G}^{i\nu}_{\aa\bb}(qz_1, qz_2) = -\frac{\pi}{2\sin(\pi\nu)} \sum\limits_{\sigma=\pm1} \sigma \int\limits_0^\infty \d p \, \frac{p}{p^2+q^2} &\Big[ e^{-i\aa\frac{\pi}{2}(\frac{d}{2}+\sigma\nu)} J_{\sigma\nu}(pz_1) \Big] \\
    \times &\Big[ e^{-i\bb\frac{\pi}{2}(\frac{d}{2}+\sigma\nu)} J_{\sigma\nu}(pz_2) \Big]. \label{eq:dressing-propagator-sector-sum}
\end{split}
\end{align}
Notice that each term in the sum over $\sigma$ only contains Bessel functions of the single order $\sigma\nu$. Compared to \eqref{eq:general_nu_factorized_propagator}, two additional features stand out: First, the SK indices $\aa$ and $\bb$ only enter through phases belonging to the vertices. Second, there is no homogeneous or oppagator term proportional to $\delta_{\aa\bb}$. 

Due to the scaling $z^{d/2}J_{\sigma\nu}(pz)\sim z^{\frac{d}{2}+\sigma\nu} $ of the eigenfunctions close to the late-time boundary, the two terms can be interpreted as carrying the conformal weight $\Delta_\sigma=\frac{d}{2}+\sigma\nu$, so that the sum over $\sigma$ is actually just the sum over the field and its shadow. 
Thus, \eqref{eq:dressing-propagator-sector-sum} is the energy representation of the two shadow propagators introduced in \cite{DiPietro:2021sjt, Sleight:2021plv}.

As a next step, we need to decide which factors we regard as belonging to the internal line and which terms are moved into the definition of vertex functions in the present description. As in section \ref{sec:SK_dressing_rules}, everything that depends on a time $z$ or on an SK index goes to the vertex. Each end of the line therefore contributes the factor in square brackets in \eqref{eq:dressing-propagator-sector-sum} to the respective vertex. This is our definition of the \textit{shadow propagator}
\begin{align}
    \mathcal{P}_{\sigma,\nu}(p,q) = -\frac{\sigma}{2\sin(\pi\nu)} \frac{1}{p^2+q^2}, \label{eq:dressing-line-rule}
\end{align}
which is integrated with the measure $\int_0^\infty \d p\, p \int \d^d q/(2\pi)^d$, where $q$ is the magnitude of the spatial momentum $\vec q$ of the line. The overall factor $\pi$ in \eqref{eq:dressing-propagator-sector-sum} cancels against the normalization of the two eigenfunctions \eqref{eq:KLF_eigenfunctions} at the ends of the line, which contribute a factor $\pi^{-1/2}$ each to the vertex factor \eqref{eq:KLF_vertex_factor}. Up to the factor $-\sigma/(2\sin(\pi\nu))$, the shadow propagator is the Euclidean flat space propagator. In contrast to section \ref{sec:SK_dressing_rules}, however, the energy $p$ is not directed and is only integrated along the positive real axis. Both ends of the line carry the same energy $p$ and the same Bessel order $\sigma$.

\subsubsection*{Step (1) - Vertex Functions}

We now collect all factors that depend on the time $z$ of a given vertex. As before, the vertex has $N$ legs in total. We denote by $E_v$ the set of its external legs and by $I_v$ the set of internal lines attached to it. The measure of \eqref{eq:KLF_vertex_function}, the Bessel $K$ functions of the external legs and the Bessel $J$ functions from the ends of the internal lines combine into the vertex function
\begin{align}
    \mathcal{I}_v(\{p_l\},\{k_s\}) = \int\limits_0^\infty \d z \, z^{\frac{d(N-2)}{2} - 1} \prod\limits_{s \in E_v} \frac{K_{\nu_s}(k_s z)}{\sqrt{\pi}} \prod\limits_{l \in I_v} J_{\sigma_l \nu_l}(p_l z)\,, \label{eq:dressing-vertex}
\end{align}
which can again be brought into the form of an auxiliary propagator\footnote{The Euler integral representation of \eqref{eq:dressing-vertex} was first obtained in collaboration with Arthur Lipstein.}
\begin{equation}
\begin{aligned}
    \mathcal{I}_v={}&e^{\frac{i\pi}{2}\Lambda_v^{(\sigma)}}\Gamma\big(\Lambda_v^{(\sigma)}\big)
    \prod_{\mathrm{external}\ s}\frac{\Gamma(\frac12-\nu_s)(2k_s)^{-\nu_s}}{2\pi}
    \prod_{\mathrm{internal}\ l}\frac{\Gamma(\frac12-\sigma_l\nu_l)(2p_l)^{-\sigma_l\nu_l}}{2\pi i\sqrt{\pi}}\\
    &\times\int\limits_{\mathbb{R}^N}\d^N t\,
    \frac{\prod\limits_{\mathrm{external}\ s}(t_s^2+k_s^2)^{\nu_s-\frac12}\prod\limits_{\mathrm{internal}\ l}\left[(t_l^2-p_l^2-i\epsilon)^{\sigma_l\nu_l-\frac12}-(t_l^2-p_l^2+i\epsilon)^{\sigma_l\nu_l-\frac12}\right]}
    {(\sum_{j=1}^{N}t_j+i\epsilon)^{\Lambda_v^{(\sigma)}}}.
\end{aligned}
\label{eq:dressing-vertex-euler-form}
\end{equation}
where $\Lambda_v^{(\sigma)}=\frac{d(N-2)}{2}+\sum_{\mathrm{external}\ s}\nu_s+\sum_{\mathrm{internal}\ l}\sigma_l\nu_l$. The derivation is given in appendix \ref{app:vertex_functions}.
This is related to the KLF vertex function \eqref{eq:KLF_vertex_function}, after replacing the Bessel function $K_{i\mu}$ of every internal leg by the Bessel function $J_{\sigma\nu}$ of the energy representation. In fact, these integrals provide exactly the individual positive and negative frequency modes of the vertex functions which were studied in \cite{Grafe:2026avi}. We will see this in more detail in section \ref{sec:shadow_dressing_massive_exchange} when we discuss the massive exchange diagram. The factors $\pi^{-1/2}$ are the normalization factors of the eigenfunctions \eqref{eq:KLF_eigenfunctions} of the external legs. Those of the internal lines have canceled against the propagator in step (0).

In contrast to the vertex integral \eqref{eq:general_z_integral} of the generalized dressing rules, $\mathcal{I}_v$ does not depend on the SK index at all. Instead, it depends on every internal energy $p_l$ and Bessel order $\sigma_l$ separately, and not simply on the net energy flowing into the vertex. The integral converges at large $z$ whenever at least one external leg is attached to the vertex whereas at small $z$, it converges if $\frac{d}{2}(N-2) > \sum_s \Re \,\nu_s - \sum_l \sigma_l \Re\,\nu_l$. For conformally coupled external legs, the Bessel functions reduce to exponentials, $K_{1/2}(kz)/\sqrt{\pi} = e^{-kz}/\sqrt{2kz}$. The vertex function then depends on the external momenta only through $K = \sum_{s \in E_v} k_s$, as in section \ref{sec:SK_dressing_rules_conformal}, and becomes a Laplace transform of a product of Bessel-$J$ functions. 

\subsubsection*{Step (2) - The Schwinger-Keldysh Sum}

The last step is to collect the phases and to perform the SK combinatorics. As in section \ref{sec:SK_dressing_rules}, there are three sources. The vertex factor \eqref{eq:KLF_vertex_factor} contributes $-i\aa e^{i\aa\pi d/2}$, and every internal line contributes the phase $e^{-i\aa\frac{\pi}{2}(\frac{d}{2}+\sigma_l\nu_l)}$ from \eqref{eq:dressing-propagator-sector-sum}. Every external leg contributes the phase $e^{-i\aa\frac{\pi}{2}(\frac{d}{2}-\nu_s)}$ of its late-time limit \eqref{eq:KLF_external}, where, as in section \ref{sec: KLF in dS}, we only keep the leading term as $\eta_0 \rightarrow 0$. Together, they form a single phase $e^{i\aa\Theta_v}$ per vertex, with
\begin{align}
    \Theta_v = \frac{\pi(d-1)}{2} - \frac{\pi}{2} \sum\limits_{s \in E_v} \Big( \frac{d}{2} - \nu_s \Big) - \frac{\pi}{2} \sum\limits_{l \in I_v} \Big( \frac{d}{2} + \sigma_l \nu_l \Big). \label{eq:dressing-vertex-phase}
\end{align}
This angle plays the role of $\psi$ in \eqref{eq:general_psi_phase}. For a vertex without oppagators attached, the two are related by $\Theta_v = \psi - \frac{\pi}{2}\sum_{l \in I_v} (\frac{1}{2} + \sigma_l \nu_l)$. In contrast to $\psi$, the angle $\Theta_v$ therefore depends on the masses of the internal lines and on their respective Bessel orders.

The SK index of a vertex now only appears in the phase $e^{i\aa\Theta_v}$, as neither the vertex function \eqref{eq:dressing-vertex} nor the shadow propagator \eqref{eq:dressing-line-rule} depend on it. The SK sum thus factorizes over the vertices, just like in \eqref{eq:SK_sum_factorization} for conformally coupled fields, and every vertex gives $\sum_{\aa=\pm} e^{i\aa\Theta_v} = 2\cos\Theta_v$. Including the additional factors $\lambda$ and $H$, the SK-summed vertex is
\begin{align}
    \mathcal{V}_v = 2\lambda \mathcal{C} H^\frac{(N-2)(d+1)}{2}\,\cos(\Theta_v) \, \mathcal{I}_v. \label{eq:dressing-sk-summed-vertex}
\end{align}
This factorization is the main advantage of the shadow representation. In section \ref{sec:SK_dressing_rules_general}, the oppagator term of the propagator tied the SK indices of neighboring vertices together, which required us to introduce clusters and the more involved sum \eqref{eq:cluster_SK_sum_explanation}. Hence, instead of the sum over SK indices, we are left with a sum over the $2^L$ assignments of Bessel orders to the $L$ internal lines of the diagram.

\subsubsection*{Feynman Rules for Shadow Dressing Rules}

The following standalone rules summarize the shadow construction and can be used to write down the shadow representation of any given diagram. As in section \ref{sec:SK_dressing_rules}, the diagrams on the right illustrate each step for the tree-level exchange, which has $L=1$ internal line.
\newdressingruleset

\begin{dressingrule}{lmuorange}{Bessel Orders}{\ruleShadowDiagram{1}}\label{rule:shadow:orders}
    Draw the corresponding flat space diagram and assign a \rulekw{Bessel order} $\sigma_l = \pm 1$ to every internal line $l$. Sum over all $2^L$ assignments, where $L$ is the total number of internal lines.
\end{dressingrule}

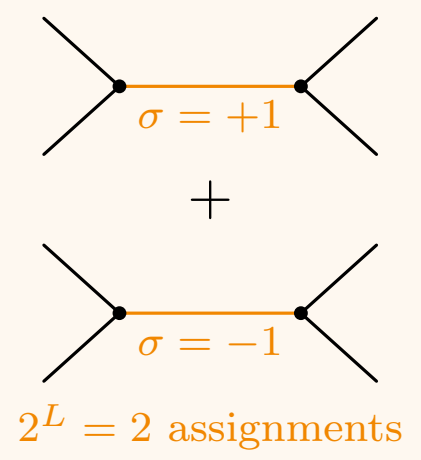
\begin{dressingrule}{lmugreen}{Propagators}{\ruleShadowDiagram{2}}\label{rule:shadow:propagators}
    Assign energies $p_l \geq 0$ and momenta $\vec q_l$ to all internal lines. For every \rulekw{propagator}, write
    \begin{align*}
        \int\limits_0^\infty \d p_l \, p_l \int \frac{\d^d q_l}{(2\pi)^d} \, \mathcal{P}_{\sigma_l,\nu_l}(p_l, q_l),
    \end{align*}
    using the shadow propagator \eqref{eq:dressing-line-rule}. The two ends of a line carry the same energy and the same Bessel order.
\end{dressingrule}

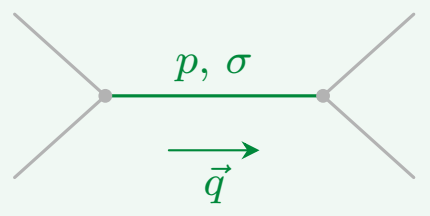
\begin{dressingrule}{lmuorange}{Vertices}{\ruleShadowDiagram{3}}\label{rule:shadow:vertices}
    For every \rulekw{vertex}, write down the SK-summed vertex
    \begin{align*}
        \mathcal{V}_v = 2\lambda H^{\frac{(N-2)(d+1)}{2}} \mathcal{C} \cos(\Theta_v)\,\mathcal{I}_v
    \end{align*}
    from \eqref{eq:dressing-sk-summed-vertex}, using the vertex function $\mathcal{I}_v$ from \eqref{eq:dressing-vertex} or equivalently \eqref{eq:dressing-vertex-euler-form}, and the phase angle $\Theta_v$ from \eqref{eq:dressing-vertex-phase}. Here, $\mathcal{C}$ is the symmetry factor.
\end{dressingrule}

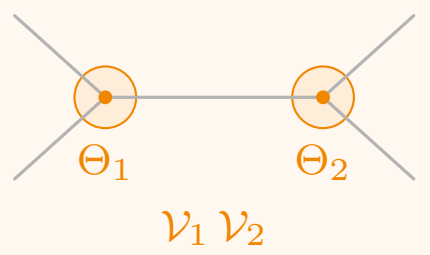
\begin{dressingrule}{lmugreen}{External Legs}{\ruleShadowDiagram{4}}\label{rule:shadow:external}
    For every \rulekw{external leg} with momentum $\vec k_s$ and mass parameter $\nu_s$, multiply by
    \begin{align}
        \mathcal{E}_{\nu_s}(k_s, \eta_0) = H^{\frac{d-3}{2}} \frac{\Gamma(\nu_s)}{2\sqrt{\pi}} \Big( \frac{2}{k_s} \Big)^{\nu_s} (-\eta_0)^{\frac{d}{2}-\nu_s}, \label{eq:dressing-external-leg-rule}
    \end{align}
    which corresponds to \eqref{eq:KLF_external} without the SK phase. As before for principal series fields, we take $\nu \rightarrow i\mu$, and shadow symmetrize with respect to all external mass parameters.
\end{dressingrule}

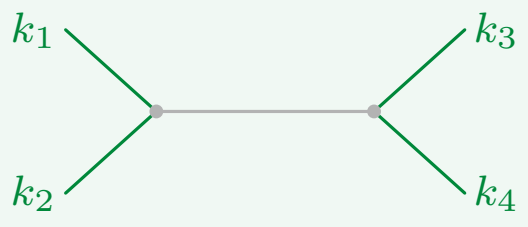
\begin{dressingrule}{lmugreen}{Momentum Conservation and Integration}{\ruleShadowDiagram{5}}\label{rule:shadow:conservation}
    Enforce \rulekw{$d$-momentum conservation} by inserting a delta function $(2\pi)^d \, \delta^{(d)}(\vec k_s + \ldots - \vec q_l + \ldots)$ at every vertex.  \rulekw{Integrate} over all energies $p_l$ and the loop momenta related to the flat space diagram. Do not impose energy conservation. The SK sum has already been taken care of. 
\end{dressingrule}
\vspace{5mm}
In a sense, the shadow dressing rules lie somewhere between the shadow formalism \cite{DiPietro:2021sjt,Sleight:2021plv} and the originally formulated dressing rules. We can think of this as rules for evaluating cosmological correlators in the shadow formulation \cite{Chowdhury:2023arc} with auxiliary propagators instead of time integrals. At fixed shadow assignment, both ends of each internal line carry the same positive energy, so this representation alone does not organize the diagram into directed energy flows as in section \ref{sec:SK_dressing_rules}. Dependence on individual energies alone is not an obstruction, since this also occurs for general masses in section \ref{sec:SK_dressing_rules_general}.

For conformally coupled fields, the sum over Bessel orders can be carried out explicitly. As a result, the vertex functions get combined into the two dressing propagators from table \ref{tab:conformally_coupled_dressings} and energy conservation is restored at each vertex, as we will show in appendix \ref{app:shadow-conformal-graph-proof}. For general masses, the two sets of dressing rules are related by a change of basis as discussed at the end of section \ref{sec:shadow_dressing_massive_exchange}.

\subsection{Massive Exchange Revisited}\label{sec:shadow_dressing_massive_exchange} 

Consider again the massive single exchange diagram from section \ref{sec:SK_dressing_massive_exchange}. Again, the conformally coupled external legs ($\nu_e=\frac12$) are coupled to a massive scalar $\chi$ with mass parameter $\nu$ via the interaction $\L \supset -\frac{\lambda}{2}\sqrt{-g}\phi^2\chi$ and we follow the conventions for the notation of momenta in section \ref{sec:SK_dressing_massive_exchange}. 
Following the definition in \eqref{eq:dressing-external-leg-rule} and setting $d=3$, the external legs contribute a total factor of
\begin{equation}
    \prod_{j=1}^4 \E_{\frac12}(k_j,\eta_0) =\frac{(-\eta_0)^{4}}{4\sqrt{k_1k_2k_3k_4}}\,.
\end{equation}
The vertices themselves contribute additional factors $\V_{L,R}^\sigma =2\lambda H^2 \cos(\Theta_\sigma) \I_{L,R}^\sigma$, where $\Theta_\sigma = -\frac{\pi}{2}(\frac{3}{2}+\sigma\nu)$ follows from \eqref{eq:dressing-vertex-phase} for the cubic vertex and the vertex integrals simplify to 
\begin{equation}
    \I_{L}^\sigma(p,k_1,k_2) =\frac{1}{2\sqrt{k_1k_2}} \int_0^\infty \d z \, z^{-\frac12} e^{-k_{12} z} J_{\sigma\nu}(pz)\,,
\end{equation}
and $k_{1,2} \leftrightarrow k_{3,4}$ for $\I_R^\sigma$. Inserting the propagator from \eqref{eq:dressing-line-rule}, the full correlator takes the compact form 
\begin{equation}
    \M_s=
    \frac{\lambda^2 H^4 (-\eta_0)^{4}}{2\sqrt{k_1k_2k_3k_4}}\int_0^\infty\frac{\d p\,p}{p^2+q^2}
    \sum_{\sigma=\pm1}
    \frac{-\sigma\,\cos^2(\Theta_\sigma)}{\sin(\pi\nu)}
    \mathcal I_L^\sigma\mathcal I_R^\sigma\,.
\end{equation}
For simplicity we discarded the overall momentum-conserving $\delta$-function of the correlator. The vertex integrals are Laplace transforms of Bessel functions. They can be evaluated using for example Mellin-Barnes representations or the method of brackets \cite{Grafe:2026hum}, or by looking them up in \cite[Eq. 6.621.1]{gradshteyn}:
\begin{equation}
    \I_L^\sigma = \frac{e^{i\frac{\pi}{4}(2\sigma\nu+1)}}{\sqrt{2\pi k_1k_2 p}} \,Q_{\sigma\nu-\frac12}\left(\frac{ik_{12}}{p}\right)\,,
\end{equation}
where $Q_{\sigma\nu-\frac12}(z)$ is a Legendre-$Q$ function. Already from here it becomes apparent that the shadow dressing representation simply corresponds to a change of basis on the solution space of the Legendre equation. Similar to section \ref{sec:SK_dressing_massive_exchange}, we can change the integration variable to $x=q/p$ and introduce the momentum ratios $u=q/k_{12}$ and $v=q/k_{34}$. Then, the correlator can be expressed, again, in terms of a dimensionless function $F(u,v)$ of these momentum ratios:
\begin{equation}
    \M_s = \frac{\lambda^2 H^4 (-\eta_0)^4}{16k_1k_2k_3k_4q}\, F(u,v)\,,
\end{equation}
where
\begin{equation}
    F(u,v) = \frac{2i}{\pi}\sum\limits_{\sigma=\pm}\frac{-\sigma(1+\sigma\sin(\pi\nu))}{\sin(\pi\nu)} e^{i\pi\sigma\nu} \int_0^\infty \frac{\d x}{1+x^2} Q_{\sigma\nu-\frac12}\left(\frac{ix}{u}\right)Q_{\sigma\nu-\frac12}\left(\frac{ix}{v}\right)\,.
\end{equation}
One can now proceed by either solving these integrals over the Legendre-$Q$ functions directly, or one can employ the connection formula between $P_\lambda$ and $Q_\lambda$ in order to show that
\begin{equation}
    Q_{\sigma\nu-\frac12} (z) = \frac12\left[Z(-z) -ie^{-i\pi\sigma\nu} Z(z)\right]
\end{equation}
for $\Im \,z >0$ and where $Z(z)$ is the vertex function from section \ref{sec:SK_dressing_massive_exchange}. With these relations, we recover \eqref{eq: massive exchange F(u,v)} and hence the two dressing rules are equivalent. In particular it has an Euler representation of the auxiliary propagator as in \eqref{eq:eu1}. However, it is worthwhile to redo the computation, because the $Q$-basis actually makes some systematics more visible than the $P$-basis. The Legendre-$Q$ functions are (up to a pre-factor $-\sigma/\sin(\pi\nu)$) precisely the positive and negative frequency modes of the vertex functions in the KLF formalism \cite{Grafe:2026avi}. If we introduce exactly these modes
\begin{equation}
    \Z_\sigma (z) = -\frac{\sigma}{\sin(\pi\nu)} \,Q_{\sigma\nu-\frac12}(z)\,,\,\,\,\,\,\,\,\,\,\,\,\, \Z_+(z)+\Z_-(z) = Z(z)\,,
\end{equation}
the full correlator acquires the representation 
\begin{equation}
    F(u,v) = \frac{1}{\pi}\int_{-\infty}^{+\infty} \frac{\d x}{1+x^2} \sum\limits_{\sigma=\pm}(1+\sigma \sin(\pi\nu))\Z_\sigma\left(\frac{ix}{u}\right)\Z_\sigma\left(\frac{ix}{v}\right)\,.
\end{equation}
To unfold the integration contour we made use of the phase shift of the Legendre functions when going to negative arguments, so that $\int_{-\infty}^{+\infty} = (1-e^{2i\pi\sigma\nu})\int_0^\infty$. In addition, the product of mode functions for opposite signs is actually odd, i.e. $\Z_\sigma(-\frac{ix}{u})\Z_{-\sigma}(-\frac{ix}{v})=-\Z_\sigma(\frac{ix}{u})\Z_{-\sigma}(\frac{ix}{v})$. Therefore, when integrating along the full real line the mixed terms vanish and we can add a zero to the integrand as follows:
\begin{equation}
    F(u,v) = \frac{1}{\pi}\int_{-\infty}^{+\infty} \frac{\d x}{1+x^2} Z\left(\frac{ix}{v}\right)\sum\limits_{\sigma=\pm}(1+\sigma \sin(\pi\nu))\Z_\sigma\left(\frac{ix}{u}\right)\,.
\end{equation}
If we compare the integrand to the homogeneous part of the solution in \eqref{eq:massive-exchange-homogeneous}, it is clear that the entire homogeneous solution stems from a single pole of the propagator $\frac{1}{1+x^2}$ when closing the integration contour downwards. For the analytic background we have to consider the branch cut of the vertex functions at $x\in[-iu,0]$ which leads to the same discontinuity integral $I(u,v)$ as in section \ref{sec:SK_dressing_massive_exchange}. 

\vskip 4pt
There are many parallels to the evaluation of the spectral $\mu$ representation. In the dressing rules case the collider signals stem from poles of the spectral function (flat-space propagator) too. The background on the other hand is related to the discontinuity along the branch cut of the Legendre functions. In the spectral gluing algorithm, the double-trace terms stem from poles of the vertex mode functions. One may therefore wonder whether there is some more fundamental connection between these two and whether there is a way to develop a gluing algorithm for dressing integrals as well. 

\section{Conclusions} \label{sec:conclusions}

The main results of this paper are the general mass dressing rules \ref{rule:gen:copies}-\ref{rule:gen:conservation}; and the alternative set of dressing rules via the shadow formulation \ref{rule:shadow:orders}-\ref{rule:shadow:conservation}. Both generalizations of the dressing rules obtained above have some complications when compared to the simple case of conformally coupled fields. The first set from section \ref{sec:SK_dressing_rules} introduces the concept of \textit{oppagators}, which break the global energy conservation structure and also feature a boundary in the energy integral. For this reason, it could prove troublesome to extend some of the flat space machinery to the general case. Despite these problems, per-vertex energy conservation stays, as well as the interpretation of the dressing propagator as lower-dimensional QFT propagator which can be expressed as an Euler integral. We have shown that the analytic structure is rich enough to work with and already computed a first diagram in section \ref{sec:SK_dressing_massive_exchange}.

The second set of dressing rules can be interpreted as an intermediate stage between the shadow formalism and flat space QFT. As such they could also be applied to wave functions rather than correlators. Only for conformally coupled and massless fields does the structure collapse to the form where an interpretation as a flat space  diagram is fully warranted. That being said, they are easier to derive than the first set and the formulation via two kinds of shadow fields may appear as more natural than the other formulation via oppagators and clusters. 

While these rules are not as simple as those for conformal or zero mass \cite{Chowdhury:2025ohm}, they are still reasonably simple to formulate and reveal useful information about the structure of these integrals.  We illustrated this for the massive tree-level exchange. The actual computation of the latter turns out to be easier in the shadow formulation. On the other hand, the calculation of the vertex function $Z$ is universal in the generalized dressing rules for a given number of external legs. In either formulation the auxiliary propagators representing time-dependent vertex functions in the Schwinger-Keldysh formulation can be formulated as Euler integrals making the whole amplitudes amenable to flat space diagrammatic tools such as Landau analysis, monodromies and hopefully differential equations. For conformal coupling this has been laid out in \cite{Sachs:2026gcr}. An extension to the present case is under investigation. 

One central complication of the original dressing rules concerning the UV-regularization of loop integrals has been addressed as well since our formulation is fully valid in $d$ dimensions, so the \textit{m\&d} regularization, as classified in \cite{Jain:2025maa}, where we do dimensional regularization while simultaneously keeping the spectral parameters $\nu$ constant, does not pose a problem. Following this construction guarantees to yield results that are consistent with the bootstrap literature, since we have derived everything from that.

In the process of deriving our dressing rules  we have also clarified the relation between the harmonic- or KLF- representation and dressing- or energy representation by identifying both approaches as different interpretations of the same Sturm-Liouville problem. 

The general challenge that is not new to this formulation, but reflected in it, are the Feynman integrals for  massive diagrams in de Sitter space. The dressing propagators can become complicated integrals themselves as we saw in section \ref{sec:SK_dressing_massive_exchange}. However, we hope that the more familiar structure may pose an invitation to the amplitudes community to push the boundary towards more complicated massive loops such as the massive triangle.

Finally, given the challenges to formulate an effective field theory in de Sitter space \cite{Li:2026xhw, Sun:2026iqd, Dorigoni:2026mbh}, one wonders whether the structural proximity of dressing rules to flat space correlators provides a pathway for importing familiar techniques (e.g. \cite{Beneke:2023wmt}) to make further progress. For this, the flat space version of the dressing rules in appendix \ref{app:flat_space_dressing} may provide a useful bridge between flat space effective field theory and its de Sitter counterpart.

\acknowledgments

We would like to thank Arthur Lipstein for an important discussion at the MIAPbP workshop on primordial cosmology (2026), as well as Chandramouli Chowdhury for comments on the manuscript and for  sharing his unpublished notes with us.  We are also grateful to Johannes Henn and Denis Werth for helpful  discussions. We further acknowledge the use of LLMs, in particular Anthropics Claude Fable 5.1, for streamlining the computations, to help with typesetting the figures, and for proofreading of the final draft. No generated content has been used in this paper without careful review by the authors.

\clearpage

\appendix
\section{Shadow Dressing Rules for Conformally Coupled Fields}
\label{app:shadow-conformal-graph-proof}

We now show that the prescribed shadow dressing rules can be simplified into the auxiliary-propagator form first illustrated in \cite{Chowdhury:2025ohm}. We will show that for $\nu=1/2$, we can combine all diagrams with edge Bessel orders $\sigma_j$ into a choice of at most two auxiliary-propagator structures connected to each vertex. This proof has been written down in \cite{Chowdhury:2025ohm} via loop-flip arguments, which results in very complicated combinatorial arguments. Here, we present a slightly altered version of this proof that uses a factorization trick in the $r_v$ variables introduced below, which we find more readable.

Let $G$ be a connected graph with $V$ vertices, $L$ internal edges, $L-V+1$ loops and $M$ external legs. At vertex $v$, let $E_v$ and $I_v$ be the sets of external and internal legs, $N_v$ the total number of legs and $\lambda_v$ its coupling. Let $\vec q_j$ be the spatial momentum in each internal edge $j$, with $q_j=|\vec q_j|$. We introduce the Bessel order proxy parameter $x_j$
\begin{equation}
    x_j=\frac{1+\sigma_j}{2}. \label{eq:shadow-conformal-proof-half-bessels}
\end{equation}
For conformally coupled fields, $x_j=0$ gives a cosine and $x_j=1$ gives a sine. The same $\sigma_j$, and hence the same $x_j$, sits at both ends of the edge, as it is a property of the propagator and not the vertices.

The Feynman rules for $\nu=1/2$ in section~\ref{sec:shadow_dressing_rules} give us (cf.~\eqref{eq:dressing-line-rule} and~\eqref{eq:dressing-external-leg-rule})
\begin{align}
    \mathcal P_{\sigma_j,1/2}(p_j,q_j) &=-\frac{\sigma_j}{2(p_j^2+q_j^2)}, \label{eq:shadow-conformal-proof-propagator-rule}\\
    \mathcal E_{1/2}(k_s,\eta_0) &=\frac{(-\eta_0)^{(d-1)/2}}{\sqrt{2k_s}} H^\frac{d-3}{2}. \label{eq:shadow-conformal-proof-external-rule}
\end{align}
The remaining vertex factor, with spatial-momentum conservation imposed, is (cf.~\eqref{eq:dressing-vertex} and~\eqref{eq:dressing-sk-summed-vertex})
\begin{align}
    \mathcal V_v ={}&2\lambda_v H^\frac{(N_v-2)(d+1)}{2} \cos\!\left[\psi_v-\frac\pi4 \sum_{j\in I_v}(1+\sigma_j)\right] \notag\\
    &\times\int_0^\infty\d z_v\,z_v^{d(N_v-2)/2-1} \prod_{s\in E_v} \frac{K_{1/2}(k_sz_v)}{\sqrt\pi} \prod_{j\in I_v}J_{\sigma_j/2}(p_jz_v), \label{eq:shadow-conformal-proof-vertex-rule}
\end{align}
where
\begin{equation}
    \psi_v=\frac{\pi(d-1)}4(2-N_v), \qquad \Theta_v=\psi_v-\frac\pi2\sum_{j\in I_v}x_j. \label{eq:shadow-conformal-proof-psi}
\end{equation}
For an edge $j$, we finally find
\begin{equation}
    \d p_j\,p_j \frac{-\sigma_j}{2(p_j^2+q_j^2)} \sqrt{\frac{2}{\pi p_j}}\sqrt{\frac{2}{\pi p_j}} =(-1)^{x_j}\frac{\d p_j}{\pi(p_j^2+q_j^2)}. \label{eq:shadow-conformal-proof-edge-measure}
\end{equation}

\subsubsection*{Vertex Parity}

We now investigate which vertices give cosine-type and which vertices give sine-type integrals. In the original proof of the dressing rules in \cite{Chowdhury:2025ohm}, it was argued that loop-flip arguments, where one singles out one loop of propagators and flips all of their Bessel order assignments from $+$ to $-$ or vice-versa, leaves this \textit{parity} invariant. However, the parity is defined as a property local to any vertex. Without referencing global properties of the graph like loops, we can therefore also define a parity variable $r_v$ at any vertex $v$ via all connected Bessel orders $x_j$
\begin{equation}
    \sum_{j\in I_v}x_j=r_v+2m_v, \qquad r_v=\left(\sum_{j\in I_v}x_j\right)\bmod2, \qquad r_v\in\{0,1\}. \label{eq:shadow-conformal-proof-rm}
\end{equation}
Then $r_v$ decides whether the vertex phase factor is a sine or a cosine:
\begin{equation}
    2\cos\!\left(\psi_v- \frac\pi2\sum_{j\in I_v}x_j\right) =(-1)^{m_v} 2\cos\!\left(\psi_v-\frac\pi2r_v\right). \label{eq:shadow-conformal-proof-local-sign}
\end{equation}
Collecting all factors of $(-1)$ from the propagators and vertices and keeping the vertex phases from the first line of \eqref{eq:shadow-conformal-proof-vertex-rule}, we find an overall
\begin{align}
    (-1)^{\sum\limits_{\mathrm{propagators} \ j} x_j \  + \sum\limits_{\mathrm{vertices} \ v} m_v} \prod\limits_{\mathrm{vertices} \ v} 2\cos\left( \psi_v - \frac{\pi}{2} r_v \right).
\end{align}
Note that the sum over propagators of $x_j$ is half of the sum over vertices $x_v$, since every propagator ends at two vertices, where the $x_v$ is the same. Therefore, we can simplify
\begin{align}
    (-1)^{\sum_j x_j+\sum_v m_v} &=(-1)^{\sum_jx_j+\left(\sum_jx_j-\frac12\sum_vr_v\right)} \notag\\
    &=(-1)^{2\sum_jx_j-\frac12\sum_vr_v} =(-1)^{\frac12\sum_vr_v}. \label{eq:shadow-conformal-proof-fixed-sign}
\end{align}

\subsubsection*{The Sum Over Bessel Orders}

Our claim is that a given parity assignment results in a specific dressing structure, i.e.\ a parity of $0$ at a vertex $v$ gives a dotted propagator and a parity of $1$ a dashed propagator (or vice-versa, depending on the interaction because of differences in notation between this work and \cite{Chowdhury:2025ohm}). To compute this, we evaluate all Bessel order assignments that result in the same parity structure and sum them. We therefore have to perform a \textit{constrained sum} over the $x_j$ such that it gives the desired parity structure. 

Performing constrained summations is difficult and the reason why the original proof in \cite{Chowdhury:2025ohm} resorts to global loop-flips. Instead, we perform a trick to remove the constraint of the sum: We need to find a \textit{selector} that is $1$ if the constraint is fulfilled and $0$ otherwise. Including this selector in the sum allows us to include those $x_j$ in the sum that do not fulfill the constraint and therefore sum over all possible $x_j$.

Fix one parity structure $\{r_v^0\}$ that we want to match with our Bessel order assignments $\{x_j\}$. The constraint then is $r_v^0 - r_v = 0$ for all vertices $v$. The aforementioned selector then can be written as
\begin{align}
    \frac12\sum_{s_v=\pm1}s_v^{\,r_v^0+ \sum_{j\in I_v}x_j} &=\frac{1+(-1)^{r_v^0+\sum_{j\in I_v}x_j}}2 \notag\\
    &=\frac{1+(-1)^{r_v^0+r_v}}2 =\begin{cases}
    1,&r_v=r_v^0,\\
    0,&r_v\ne r_v^0.
    \end{cases}
    \label{eq:shadow-conformal-proof-selector}
\end{align}
Here, $s_v$ is a new variable local to any given vertex.
The relevant piece of the diagram at this point is the Bessel function $J$ inside the vertex function, which in the conformally coupled case reduces to
\begin{align}
    J_{\sigma_j/2}(p_j z) =\sqrt{\frac{2}{\pi p_j z}} \cos\!\left(p_j z-\frac\pi2x_j\right).
\end{align}
Introducing the selector, we can write the constrained sum over these Bessel functions as
\begin{align}
    &\sum\limits_{\{r_v\} = \{r_v^0\}} \prod_{\mathrm{propagator} \ j} \cos\!\left(p_jz_v-\frac\pi2x_j\right) \cos\!\left(p_jz_w-\frac\pi2x_j\right) \notag\\
    &\quad=\frac1{2^V}\sum_{\{s_v=\pm1\}} \left(\prod_vs_v^{r_v^0}\right) \prod_{j}\left[ \sum_{x_j=0,1}(s_vs_w)^{x_j} \cos\!\left(p_jz_v-\frac\pi2x_j\right) \cos\!\left(p_jz_w-\frac\pi2x_j\right)\right]. \label{eq:shadow-conformal-proof-unconstrained-sum}
\end{align}
This can now be done easily:
\begin{align}
    &\sum_{x_j=0,1}(s_vs_w)^{x_j} \cos\!\left(p_jz_v-\frac\pi2x_j\right) \cos\!\left(p_jz_w-\frac\pi2x_j\right) \notag\\
    &\quad=\cos(p_jz_v)\cos(p_jz_w) +s_vs_w\sin(p_jz_v)\sin(p_jz_w) \notag\\
    &\quad=\cos\!\left[p_j(s_vz_v-s_wz_w)\right]. \label{eq:shadow-conformal-proof-edge-sum}
\end{align}

\subsubsection*{Directed Energies and Dressing Propagators}

For each propagator, we now have the final energy integral
\begin{align}
    &\int_0^\infty\frac{\d p_j}{\pi(p_j^2+q_j^2)} \cos\!\left[p_j(s_vz_v-s_wz_w)\right] \notag\\
    &\quad=\int_0^\infty\frac{\d p_j}{2\pi(p_j^2+q_j^2)} \left[e^{ip_j(s_vz_v-s_wz_w)} +e^{-ip_j(s_vz_v-s_wz_w)}\right] \notag\\
    &\quad=\int_{-\infty}^{\infty} \frac{\d p_j}{2\pi(p_j^2+q_j^2)} e^{ip_j(s_vz_v-s_wz_w)}. \label{eq:shadow-conformal-proof-full-line}
\end{align}
The energy variable now runs over the full real line. As before in \eqref{eq:conformal_propagator_p_space}, there is a relative sign in the exponent in the last line, which we can interpret as a direction of the energy flow. We can therefore define $P_v$ as in the main text.

The last remaining piece is the $s_v$ sum, which can be performed easily, as only two terms depend on any given $s_j$: The phase in front of \eqref{eq:shadow-conformal-proof-unconstrained-sum} and the exponential in \eqref{eq:shadow-conformal-proof-full-line}:
\begin{align}
    \frac1{2^V}\sum_{\{s_v=\pm1\}} \left(\prod_vs_v^{r_v^0}\right)e^{i\sum_vs_vP_vz_v} &=\prod_v\left[\frac12\sum_{s_v=\pm1} s_v^{r_v^0}e^{is_vP_vz_v}\right], \label{eq:shadow-conformal-proof-local-sums}
\end{align}
where each individual $s_v$ sum is
\begin{align}
    \frac12\sum_{s_v=\pm1}s_v^{r_v^0}e^{is_vP_vz_v} &=\frac12\left[e^{iP_vz_v}+(-1)^{r_v^0}e^{-iP_vz_v}\right] =\begin{cases}
    \cos(P_vz_v),&r_v^0=0,\\
    i\sin(P_vz_v),&r_v^0=1.
    \end{cases}
    \label{eq:shadow-conformal-proof-local-parity}
\end{align}
Multiplying by the vertex phase gives
\begin{equation}
    2\cos\!\left(\psi_v-\frac\pi2r_v^0\right) \frac12\sum_{s_v=\pm1}s_v^{r_v^0}e^{is_vP_vz_v} =(-i)^{r_v^0}\begin{cases}
    2\cos\psi_v\cos(P_vz_v),&r_v^0=0,\\
    -2\sin\psi_v\sin(P_vz_v),&r_v^0=1.
    \end{cases}
    \label{eq:shadow-conformal-proof-local-phase}
\end{equation}
Together with the signs already collected, the net factor from the sine vertices is
\begin{equation}
    (-1)^{\frac12\sum_vr_v^0}(-i)^{\sum_vr_v^0}=1, \label{eq:shadow-conformal-proof-phase-cancellation}
\end{equation}
as $\sum_v r_v^0$ is always even by definition \eqref{eq:shadow-conformal-proof-rm}, after rearranging the sum over vertices as twice the sum over lines.

It remains to define the two vertex integrals
\begin{align}
    C_v(P_v,K_v)&=\int_0^\infty\d z_v\, z_v^{\frac{N_v(d-1)}2-d-1}e^{-K_vz_v}\cos(P_vz_v), \notag\\
    S_v(P_v,K_v)&=\int_0^\infty\d z_v\, z_v^{\frac{N_v(d-1)}2-d-1}e^{-K_vz_v}\sin(P_vz_v). \label{eq:shadow-conformal-proof-CS}
\end{align}
The two dressings can be read off as
\begin{equation}
    D_v^{(r_v^0)}(P_v,K_v) =\begin{cases}
    2\cos\psi_v\,C_v(P_v,K_v),&r_v^0=0,\\
    -2\sin\psi_v\,S_v(P_v,K_v),&r_v^0=1.
    \end{cases}
    \label{eq:shadow-conformal-proof-dressings}
\end{equation}
This agrees exactly with the construction in section \ref{sec:SK_dressing_rules_conformal}.

\section{Integral Representations of Vertex Functions}\label{app:vertex_functions}

Here we will provide the derivation of \eqref{eq:general-Z-flatspace-form} starting from the time integral representation of the vertex function, i.e. the definition of $Z_\aa$ as a multiple-$K$ integral (see also \cite{Grafe:2026hum} for related discussions). Re-inserting the modified Bessel functions into \eqref{eq:general_z_integral}, we find
\begin{equation}
    Z_\aa(\{\sigma_lp_l\},\{k_s\}) = \int\limits_0^\infty \d z\,z^{\frac{d(N-2)}{2}-1} \prod_{\text{legs } j} \sqrt{\frac{2\beta_j}{\pi}} K_{\nu_j}(\beta_j z)\,.
\end{equation}
We will insert Basset's integral representation
\begin{equation}
    K_{-\nu}(\beta z) = \frac{\Gamma(\frac12-\nu)(2\beta)^{-\nu}}{\sqrt{\pi}\,z^{-\nu}} \int\limits_0^\infty \d t\, \frac{\cos (zt)}{(t^2+\beta^2)^{\frac12-\nu}}\,,
\end{equation}
that is valid for $\Re \,\nu <\frac12$ and $\Re\,\beta > 0$ and use $K_\nu(z) = K_{-\nu}(z)$. The cosine can be written as a sum of two exponential functions which allows us to extend the integration contour to the full real line:
\begin{equation}
    \sqrt{\frac{2\beta}{\pi}} K_\nu(\beta z) = \frac{\Gamma(\frac12-\nu)(2\beta)^{\frac12-\nu} z^\nu}{2\pi} \int\limits_{-\infty}^{+\infty} \d t \, (t^2+\beta^2)^{\nu-\frac12} e^{izt}\,.
\end{equation}
For the external legs, this representation can be used directly. However, for internal legs the parameter $\beta = -i\aa \sigma p$ is imaginary and hence sits at the boundary of the validity range with $\Re\,\beta =0$. We thus introduce an $i\epsilon$-prescription by first setting $\beta = -i\aa \sigma p + \delta$ with $\delta\to 0$ and let $t^2+\beta^2 = t^2-p^2 -i\aa \sigma \epsilon$. Then, around $t=\pm p$ the integrand behaves as $|t^2- p^2|^{\nu-\frac12}$ which is integrable for $\Re\,\nu > -\frac12$. \\
The remaining step is to interchange the parametric integrals with the time integral over $z$. To do so, we have to introduce a small damping factor $e^{-\epsilon z}$ with $\epsilon > 0$. Since the initial time integral converges absolutely, this modification can change the integral only by a term that will vanish when taking the limit $\epsilon\to 0$ at the end. Collecting all powers of $z$ into $\Lambda_v = \frac{d(N-2)}{2}+\sum \nu_j$ leaves us with 
\begin{equation}
    \int\limits_0^\infty \d z \, z^{\Lambda_v -1} e^{-z(\epsilon - i\sum_j t_j)} =\Gamma(\Lambda_v) \left(\epsilon -i\sum_j t_j\right)^{-\Lambda_v} = e^{\frac{i\pi}{2}\Lambda_v}\Gamma(\Lambda_v)\left(\sum_j t_j +i\epsilon\right)^{-\Lambda_v}\,.
\end{equation}
Combining this with the parametric integrals, we arrive at
\begin{equation}
    Z_\aa = e^{\frac{i\pi}{2}\Lambda_v}\Gamma(\Lambda_v) \prod_{\mathrm{legs} \ j} \frac{\Gamma(\frac12-\nu_j)(2\beta_j)^{\frac12-\nu_j}}{2\pi} \int\limits_{\mathbb{R}^N} \d^N t\, \frac{\prod\limits_{\mathrm{legs} \ j^\prime} (t_{j^\prime}^2+\beta_{j^\prime}^2)^{\nu_{j^\prime}-\frac12}}{\Big(\sum\limits_{\mathrm{legs} \ j^\prime} t_{j^\prime} +i\epsilon\Big)^{\Lambda_v}}\,.
\end{equation}
Let us quickly investigate what happens when we consider conformally coupled external legs with $\nu_s=-\frac12$. In this case, the propagator $(t^2+k_s^2)^{\nu_s-\frac12} =(t^2+k_s^2)^{-1}$ does not have a branch cut anymore, but instead has simple poles at $t=\pm i k_s$. We can close the contour of Basset's integral and pick up the remaining integrand at the location of the pole, which is just another contribution to the exponential factor. This is of course equivalent to directly plugging the plane wave representation of $K_\frac12(z)$ into the time integral and evaluating it as part of the $z$-integral directly. This will simply shift the total-energy pole $(\sum_j t_j +i\epsilon)$ by the respective energy of the conformally coupled field.\\
Explicitly, for $d=3$ and the massive exchange discussed in sections \ref{sec:SK_dressing_massive_exchange} and \ref{sec:shadow_dressing_massive_exchange}, the vertex functions read
\begin{equation}
    Z_\aa = \Gamma(\tfrac12+\nu)\Gamma(\tfrac12-\nu)\frac{(-2i\aa\sigma p)^{\frac12-\nu}}{2\pi} \int\limits_{-\infty}^{+\infty} \d t\frac{(t^2-p^2 -i\aa\sigma \epsilon)^{\nu-\frac12}}{(k_{12}-it)^{\nu+\frac12}}\,.
\end{equation}
When rotating the contour $t\to i t$ and simplifying the pre-factors, we get
\begin{equation}
    Z_\aa = \frac{(2p)^{\frac12-\nu}(-i\aa\sigma)^{\frac12+\nu}}{2\cos \pi\nu} \int\limits_{-\infty}^{+\infty} \d t\frac{(t^2+p^2)^{\nu-\frac12}}{(t+k_{12}-i\aa\sigma\epsilon)^{\nu+\frac12}}\,.
\end{equation}
This representation was used to obtain the form \eqref{eq:F(u,v) flat space form} of the full tree-level exchange correlator.

The same construction applies to the shadow vertex function \eqref{eq:dressing-vertex}, using $J_{\sigma\nu}(pz)=\frac{1}{\pi i}\left[e^{-i\pi\sigma\nu/2}K_{\sigma\nu}(-ipz)-e^{i\pi\sigma\nu/2}K_{\sigma\nu}(ipz)\right]$. Applying Basset's representation with $\beta=\delta\mp ip$, the phases cancel as $\delta\to0^+$, giving

\begin{equation}
    J_{\sigma\nu}(pz)=\frac{\Gamma(\frac12-\sigma\nu)(2p)^{-\sigma\nu}z^{\sigma\nu}}{2\pi i\sqrt{\pi}}\int\limits_{-\infty}^{+\infty}\d t\,e^{izt}\left[(t^2-p^2-i\epsilon)^{\sigma\nu-\frac12}-(t^2-p^2+i\epsilon)^{\sigma\nu-\frac12}\right].
    \label{eq:shadow-basset-J}
\end{equation}
For the external legs, $\beta_s=k_s$, and collecting the powers of $z$ results in $\Lambda_v^{(\sigma)}=\frac{d(N-2)}{2}+\sum_{ s}\nu_s+\sum_{l}\sigma_l\nu_l$, so that after evaluating the $z$ integral we are left with
\begin{equation}
\begin{aligned}
    \mathcal{I}_v={}&e^{\frac{i\pi}{2}\Lambda_v^{(\sigma)}}\Gamma\big(\Lambda_v^{(\sigma)}\big)
    \prod_{\mathrm{external}\ s}\frac{\Gamma(\frac12-\nu_s)(2k_s)^{-\nu_s}}{2\pi}
    \prod_{\mathrm{internal}\ l}\frac{\Gamma(\frac12-\sigma_l\nu_l)(2p_l)^{-\sigma_l\nu_l}}{2\pi i\sqrt{\pi}}\\
    &\times\int\limits_{\mathbb{R}^N}\d^N t\,
    \frac{\prod\limits_{\mathrm{external}\ s}(t_s^2+k_s^2)^{\nu_s-\frac12}\prod\limits_{\mathrm{internal}\ l}\left[(t_l^2-p_l^2-i\epsilon)^{\sigma_l\nu_l-\frac12}-(t_l^2-p_l^2+i\epsilon)^{\sigma_l\nu_l-\frac12}\right]}
    {\left(\sum\limits_{j=1}^{N}t_j+i\epsilon\right)^{\Lambda_v^{(\sigma)}}}.
\end{aligned}
\label{eq:shadow-vertex-flatspace-form}
\end{equation}

\section{Flat Space Dressing Rules} \label{app:flat_space_dressing}

The computation of correlation functions, and by extension the application of the in-in formalism to them, is usually directly connected to de Sitter QFT. However, it can be pedagogical to also consider the same techniques in flat space in order to investigate aspects of the formalism in a simpler and more well-understood framework. In this appendix, we briefly apply our entire construction, from the Sturm-Liouville problem to the final dressing rules, to flat space QFT.

The EoM in flat space in partial momentum space (i.e.\ in the variables $t \in (-\infty, 0)$ for time, where w.l.o.g. $t = 0$ is the time at which the correlators are evaluated, and $\vec k$ for spatial momentum) is
\begin{align}
    \partial_t^2 \phi = -(k^2 + m^2) \phi.
\end{align}
Technically, the same distinction of mass $m$ and momentum $k$ as a parameter vs.\ as an eigenvalue exists here. However, since the potential and weight functions are both simply constants, this distinction is manifestly degenerate and it does not make a difference which one is chosen as what. The eigenfunctions are therefore always
\begin{align}
    \phi_{k, m} = e^{\pm i \sqrt{k^2 + m^2} t}, \label{eq:flat_space_mode_massive}
\end{align}
as known from every QFT textbook. From now on, we follow the same structure as section \ref{sec:SK_dressing_rules_conformal} for the derivation. We choose the unique vacuum state at $t \rightarrow -\infty$ to fix the sign.

\subsubsection*{Step (0) - Energy Representation}

With the mode functions \eqref{eq:flat_space_mode_massive}, the SK propagators and Wightman functions in Euclidean signature are
\begin{subequations}
\begin{align}
    \mathcal{G}_{\aa\aa} &= \frac{1}{2\sqrt{q^2 + m^2}} e^{-\sqrt{q^2 + m^2} |t_1 - t_2|}, \\
    \mathcal{G}_{\aa,(-\aa)} &= \frac{1}{2\sqrt{q^2 + m^2}} e^{-\sqrt{q^2 + m^2} |t_1 + t_2|}.
\end{align}
\end{subequations}
The above expression looks unusual since $t < 0$. Just like in section \ref{sec:SK_dressing_rules_conformal}, both can be combined into a single form under a $p$ integral
\begin{align}
    \mathcal{G}_{\aa\bb} = \int\limits_{-\infty}^\infty \frac{\d p}{2\pi} \frac{e^{i\aa p t_1}e^{-i\bb p t_2}}{p^2+q^2+m^2}.
\end{align}
Up to the mass in the denominator, some phases and the conformal time factors, this is exactly the same propagator as \eqref{eq:conformal_propagator_p_space}. This should not be surprising, since conformally coupled fields in de Sitter are essentially massless fields in flat space. Since the entire propagator in the above expression is directed, there are no oppagators in flat space, even for general masses.

External propagators are Wightman functions. The external half-leg is a constant, since we evaluate at $t=0$, where the expression is regular. The internal half-leg gives a falling exponential, like for conformally coupled fields in de Sitter.

\subsubsection*{Step (1) - Performing the $t$ Integral}

The time integrals are particularly simple in flat space, since
\begin{align}
    Z_\aa(P, K) &= \int\limits_{-\infty}^0 \d t \prod\limits_{\mathrm{external} \ s} e^{\sqrt{k_s^2 + m_s^2} t} \prod\limits_{\mathrm{internal} \ l} e^{i\aa\sigma_l p_l t} = \frac{1}{K + i \aa P}, \label{eq:flat_space_Z}
\end{align}
where we used the adapted definitions of $K = \sum_s \sqrt{k_s^2 + m_s^2}$ and $P = \sum_l \sigma_l p_l$ as before. There are no SK phases, so the dressing propagator is already complete.

\subsubsection*{Step (2) - The SK Sum}

The SK sum is trivial, as we simply sum over $\aa$ in \eqref{eq:flat_space_Z}, which gives
\begin{align}
    \sum\limits_{\aa = \pm} Z_\aa(P,K) = \frac{2K}{K^2 + P^2}.
\end{align}
This is the one and only dressing in flat space. We could have also guessed this, since the $\phi^4$ coupling is the only truly conformal self-coupling and therefore flat space should inherit exactly its dressing.

\subsubsection*{Step (3) - Dressing Propagators}

We perform the same shuffling of terms as in de Sitter and write the vertex function as a propagator such that energy is conserved at the vertices. The energy is also automatically conserved at the dressing point.

\subsubsection*{Flat Space Dressing Feynman Rules}

With this, we can formulate the flat space dressing Feynman rules for correlation functions:
\begin{enumerate}
    \item Draw the flat space diagram.
    \item Write a massive Euclidean flat space propagator for every line: $\int\d^{d+1}Q/(2\pi)^{d+1} \frac{1}{Q^2 + m^2}$.
    \item Attach a dressing line to every vertex, such that they meet at the dressing point: $\int_{-\infty}^\infty \d P/(2\pi) \frac{2K}{K^2 + P^2}$.
    \item Write the factor $1/(2\sqrt{k^2 + m^2})$ for every external leg and the usual factor $\lambda \, \mathcal{C}$ for every vertex.
    \item Enforce $d$-momentum conservation at every vertex, and enforce energy conservation at every vertex, writing the respective delta functions and considering the relevant line types.
\end{enumerate}
These Feynman rules may be useful when comparing specific computations between flat space and de Sitter space, as they present correlators on the same footing. Similar formulations have already been derived in \cite{Donath:2024utn,Arkani-Hamed:2025mce}.

\bibliographystyle{JHEP}
\bibliography{references.bib}

\end{document}